\documentclass[acmsmall]{acmart}
\usepackage{subfigure}
\usepackage{lscape}
\usepackage{threeparttable}
\usepackage{multirow}
\AtBeginDocument{%
  \providecommand\BibTeX{{%
    \normalfont B\kern-0.5em{\scshape i\kern-0.25em b}\kern-0.8em\TeX}}}

\setcopyright{acmcopyright}
\copyrightyear{2022}
\acmYear{2022}
\acmDOI{01.1234/1234567.1234567}

\acmJournal{CSUR}
\acmVolume{00}
\acmNumber{0}
\acmArticle{11}
\acmMonth{8}

\begin{document}

\title{A Survey on Perceptually Optimized Video Coding}

\author{Yun Zhang}
\email{zhangyun2@mail.sysu.edu.cn}
\affiliation{%
  \institution{School of Electronics and Communication Engineering, Sun Yat-Sen University}
  \city{Shenzhen}
  \state{Guangdong}
  \country{China}
  \postcode{518107}
}
\author{Linwei Zhu}
\email{lw.zhu@siat.ac.cn}
\affiliation{%
  \institution{Shenzhen Institutes of Advanced Technology, Chinese Academy of Sciences}
  \city{Shenzhen}
  \state{Guangdong}
  \country{China}
  \postcode{518055}
}

\author{Gangyi Jiang}
\affiliation{%
  \institution{Faculty of Information and Engineering, Ningbo University}
  \city{Ningbo}
  \state{Zhejiang}
  \country{China}
  \postcode{315211}
  }
\email{jianggangyi@nbu.edu.cn}

\author{Sam Kwong}
\affiliation{%
  \institution{Department of Computer Science, City University of Hong Kong}
  \city{Hong Kong}
  \country{China}
}
\email{cssamk@cityu.edu.hk}

\author{C. -C. Jay Kuo}
\affiliation{%
 \institution{University of Southern California}
 \city{Los Angeles}
 \state{California}
 \country{USA}}
\email{cckuo@sipi.usc.edu}

\thanks{This work was supported in part by the National Natural Science Foundation of China under Grants 62172400, 61901459 and 62271276, in part by the Shenzhen Science and Technology Program under Grant JCYJ20200109110410133, in part by the Guangdong Natural Science Foundation under Grant 2022A1515011351, in part by the CAS President's International Fellowship Initiative (PIFI) under Grant 2022VTA0005, in part by the Hong Kong Innovation and Technology Commission (InnoHK Project CIMDA), in part by the Hong Kong GRF-RGC General Research Fund under Grants 11209819 (CityU 9042816) and 11203820 (CityU 9042598).}

\authorsaddresses{Authors' addresses: Yun Zhang (corresponding author), zhangyun2@mail.sysu.edu.cn, School of Electronics and Communication Engineering, Sun Yat-Sen University, Shenzhen, Guangdong, China, 518107; Linwei Zhu, lw.zhu@siat.ac.cn, Shenzhen Institutes of Advanced Technology, Chinese Academy of Sciences, Shenzhen, Guangdong, China, 518055; Gangyi Jiang, Faculty of Information and Engineering, Ningbo University, Ningbo, Zhejiang, China, 315211, jianggangyi@nbu.edu.cn; Sam Kwong, Department of Computer Science, City University of Hong Kong, Hong Kong, China, cssamk@cityu.edu.hk; C.C. Jay Kuo, University of Southern California, Los Angeles, California, USA, cckuo@sipi.usc.edu.}

\renewcommand{\shortauthors}{Y. Zhang et al.}

\begin{abstract}
 To provide users with more realistic visual experiences, videos are developing in the trends of Ultra High Definition (UHD), High Frame Rate (HFR), High Dynamic Range (HDR), Wide Color Gammut (WCG) and high clarity. However, the data amount of videos increases exponentially, which requires high efficiency video compression for storage and network transmission. Perceptually optimized video coding aims to maximize compression efficiency by exploiting visual redundancies. In this paper, we present a broad and systematic survey on perceptually optimized video coding. Firstly, we present problem formulation and framework of the perceptually optimized video coding, which includes visual perception modelling, visual quality assessment and perceptual video coding optimization. Secondly, recent advances on visual factors, computational perceptual models and quality assessment models are presented. Thirdly, we review perceptual video coding optimizations from four key aspects, including perceptually optimized bit allocation, rate-distortion optimization, transform and quantization, filtering and enhancement. In each part, problem formulation, working flow, recent advances, advantages and challenges are presented. Fourthly, perceptual coding performances of the latest coding standards and tools are experimentally analyzed. Finally, challenging issues and future opportunities are identified.
\end{abstract}

\ccsdesc[500]{Computing methodologies~Image compression}
\ccsdesc[500]{General and reference~Surveys and overviews}

\keywords{Perceptual video coding, quality assessment, human visual system, visual attention, just noticeable difference, rate-distortion optimization}

\maketitle

\section{Introduction}
With the development of capturing, display and computing technology, video applications are developing in the trends of providing more realistic and immersive visual experiences. For example, they develop from Standard Definition (SD) to High Definition (HD) and 4K/8K Ultra-HD (UHD) for higher spatial resolution, from low dynamic range, color gammut and frame rate to High Dynamic Range (HDR), Wide Color Gammut (WCG), and High Frame Rate (HFR) for higher color and temporal fidelity, from 2D to stereo, multiview and 3D for providing depth perception, and from low clarity to high clarity. In addition, Virtual Reality (VR) and Augmented Reality (AR) applications based on volumetric video are also boosting as they provide more immersive visual experiences and interactive functionalities between real and virtual objects. However, data volume of these videos increases to dozens or even hundreds of times, which has been a critical challenge for video transmission/streaming \cite{SV_DNN}, storage and computing. To tackle this problem, developing highly effective video coding algorithms that compress videos into smaller ones is highly desired.


 The worldwide researchers and organizations have made significant contributions to the development of video coding technologies. In the past three decades, video coding standards have been developed for four generations, where leading standards in each generation are MPEG-1/2, H.264/Advanced Video Coding (AVC), High Efficiency Video Coding (HEVC) \cite{OV_HEVC} and Versatile Video Coding (VVC) \cite{OV_VVC}, respectively. In 2003, experts from ITU-T Video Coding Experts Group (VCEG) and ISO/IEC Moving Picture Experts Group (MPEG) established Joint Video Team (JVT) to develop the H.264/AVC, which has been widely used in SD/HD video applications nowadays. Since 2013, VCEG and MPEG gathered together again and founded Joint Collaborative Team on Video Coding (JCT-VC) to standardize HEVC for UHD video, which doubled the compression ratio as compared with H.264/AVC. Meanwhile, 3D/multiview, scalable and screen content extensions of HEVC, called 3D/MV-HEVC\cite{OV_3DVC}, Scalable HEVC (SHVC) and Screen Content Coding (SCC), respectively, were investigated for specific video applications. Beyond HEVC, Joint Video Experts Team (JVET) was established to standardize VVC \cite{OV_VVC_tools}, which targeted to encode 8K UHD video and beyond. To include a variety of video sources and applications, JVET also launched standardization activities for representation and coding of immersive media, including $360^\circ$ omnidirectional video coding and point cloud compression \cite{OV_IVC}.
Since 2003, Audio Visual coding Standard (AVS) workgroup was established in China to develop video coding standards, called AVS-1/2/3 \cite{OV_AVS}. In the past few decades, many advanced coding technologies have been developed to improve the compression efficiency further and further, which becomes saturated nowadays. A little further improvement may require extremely high cost, which is more challenging than ever.

To compress videos more effectively, perceptual video coding is one of the most promising research directions, which tends to minimize visual redundancies in videos by exploiting properties of Human Visual System (HVS). This is a systematic research not only includes image/video signal processing, but also involves video representation, psychovisual and neurophysiological researches. World-wide researchers have devoted their efforts on this perceptual video coding field. Wu \emph{et al.} \cite{SV_JND} presented a survey on visual Just Noticeable Difference (JND) estimation that affected by luminance adaptation, contrast masking, pattern masking, and visual sensitivity. Lin \emph{et al.} \cite{SV_JND_Lin} reviewed handcrafted modeling and machine learning approaches for JND.
Lin \emph{et al.} \cite{SV_PVQA} presented an overview on Perceptual Visual Quality Metrics (PVQM), which included basic computational models for key factors of human perception, model based and signal driven PVQM models. Athar \emph{et al.} \cite{SV_IQA} presented a survey on Image Quality Assessment (IQA) metrics for 2D images and reported their performances over different datasets. Min \emph{et al.} \cite{SV_SCIQA} analyzed characteristics of screen content images from human, system, and context perspectives, and then reviewed quality assessment methods and datasets for them.
However, these overviews focus on perceptual models for image/video quality assessment, but video coding optimizations have not been considered. Zhang \emph{et al.} \cite{SV_MLVC} reviewed machine learning based video coding optimizations from three perspectives, including low complexity, high compression ratio and high visual quality. To further improve video compression efficiency, Liu \emph{et al.} \cite{SV_DLVC} reviewed representative Deep Learning based Video Coding (DLVC) schemes for key coding modules, including deep intra/inter prediction, cross channel prediction, transform, in/post-loop filtering, probability prediction and up/down-sampling, an so on. They utilized learning algorithms, especially deep learning, to promote video coding performances. However, visual redundancies have not been well considered.
Yuan \emph{et al.} \cite{SV_PQA_VC} reviewed recent advances on visual JND models and their applications to Rate-Distortion Optimization (RDO) and quantization in video coding, where machine learning algorithms were used to improve JND prediction. Lee \emph{et al.} \cite{SV_PVC} reviewed perceptual video compression from perceptual model, coding implementation and performance evaluation. Chen \emph{et al.} \cite{SV_PVC2} briefly reviewed visual attention and visual sensitivity factors, which were used to guide perceptual quality allocation for constrained video coding. However, these works mainly focused on H.264/AVC. Perceptual factors on UHD/HDR/WCG video and coding algorithms on more recent HEVC/VVC/AVS standards have not been addressed. Also, learning based optimizations have not been addressed.

Therefore, an in-depth survey on recent advances of perceptual video coding is highly desirable at this time due to the following five reasons: 1) Since video representation develops in the trends of providing more realistic visual experiences, such as HD/UHD and 3D, visual properties are different. In-depth analyses are required for possible extensions of existing perceptual models in improving their applicabilities. 2) More advanced perceptual factors and mechanism have been revealed with the developments of neural and computer sciences. Consequently, the perceptual models and quality assessment methods have been significantly advanced. In addition, the concept of perceptual quality has been extended from clarity to a wider Quality-of-Experience (QoE). 3) Computational models in modelling visual factors and assessing visual quality have been significantly improved, which can be used to exploit visual redundancies in videos. 4) In the latest coding standards, such as AVS-3 and HEVC/VVC, many novel coding algorithms have been proposed. It becomes more challenging to further improve the latest coding technologies in exploiting the visual redundancies. 5) Machine learning algorithms, especially the deep learning, have been explored in enhancing video coding algorithms\cite{SV_MLVC}\cite{SV_DLVC}, which bring new opportunities to boost perceptual coding optimizations with learning techniques. By taking these opportunities, an in-depth review on perceptual video coding will provide helpful guidelines for future researches in promoting video coding.


In this survey, we firstly analyze visual properties of HVS from neurophysiological perspectives, and review computational perceptual models and visual quality assessment algorithms. Then, we review on recent advances of perceptually optimized video coding algorithms, where problem formulation, key features, performances, advantages and disadvantages are analyzed. Thirdly, coding performance of latest coding standards are experimentally analyzed. Finally, challenges and opportunities in perceptual coding are identified.

\section{Problem Formulation and Framework of Perceptual Video Coding}
\subsection{Problem Formulation}
\begin{figure*}[!t]
	\centering
	\includegraphics[width=1.0\linewidth]{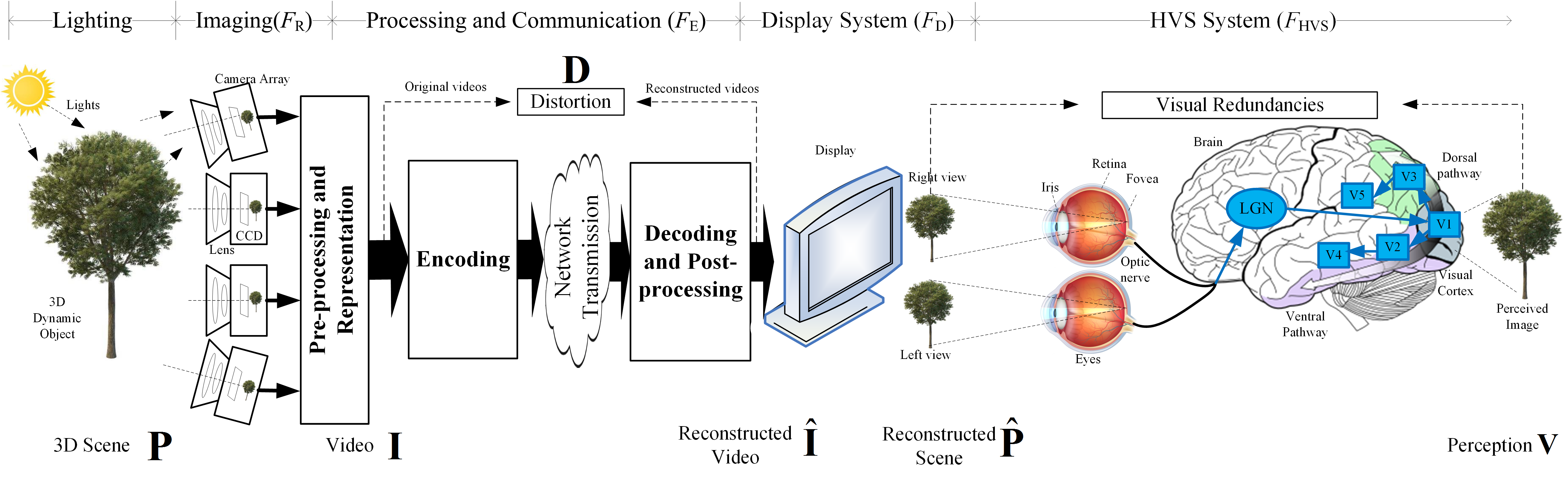}

	\caption{End-to-end chain of a visual communication system.}
	\label{fig:Framework}
\end{figure*}

\begin{figure}[!t]
	\centering
	\includegraphics[width=0.7\linewidth]{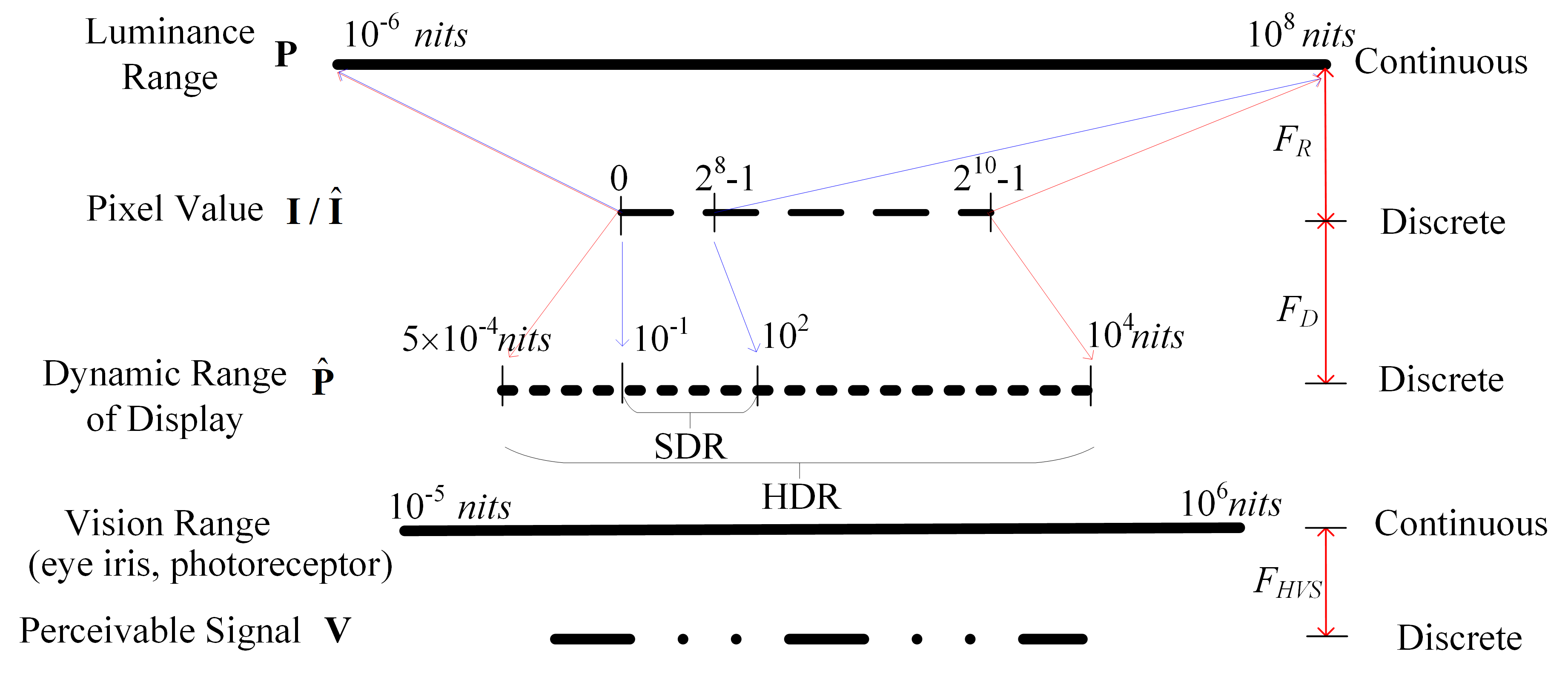}	
	\caption{Luminance dynamic range mappings among nature light, digital signal, display and HVS.}
	\label{fig:HDRBitMap}
\end{figure}

Fig.\ref{fig:Framework} shows an end-to-end chain of a visual communication system, which consists of five key components, light and scene $\mathbf{P}$, imaging and representation ($F_R$) \cite{Imaging}, visual processing and communication $(F_E)$, display and viewing condition $(F_D)$, and HVS $(F_{HVS})$.
Dynamic scene ($\mathbf{P}$) in 3D world is firstly captured by cameras with different view angles or positions. Then, captured successive images are processed and organized to be an effective video representation, denoted as $\mathbf{I}=F_R(\mathbf{P})$. $F_R$ is an Opto-Electronic Transfer Function (OETF) \cite{HDR_Francois_CSVT20} that converts captured lights into digital signal, e.g., 8-bit depth RGB color space based on BT.709 \cite{BT709}. Perceptual Quantization (PQ) and Hybrid Log-Gamma (HLG) transfer functions \cite{HDR_Francois_CSVT20} were developed to represent higher dynamic range light with 10$\sim$12-bit depth signal according to BT.2100 \cite{BT2100}.
Then, digital videos are encoded to bitstreams and then transmitted to client through network. At remote clients, bitstreams are decoded and reconstructed as $\mathbf{\hat{I}}=F_E(\mathbf{I})$. Then, $\mathbf{\hat{I}}$ is shown with display $F_D$, which transfers electronic signal back to visual light $\mathbf{\hat{P}}$ with Electro-Optical Transfer Function (EOTF). Finally, visual light $\mathbf{\hat{P}}$ representing image $\mathbf{\hat{I}}$ is shown to human eyes. As shown in Fig.\ref{fig:HDRBitMap}, there are mismatches among the numbers of levels in representing natural scene ($10^{-6}$$\sim$$10^8$ $nits$), digital pixel (0$\sim$$2^{8}$ for SDR and 0$\sim$$2^{10}$ for HDR) and display light ($10^{-1}$$\sim$$10^2$ $nits$ for SDR, $5\times10^{-4}$$\sim$$10^4$ $nits$ for HDR).


In HVS $F_{HVS}$, visible light $\mathbf{\hat{P}}$ is perceived by retina and then converted to neural visual signal with rods and cones cells, known as photoreceptive cells. These neural impulses are transmitted to visual cortex through optic nerve and Lateral Geniculate Nucleus (LGN), which generates primary visual perceptual image $(\mathbf{V})$, denoted as $\mathbf{V}=F_{HVS}(\mathbf{\hat{P}})$. Then, $\mathbf{V}$ goes through two functional pathways for further perception and recognition tasks, which are a ventral pathway to V3 and V5 for ``Where/How'' information and a dorsal pathway to V2 and V4 for ``What'' information. There is a mismatch between the number of levels for vision range ($10^{-5}$$\sim$$10^6$ $nits$) and perceivable scales (about $10^2$) at one time, as shown in the bottom part of Fig.\ref{fig:HDRBitMap}. Due to the visibility threshold and properties of HVS, not every distortion is perceivable, i.e., visual redundancy.


Conventionally, error based visual metric, such as Mean-Squared-Error (MSE) and Peak Signal-to-Noise Ratio (PSNR), measures the signal distortion $\mathbf{D}$ as $\mathbf{D}={\mathbf{I}-\hat{\mathbf{I}}}$, which may overestimate or underestimate visual responses. While taking the whole end-to-end visual system into account, perceptual distortion $(\mathbf{D}_V)$ is expressed as
\begin{equation}
 \begin{cases}
	\mathbf{D}_V=\mathbf{V}-\mathbf{V}_{Org} \\
   \mathbf{V}=F_{HVS}(F_D(F_E(F_R(\mathbf{P})))) \\
   \mathbf{V}_{Org}=F_{HVS}(F_D(F_R(\mathbf{P})))
\end{cases},
\label{formula:VisualD}
\end{equation}
where $\mathbf{V}$ and $\mathbf{V}_{Org}$ are perceived images with or without the coding distortion $\mathbf{D}$ in visual communication. It has been widely acknowledged that $\mathbf{D}\neq\mathbf{D}_V$ due to visual redundancies \cite{SV_JND,SV_IQA,SV_PVQA} and non-linear properties in HVS. Based on Eq.\ref{formula:VisualD}, we shall not only consider processing and communication ($F_E$), but also jointly consider imaging ($F_R$), display ($F_D$) and HVS ($F_{HVS}$), while measuring the perceptual distortion $\mathbf{D}_V$. Due to visibility threshold in HVS, not every distortion $\mathbf{D}$ is perceivable. In addition, people is more interested in semantic contents, such as human face, sign and characters. Distortion that causes misunderstanding and false recognition will severely degrade QoE. In this work, we mainly analyze the perceptual distortion $\mathbf{D}_V$ by jointly considering $\mathbf{P}$,$F_R$,$F_E$,$F_D$, and $F_{HVS}$. Then, based on findings in $\mathbf{D}_V$ and $F_{HVS}$, perceptual redundancies are exploited to optimize video compression in $F_E$, called perceptually optimized video coding.


\begin{figure*}[!t]
	\centering
	\includegraphics[width=0.85\linewidth]{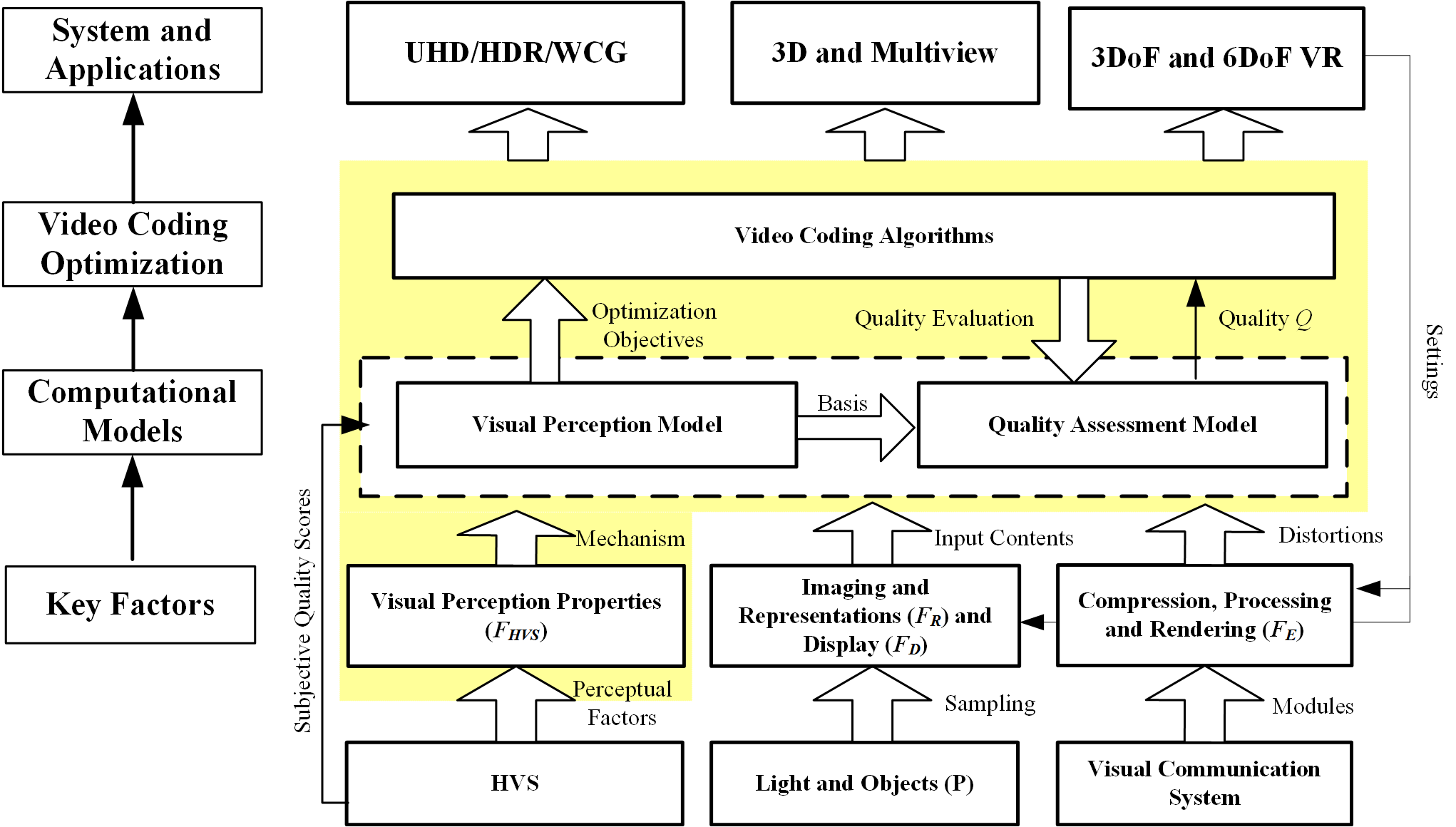}
	\caption{Framework of the perceptually optimized video coding.}
	\label{fig:PVC_Framework}
\end{figure*}

\subsection{Framework of the Perceptually Optimized Video Coding}

Generally, a framework of perceptually optimized video coding can be divided into four levels, which are principal factors, computational models, video coding optimization and video applications, as shown in Fig.\ref{fig:PVC_Framework}. There are four categories of key factors in the first level, which are HVS properties ($F_{HVS}$), imaging ($F_R$) and display ($F_D$), and visual processing and communication system ($F_E$) including compression, transmission and rendering. The second level includes computational visual perception models to model visual properties and responses of HVS, and computational visual quality assessment model to evaluate visual quality. Above this level is video coding optimization by exploiting the visual redundancies. Then, qualities of compressed videos are evaluated by the quality assessment models. Finally, video coding algorithms are applied to various applications to minimize bit rate while maintaining the visual quality. In this paper, we firstly review the physiological visual factors, the computational visual perception models and the visual quality assessment. Then, perceptually optimized video coding algorithms are reviewed.

\begin{figure*}[!t]
	\centering
	\includegraphics[width=0.95\linewidth]{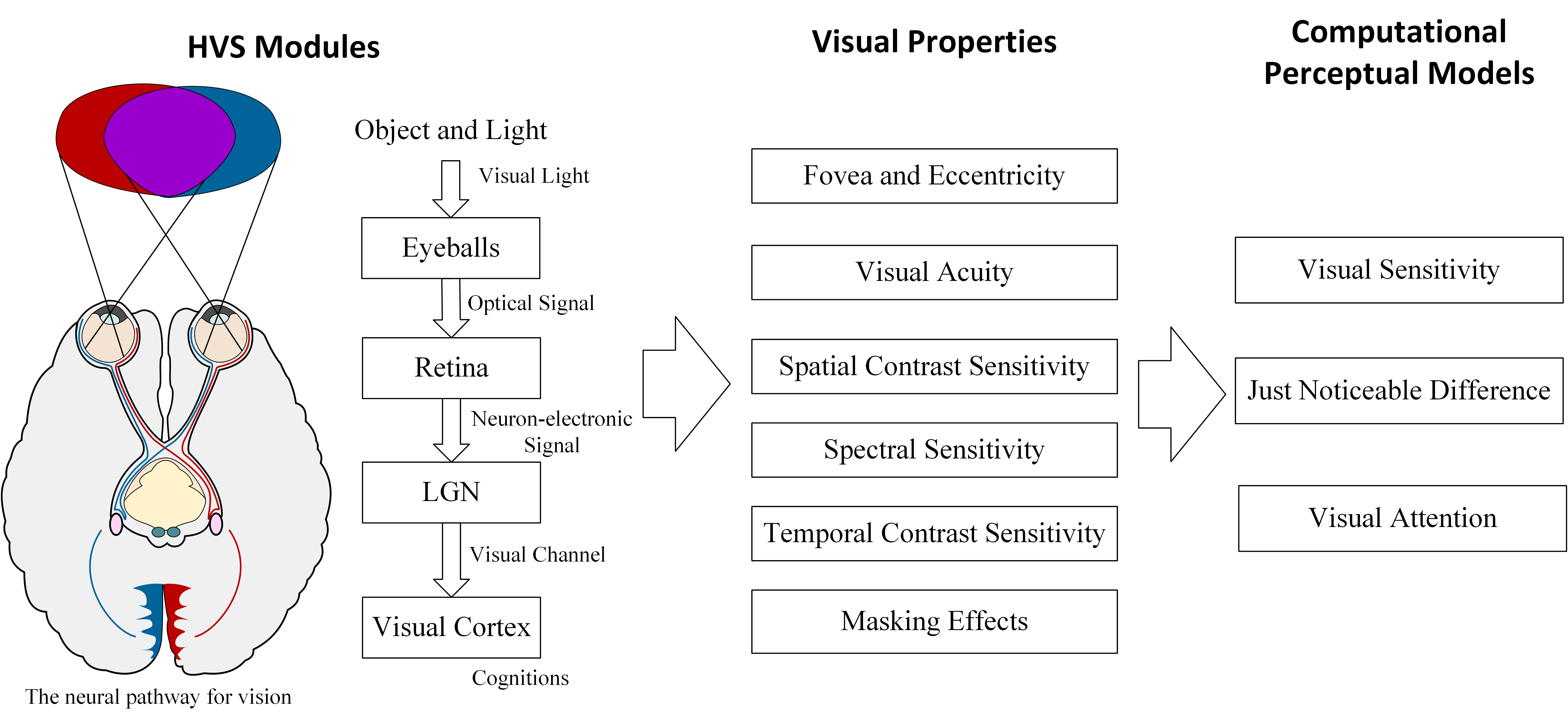}

	\caption{The neural pathway and key perceptual factors of HVS, $F_{HVS}$.}
	\label{fig:VisualFactors}
\end{figure*}

\section{Physiological Visual Factors and Computational Perceptual Models of HVS}
\label{sec:PerceptualFactors}
In HVS, visual light perceived by photoreceptors will be converted to neuron-electronic visual signal and then transmitted to LGN via nervous system. Then, these visual signals go to visual cortex for cognitions via visual channels, as shown in the left column of Fig. \ref{fig:VisualFactors}.
Based on these bio-mechanism of HVS, we present visual properties of monocular vision into six key categories, including fovea and eccentricity, angular resolution, spatial and temporal contrast sensitivity, spectral sensitivity and masking effects, as shown in the middle column of Fig. \ref{fig:VisualFactors}. Then, three higher-level computational models, including visual sensitivity, JND and visual attention, are reviewed based on the visual properties. Finally, visual quality assessment models are reviewed.

\subsection{Visual Properties and Models}
\subsubsection{Retinal Fovea and Visual Acuity}
Field of View (FoV) of human eyes covers 200$^{\circ}$ in width and 135$^{\circ}$ in height. However, visual acuity is not evenly distributed in FoV. Since the photoreceptors and ganglion cells distributed extremely dense at the retinal fovea, whose radius is about 1.5 $mm$ covering 1\% of the retina, the fovea becomes the most sensitive visual area. As the densities of the photoreceptors and ganglion cells decrease rapidly from the fovea to the peripheral, visual acuity progressively decreases as the eccentricity increases \cite{CM_PeriAcuity}. In other word, visibility threshold increases as the eccentricity increases \cite{CM_CSF_Rovamo}. Recently, it was found the visual acuity at isoeccentric locations was asymmetric, which was better along horizontal meridian than along the vertical \cite{CM_AsyFoveatedJND}. 

The visual acuity of the fovea can be modelled as $\theta$=1.22($\lambda/D$) rad, where $\lambda$ is wavelength of light and $D$ is pupil diameter, which varies from 3 $mm$ at day time to 9 $mm$ at night vision. Accordingly, the visual acuity decreases from day to night. The optimal acuity is about 0.0128$^{\circ}$ for 0.55 ${\mu}m$ V-band at day time according to the Rayleigh's criterion. When we watch a 4K@3840$\times$1920 video on a 75 inch UHDTV with 2 $m$ distance, the fidelity provided by each pixel is about 0.0104$^{\circ}$ in horizonal and 0.0139$^{\circ}$ in vertical, which close to the optimal acuity of the retinal fovea. 

\subsubsection{Spectral Sensitivity}
There are two kinds of photoreceptors: rods and cones. The rods are the basis for scotopic vision sensing monocular luminance and shape stimuli, and the cones are the basis for photopic vision sensing color stimuli. Because the photoreceptrtors are selectively sensitive to wavelengths and have non-linear responses, there exists visual sensitivities to spectrum, i.e., luminance and chromatic sensitivities.
\begin{itemize}

\item Luminance Sensitivity: Natural luminance is continuous ranging from $10^{-6}$$\sim$$10^8$ $nits$. Human eye can perceive a high dynamic range of luminance that changes from scotopic ($10^{-5}$$\sim$10 $nits$) to photopic (10$\sim$$10^6$ $nits$), as shown in Fig.\ref{fig:HDRBitMap}. However, it is much narrower when human eyes distinguish the relative brightness differences at one time, which is about $10^2$ scales. HVS requires to adjust the iris, photoreceptors and neurons to adapt a wide range luminance. Thus, only 256 scales with 8-bit depth are used to representing luminance of each pixel in conventional video. Meanwhile, SDR display provides 0.1 to 300 $nits$ brightness with 3000 contrast ratio. Recently, more bits (10$\sim$12 bits) are used in representing HDR video and HDR display provides a wider luminance range from 0.05 to 1000 $nits$ with 20000 contrast ratio. The mismatch between vision range and the narrower discrimination scales of HVS motivates visibility threshold and JND in $F_{HVS}$, which depends on image background, surround, peak and dynamic range of luminance, as well as image contents \cite{TQ_Kim}.
\item Chromatic Sensitivity: The visual sensitivity of human eye varies strongly for the light with wavelengths between 380 $nm$ and 800 $nm$. Three kinds of cones (S,M,L-cone) are sensitive to blue (437 $nm$), green (533 $nm$) and red (564 $nm$) lights, respectively \cite{CM_Stochman}, which motivates the RGB primary representation. Meanwhile, the photoreceptors have a relatively higher sensitivity in green than in the other two primary colors. Since the number of rods (1.3$\times$$10^8$) is much larger than that of cones (6.5$\times$$10^6$), HVS is less sensitive to color than luminance and scotopic vision is colorless. Also, color components are usually downsampled in video representation, such as YUV422 and YUV420. 
\end{itemize}

\subsubsection{Spatial Contrast Sensitivity}
Spatial contrast sensitivity describes the luminance discriminability between brightness and darkness regions in spatial domain. Given an image with background luminance $I$ and a centered brightness spot with luminance $I+\Delta I$, there is a ratio defined as $K =\Delta I/I$, called Weber ratio \cite{CM_ContrastSen}. The minimum distinguishable contrast threshold between the spot and background is found when $K$ is 0.02. Also, a larger $K$ enables a stronger visual response.

\begin{figure}[!t]
	\centering
	\subfigure[]{
		\label{fig:SCSF-brt}
	\includegraphics[width=0.33\linewidth]{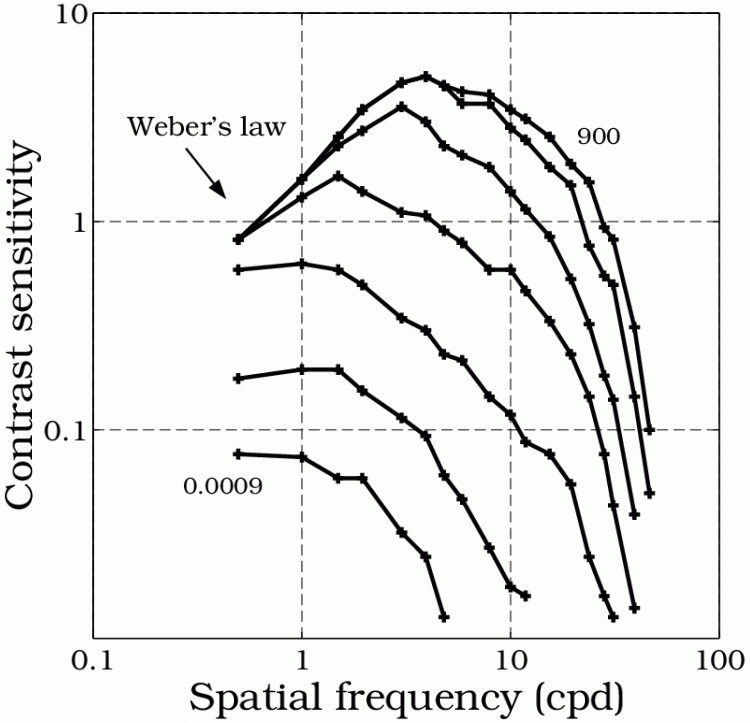}}
	\subfigure[]{
	   \label{fig:TCSF-brt}
	\includegraphics[width=0.325\linewidth]{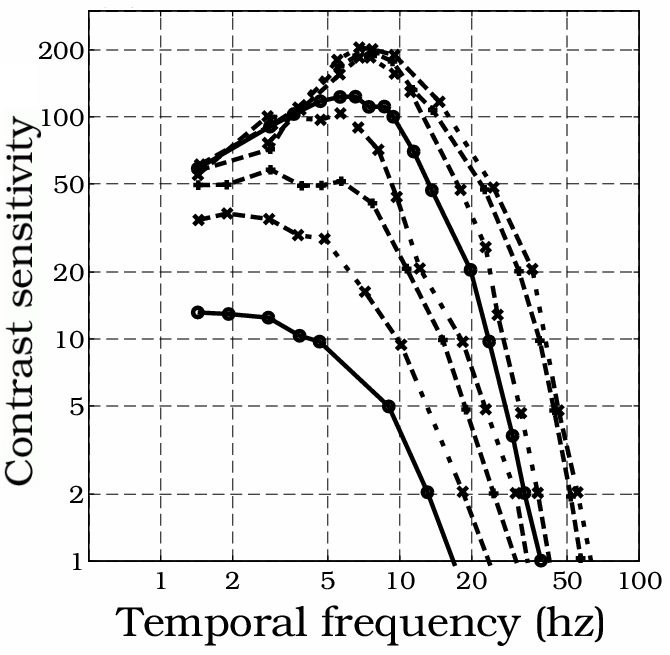}}
	\caption{Spatial and temporal CSFs with luminance adaptation. (a) Spatial \cite{CM_ContrastSen}, (b) Temporal \cite{CM_TCSF_Kelly}.}
	\label{fig:brtCSF}
\end{figure}

The spatial contrast sensitivity provides a characterization of frequency response in HVS, which is a bandpass filter in low spatial frequency and a lowpass filter in high spatial frequency \cite{CM_ContrastSen}, as shown in Fig. \ref{fig:SCSF-brt}. It reduces at extremely low or high spatial frequency and reaches its peak around 10 $cpd$. In addition to the frequency, the spatial contrast sensitivity is affected by many other factors, including background luminance, human age, eccentricity, spectral, patterns and types of stimuli \cite{CM_Barten-CSF}. It was found that the contrast sensitivities for chromatic (blue-yellow, red-green) are low-pass filters \cite{CM_CSF_HDR} and their peaks are relatively lower than those of achromatic \cite{CM_Color-CSF}. Generally, the spatial contrast sensitivity increases with the background luminance increases\cite{CM_S_CSF}, and reduces as age and distance from eccentric \cite{CM_CSF_Rovamo} increase. Moreover, the sensitivity falls for test pattern whose spatial frequency is similar to adaptation patterns and orientations, and also falls for sinewave stimuli as compared with squarewave stimuli.

\subsubsection{Temporal Contrast Sensitivity}
Temporal contrast sensitivity measures the discriminability of the luminance difference over time, which relates to temporal frequency, level of contrast, patterns and so on \cite{CM_TCSF_Kelly}. It was found that this sensitivity increased as input contrast stimulus increased. In this case, the photoreceptors (rods and cones) are easier to reach their perceptual threshold and critical duration is shorter. The temporal contrast sensitivity is a bandpass filter and reaches its peak around 5Hz \cite{CM_TCSF_Kelly}, as shown in Fig. \ref{fig:TCSF-brt}, which can be modelled as an addition of two log-scaled Gaussian filters \cite{CM_ST-CSF}. Also, it was found that complex cells in V1 and V2 were usually more sensitive in spatial and temporal contrasts as compared with simple cells \cite{CM_ST-Nature}. The temporal contrast sensitivity is higher than the spatial contrast sensitivity \cite{CM_STCSF_Lambrecht}, as shown in Fig. \ref{fig:brtCSF}, which indicates that the temporal distortion is more perceivable.

When the temporal frequency reaches 50Hz, the temporal contrast sensitivity is extremely low, as shown in Fig.\ref{fig:TCSF-brt}. So, the refresh rate of a display is higher than 60Hz to avoid perceiving flickering. Also, the temporal contrast sensitivity degrades when background luminance is extremely bright or dark \cite{CM_TCSF_Kelly}. 
Since HDR display provides a higher dynamic range of brightness, HDR video requires a higher frame rate, e.g., 60 to 120 fps, to reduce motion judder and smooth motion transition.

\subsubsection{Visual Masking Effects}
The LGN and primary visual cortex have different sensitivity responses when receiving neuro-electronic signals from retina over multiple channels, such as frequency, luminance, color, motion and so on. Their responses are bandpass filters whose peaks and bands vary with stimuli. Meanwhile, interferences among multiple visual stimuli will be caused due to their co-existence. When these stimuli present simultaneously, the presence of one stimulus may weaken or enhance the responses of other excitations, called visual masking effects. The masking effects can be categorized as Luminance Masking (LM), spatial and temporal contrast masking \cite{TQ_Kim}, binocular masking \cite{MVC_Jaballah,CM_BJND_Zhao} and pattern masking \cite{CM_JND_Pattern} depending on the input stimulus. Response from the strongest stimulus weakens other responses. For example, the perceived brightness of an object not only depends on brightness intensity of the object, but also depends on its surrounding background.
Moreover, distortion in regions with regular pattern, such as parallel lines, is more perceivable than that in chaos textural regions, such as grasses \cite{CM_JND_Pattern,IQA_PWMSE}.

\subsection{Computational Perceptual Models}
Computational models were proposed to model the perceptual factors of HVS, i.e., $F_{HVS}$. Three kinds of high-level models, including visual sensitivity, JND, and visual attention, are reviewed.
\subsubsection{Visual Sensitivity Models}
Visual sensitivity is affected by many factors including spatial contrast \cite{CM_Barten-CSF,CM_S_CSF,IQA_PWMSE,CM_Bosse}, temporal contrast \cite{VQA_PWMSE,CM_ST-CSF,CM_STCSF_Lambrecht}, input stimuli pattern \cite{CM_JND_Pattern}, brightness \cite{CM_ContrastSen,CM_CSF_HDR}, color sensitivity \cite{CM_Color-CSF,IQA_CSPSNR}, type of visual cells, fovea and eccentricities \cite{CM_AsyFoveatedJND}, binocular masking \cite{MVC_Jaballah} and so on. Barten \emph {et al.} \cite{CM_Barten-CSF} built a spatial Contrast Sensitivity Function (CSF), called Barten's CSF, as an isotropic bandpass shaped function of image size, luminance level and pupil size, which is \cite{CM_Barten-CSF}

\begin{equation}
 \begin{cases}
	{S(u)}=\frac{M_{O}(u)}{k\sqrt{\frac{2}{T}(\frac{1}{X^2_O}+\frac{1}{X^2_{max}}+\frac{u^2}{N^2_{max}})(\frac{1}{\eta{pE}}+\frac{\Phi{_0}}{1-e^{{-(u/u_0)}^2}})}}\\
    M_{O}(u)=e^{-2\pi^2(\sigma^2_0+C_{ab}^2d^2)u^2}
 \end{cases},
\label{formula:Barten_CSF}
\end{equation}
where $S(u)$ is the sensitivity, $M_{O}(u)$ is the optical Modulation Transfer Function (MTF) of eye, $E$ is retinal luminance, $d$ is pupil diameter, $k$, $p$, $T$, $\eta$, $\sigma_0$, $X_{max}$, $\Phi_0$, $C_{ab}$, $N_{max}$ and $u_0$ are parameters, whose typical values and simplified form of Eq.\ref{formula:Barten_CSF} can be found in \cite{CM_Barten-CSF}. Kim \emph{et al.}\cite{CM_CSF_HDR} modelled spatio-chromatic contrast sensitivities as low-pass filters, whose peaks increased with the background luminance in HDR display. Lambrecht and Kent \cite{CM_STCSF_Lambrecht} modelled a joint spatial and temporal CSF in a 3D non-separable form,
which took the difference between excitatory and inhibitory mechanism.
Hu \emph{et al.} modelled the visual sensitivity with spatial randomness \cite{IQA_PWMSE} and temporal randomness \cite{VQA_PWMSE}. Kim \emph{et al.} \cite{TQ_Kim} modelled the visual sensitivity by jointly exploiting temporal masking from motion vector, contrast masking from edge, and luminance masking from mean pixel values in images. Furthermore, a stereo visual sensitivity model \cite{MVC_Jaballah} was proposed by considering the binocular masking effect.
Since HVS is complicated and many visual factors are not fully understood yet, it is challenging to model $F_{HVS}$ mathematically. Data-driven sensitivity models \cite{CM_Bosse,CM_Deep_CSF} were developed to explore visual responses to image distortion. Bosse \emph{et al.} \cite{CM_Bosse} found that the spatial visual sensitivity to MSE distortion decreased as the image texture increased. Then, a neural network was trained to predict the perceptual weights for each MSE based distortion. Hosseini \emph{et al.} \cite{CM_Deep_CSF} tried to synthesize a convolutional filter to mimic the falloff sensitivity response of HVS for image sharpness, i.e., $\mathbf{I}_O=\mathbf{I}*F_{HVS}$, which was approximated as an inverse generalized Gaussian distribution and implemented with MaxPol filter library.

\subsubsection{JND Models}
Due to visual sensitivity and masking effects in HVS, not every distortion is perceivable. The minimum visibility threshold of pixel intensity change is denoted as JND, such as the Weber's law in CSF. Many pixel-wise JND and sub-band domain JND models for 2D images/videos have been developed. In \cite{TQ_Kim,RDO_Bae}, video JND was modelled by jointly exploiting spatial CSF, temporal masking from motion vectors, contrast masking from texture edges and luminance masking. Based on the finding that random pattern region has higher spatial masking, this pattern masking effect was modelled and incorporated in modelling JND \cite{CM_JND_Pattern}. Jiang \emph{et al.} \cite{CM_JND_JiangTIP22} predicted a Critical Perceptual Lossless (CPL) threshold of an image with Karhunen-Loeve Transform (KLT), and calculated the difference between the input image and its CPL image as a JND map. In \cite{CM_JND_WangTIP21}, JND prediction was modelled as maximizing the difference between the original and distorted image subject the same perceptual quality. Then, a JND estimation scheme was proposed based on the hierarchical predictive coding model of visual cortex.
More JND models can be referred in \cite{SV_JND,SV_JND_Lin,SV_IQA}. Most of the existing JND models are perceptual model based and process the pixel or block as basic unit. However, HVS perceives image/video entirely rather than pixel or block individually, which is more complicated.

\begin{figure}[!t]
	\centering
	\includegraphics[width=0.55\linewidth]{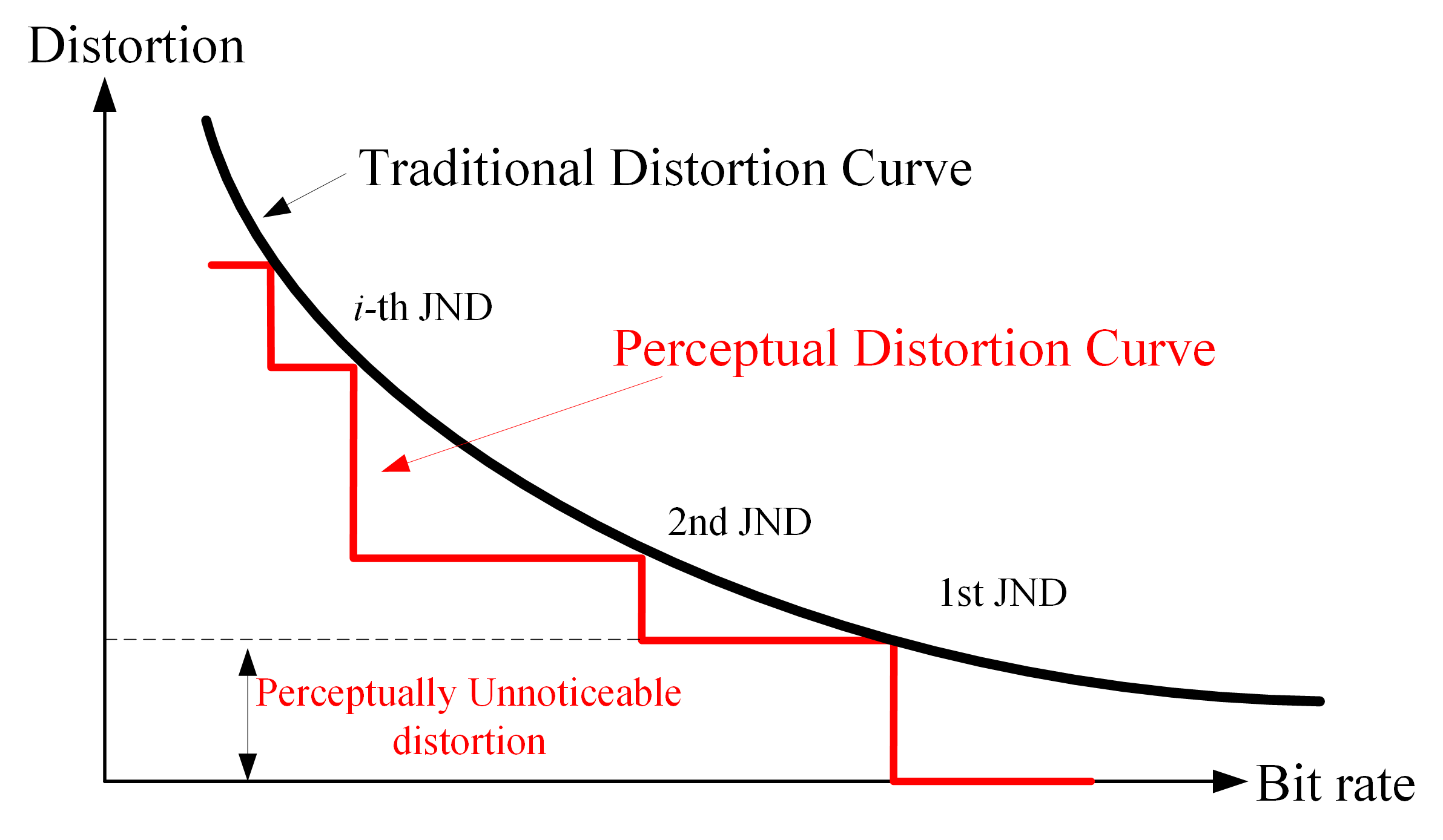}

	\caption{PWJND and VWJND for compressed images/videos.\cite{CM_VideoJND_Wang,CM_PWJND}}
	\label{fig:JND}
\end{figure}

Picture and Video Wise JND (PWJND/VWJND) models were studied. Although there are 100 quality scales or more for image compression with JPEG/JPEG2000 and 52 quality scales for video compression with H.264/HEVC, Jin and Wang \emph{et al.} found that HVS can only distinguish 5 to 7 quality scales \cite{CM_ImageJND_Jin,CM_VideoJND_Wang} and built a PWJND dataset, called MCL\_JCI, for JPEG compressed images. As shown in Fig.\ref{fig:JND}, the real perceptual distortion in compressed image/video does not decrease continuously as bit rate increases, but decreases as a discrete staircase function. The jump point between two quality scales is the visibility threshold that subjects can just distinguish the difference between distorted image/video and its reference, called PWJND/VWJND, which slightly varies with image/video contents, resolutions and displays. To analyze the VWJND in compressed videos, Wang \emph{et al.} \cite{CM_VideoJND_Wang} built a large-scale compressed video quality dataset based on JND, called VideoSet, where the first three perceptual JND points are labeled among 52 quality levels of H.264 compressed videos. It has 220 source videos with four resolutions. Liu \emph{et al.} \cite{CM_PWJND} developed a deep learning based JND model to predict the perceptual difference, where the JND prediction was modelled as a binary classification problem and then solved. Shen \emph{et al.} \cite{CM_PatchJND} proposed a deep learning-based structural degradation estimation model to predict structural visibility at patch-level, and then predicted the PWJND. Fan \emph{et al.} \cite{CM_SUR_JND_Stereo} modelled the PWJND for symmetrically and asymmetrically compressed stereo images with H.265 Intra coding and JPEG2000, where Scale-Invariant Feature Transform (SIFT), cyclopean image quality, rivalry quality were additionally introduced, and gradient boosting decision tree was used for feature selection and fusion. Then, based on VMAF features \cite{VQA_VMAF} and Support Vector Regression (SVR), Zhang \emph{et al.} \cite{CM_SUR_SVR} developed a Satisfied User Ratio (SUR) prediction method for compressed videos, which indicated the ratio of subjects who cannot perceive the visual difference between the distorted video and its reference. JND point was derived at 75\% SUR. In \cite{CM_DL_JND_Video}, a deep learning based Video Wise Spatial-Temporal SUR (VW-STSUR) model was proposed to predict the SUR and VWJND using two-stream CNN, where spatio-temporal features were fused at score and feature levels. These PWJND/VWJND and SUR models were developed in a data-driven way, and high dimensional handcrafted features or deep learning features were used to improve the JND prediction accuracy. Compared with conventional pixel-wise JND models in \cite{SV_JND}, more distortion is allowed in PWJND/VWJND due to compound masking effects in entire image/video \cite{CM_JND_TOMM}. However, large scale JND datasets are required, which are time-consuming and laborious to create. Further subjective studies and JND modelling related to human, content and system factors will be interesting.

\subsubsection{Visual Attention Models}
Visual Attention (VA) or saliency is a high-level cognitive mechanism that drives the retinal fovea in eye to attentional contents for higher fidelity, which is also noted as Region-Of-Interest (ROI). Usually, HVS is easier to be attracted by regions with high contrast, such as luminance, texture, orientation, temporal motion and color contrasts. To mimic the decomposition happened in the visual cortex, multi-scale image decomposition was usually employed. So, these inputs were represented with multi-scale center-surround contrasts of texture, optical flow or (blue-yellow, red-green) chromatic maps \cite{CM_VA_Itti}.
At higher level, cognitive features, such as shape, sign, faces, skin, and characters, draw attention. However, these high-level features vary from person to person due to their different knowledge backgrounds and preferences, while low-level features are fundamental and stable among people. In \cite{CM_DL_StereoVA}, Zhang \emph{et al.} explored stereoscopic video saliency with deep learning, where 3D Convolutional Neural Network (CNN) was used to extract spatio-temporal saliency feature and Convolutional Long Short-Term Memory (Conv-LSTM) was used to fuse spatio-temporal and depth attributes. More recent advances on VA models can be referred in \cite{SV_VA,CM_Stereo_VA}.

\subsection{Discussions}

With the advances of visual-psychological and physiological research, more perceptual mechanism and models have been revealed. Conventional visual-psychological studies usually analyzed a perceptual factor and its visual response by adjusting one stimulus and fixing the rest. However, multiple stimuli co-exist and mutually interact, such as binocular fusion and rivalry in LGN. HVS is a compound non-linear binocular vision system, which is more complicated than a simple combination of multiple models. Modeling the joint effects among multiple stimuli and binocular vision is more challenging. Thanks to learning algorithms, especially the deep learning, which are able to discover statistical relationships in massive data, data-driven approach to computational perceptual model is a promising direction for scenarios with sufficient labels. On the other hand, the found perceptual mechanisms, such as multi-scale and visual attention, are also able to improve the design of learning models.

\section{Computational Models for Visual Quality Assessment}
\label{sec:SecVQA}
Subjective test \cite{BT500,P910} is the ultimate way to evaluate image or video quality. The main subjective quality methods include Degradation Category Rating (DCR), Pair Comparison (PC) and Absolute Category Rating (ACR)\cite{BT500}. The reference and distorted sequences are shown to a number of human subjects, 16 or more, under controlled viewing conditions. Then, the subjects are asked to assess the overall quality of the distorted sequences with respect to the reference, and score it on a five or nine-grade scale corresponding to their perceived quality in mind. Finally, Mean Opinion Score (MOS) is calculated based on the average rating scores of the subjects, while Differential MOS (DMOS) calculates the average difference between scores of the distorted and reference sequences. However, subjective test is expensive, time-consuming, and often impractical. Computational models for visual quality assessment that measure the quality of video are required.

To evaluate the perceptual quality of visual signal $\mathbf{\hat{I}}$, visual quality assessment is to model the relationship between $\mathbf{V}$ and $\mathbf{\hat{I}}$, which is mathematically expressed as
\begin{equation}
	\mathbf{V}=F_{VQA}(\mathbf{\hat{I}}),
\label{formula:VQA}
\end{equation}
where $F_{VQA}()$ is a relationship function defining non-reference visual quality assessment. According to Eq.\ref{formula:VisualD}, visual quality $\mathbf{V}=F_{HVS}(F_D(\mathbf{\hat{I}}))$. Thus, $F_{VQA}()=F_{HVS}(F_D())$.
In video coding, reference signal $\mathbf{I}$ is usually available and visual distortion $\mathbf{D}_V$ caused by signal distortion $\mathbf{D}$ really matters. So, quality assessment with the reference $\mathbf{I}$ is
\begin{equation}
	\mathbf{D}_V=F_{VQA}(\mathbf{D}).
\label{formula:VQA_D}
\end{equation}
where $\mathbf{D}_V$ is the visual quality degradation of $\mathbf{\hat{I}}$ whose groudtruth is obtained through subjective experiments, signal distortion $\mathbf{D}=\mathbf{I}-\mathbf{\hat{I}}$. Based on whether the temporal information is exploited or not, the full-reference visual quality assessment $F_{VQA}()$ is categorized as IQA and Video Quality Assessment (VQA) models.

\begin{table*}[!t]
\tiny
	\renewcommand{\arraystretch}{1.3}
	\caption{Key Features and Fusion Schemes of the Featured IQA/VQA Schemes for video compression.}
	\label{table_2DIQA}
	\centering

	\begin{tabular}{|c|c|c|c|c|}
		\hline
		 Methods& Types & Key Features  & Fusion Schemes \\
        \hline
        PSNR/PSNR-HVS  \cite{IQA_PSNR-HVS} & IQA & MSE between the reference and distorted images  & / \\
       \hline
        CSPSNR \cite{IQA_CSPSNR} &IQA& MSEs of Y, Cb and Cr &{weighted summation with empirical weights}\\
   		\hline
        HDR-VDP-2 \cite{IQA_HDR_VDP2}&IQA&threshold-normalized difference for each band & weighted logarithmical summation\\
		\hline

		SSIM/MS-SSIM \cite{IQA_MSSSIM} &IQA & Similarity for luminance, contrast and structure & multiplication  \\
		\hline
        VSI\cite{IQA_VSI}& IQA & \parbox[c]{5.0cm} {Visual saliency, contrast sensitivity and chroma saturation are used as perceptual weight } & / \\
        \hline
        PWMSE\cite{IQA_PWMSE}&IQA&\parbox[c]{5.0cm} { Masking from spatial randomness,low-pass filtered MSE}& \parbox[c]{5.0cm}{logarithmical summation of MSE modulated with exponentially tuned randomness} \\
        \hline
        SDS\cite{IQA_SDS_Zhang20} & IQA& \parbox[c]{5.0cm}{Divisively normalized sparse-domain similarity for luminance energy and sparse coefficient} & multiplication \\
        \hline
        DeepQA\cite{IQA_DeepQA} &IQA&\multicolumn{2}{c|}{ CNN for visual sensitivity and fully connected layers for quality regression}\\
        \hline
        Fan\cite{IQA_MultiCNN_QA} &IQA& \multicolumn{2}{c|}{One CNN to classify the distortion types and multiple CNNs to predict the quality of each distortion types}\\
        \hline
       VQM  \cite{VQA_VQM}  &VQA&  \parbox[c]{5.0cm} {spatial information loss, spread of chroma, spatial gain, temporal impairments, local color impairments, edge shift}&  weighted summation with empirical weights \\

       \hline
       MOVIE \cite{VQA_MOVIE}& VQA&\parbox[c]{5.0cm} {Difference between spatial and temporal coefficients from Gabor filter banks}&  multiplication\\
      \hline
       HDR-VQM\cite{VQA_HDR_VQM} &VQA&Subband error from log-Gabor filters &{spatial pooling, short and long temporal error pooling}\\

       \hline
      PWMSE-V\cite{VQA_PWMSE} &VQA&\parbox[c]{5.0cm}{Contrast sensitivity from spatial, temporal and foveated low-pass filters, spatial and temporal randomness} & perceptually weighted MSE\\
      \hline
      VMAF  \cite{VQA_ST_VMAF}&VQA&  AN-SNR,DLM,VIF,MCPD& SVR\\
      \hline
      ST-VMAF \cite{VQA_VMAF}&VQA&\parbox[c]{5.0cm}{12 features from VMAF (5), T-SpEED (3) and T-VIF (4)}& SVR and hysteresis temporal pooling \\
      \hline
      E-VMAF \cite{VQA_VMAF}&VQA& ST-VMAF, VMAF& ensembled with SVM, hysteresis pooling over time\\
      \hline
      DeepVQA\cite{VQA_DeepVQA}&VQA&\multicolumn{2}{c|}{CNN extracts spatiotemporal sensitivity and CNAN pools qualities in temporal}\\
        \hline
      C3DVQA\cite{VQA_C3DVQA}&VQA&\multicolumn{2}{c|}{CNN extracts spatial features and 3D CNN extracts spatiotemporal features, regression layers to predict quality}\\
        \hline
	\end{tabular}
\end{table*}

\subsection{IQA Models}
PSNR and MSE measure image quality with mean squared difference between the source and distorted images, i.e., $\mathbf{D}_V$ is approximated with $\mathbf{D}$, which has been widely used in video coding due to its simplicity. However, PSNR only measures the signal difference without considering visual properties, which cannot truly reflect the perceived quality in HVS. To handle this problem, developing a evaluator to qualitatively measure the perceptual quality of visual contents has been a key topic for visual signal processing.

\begin{figure}[!t]
	\centering
	\subfigure[]{
		\label{fig:PerVQA}
	\includegraphics[width=0.65\linewidth]{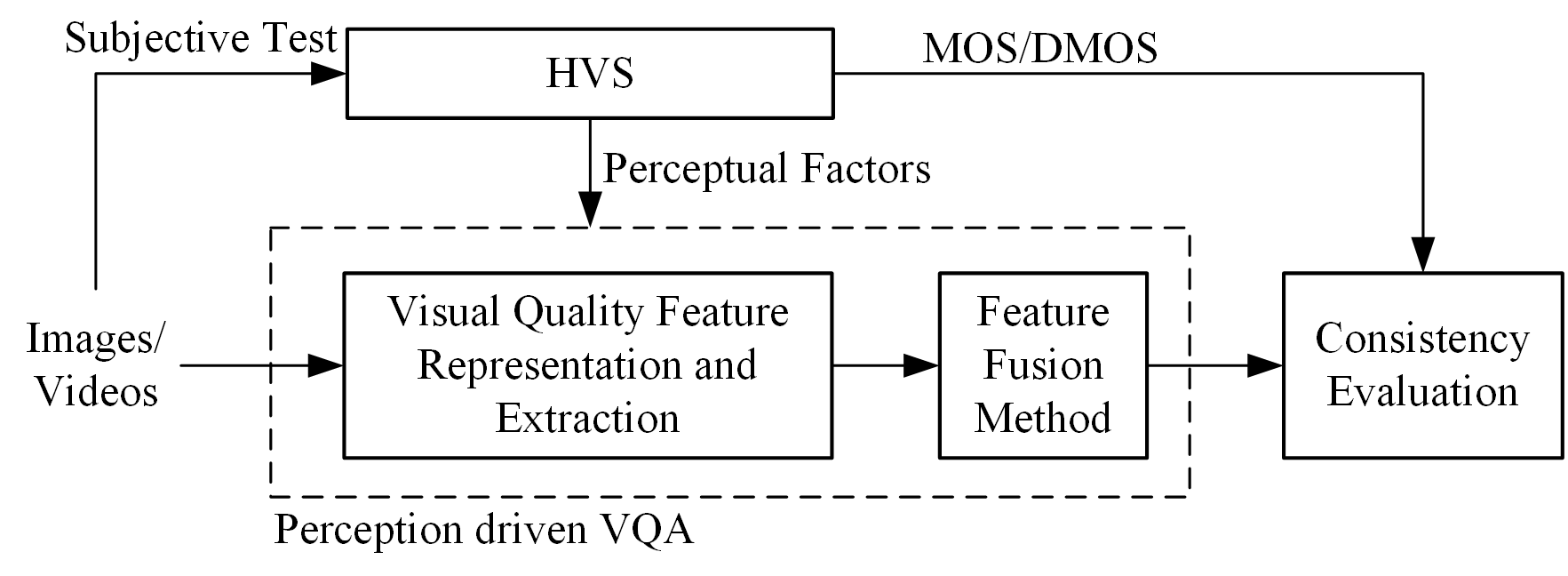}}
	\subfigure[]{
	   \label{fig:LearnVQA}
	\includegraphics[width=0.65\linewidth]{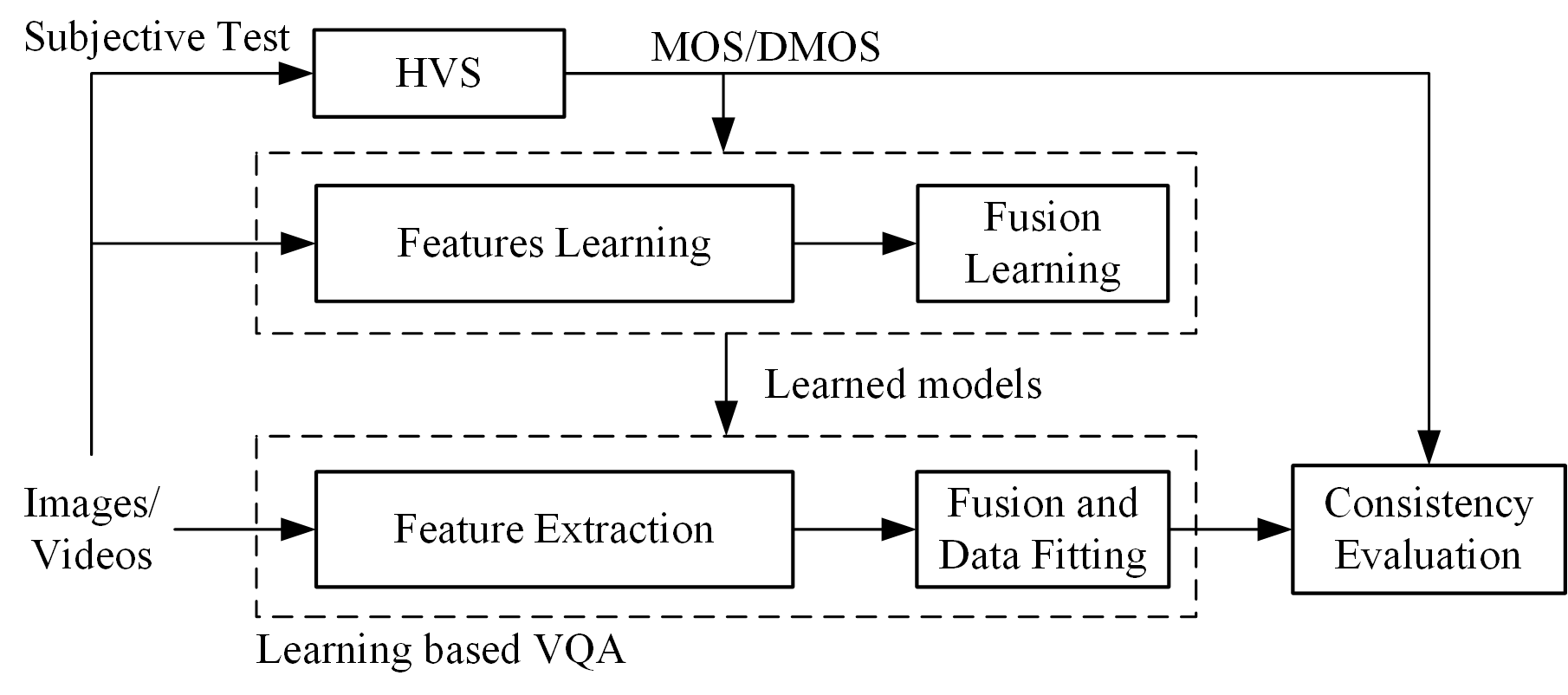}}

	\caption{Framework of the IQA/VQA computational models.(a) Perception based IQA/VQA. (b) Learning based IQA/VQA.}
	\label{fig:VQA_framework}
\end{figure}

A large number of IQAs have been proposed in the past few decades. Image artifacts, including blocking, blurring, ringing, and color bleeding, will be caused in compression. Based on the PSNR, an extension named PSNR-HVS \cite{IQA_PSNR-HVS} was developed with weighted summation of three PSNRs, which were error sensitivity from RGB channels, structural distortion from mean, max and min values, and edge distortion from edge, texture and flat regions. For compressed color images, Shang \emph{et al.} \cite{IQA_CSPSNR} proposed a Color-Sensitivity-based PSNR (CSPSNR) where MSEs of Y, Cb and Cr components were combined with weights 0.695, 0.130 and 0.175 from subjective test. It had a better consistency than the empirical 6:1:1 combination in PSNR \cite{IQA_WTPSNR} and 4:1:1 combination in MSE in the HEVC reference software. Moreover, to access HDR images with a broader luminance range, distortion of each frequency and orientation band was threshold-normalized and algorithmically summed \cite{IQA_HDR_VDP2}. Since HVS is more sensitive to structural distortion, Structural SIMilarity (SSIM) \cite{IQA_SSIM} and its variants, such as Gradient based SSIM (GSSIM), Multi-scale SSIM (MS-SSIM) \cite{IQA_MSSSIM}, have been developed by measuring the mean and variance similarities between the reference and distorted images. Zhang \emph{et al.} \cite{IQA_VSI} proposed a Visual Saliency-Induced (VSI) Index for IQA, where visual saliency, contrast sensitivity and chroma saturation features were exploited and multiplicatively fused. Hu \emph{et al.} \cite{IQA_PWMSE} found that HVS was more sensitive to the distortion in regular texture than that in disordered texture. They proposed a Perceptually Weighted MSE (PWMSE) by exploiting the masking effects in spatial and pattern. There are many classical IQA metrics that exploit the perceptual factors mentioned in Section \ref{sec:PerceptualFactors}. Their general framework includes feature representation to extract effective visual features and fusion model to fit the quality score, as shown in Fig.\ref{fig:PerVQA}, where feature representation is motivated from the perceptual factors of HVS. Finally, consistency indices, such as Pearson Linear Correlation Coefficient (PLCC), Spearman Rank-Order Correlation Coefficient (SROCC) and MSE, were measured between the predicted scores from IQA and MOS/DMOS from subjective tests, to validate the effectiveness of an IQA model.

To improve the accuracy of IQAs, learning algorithms have been exploited to discover statistical knowledge in massive data. Fig.\ref{fig:LearnVQA} shows a framework of the learning based IQAs, where feature extraction and fusion models are learned. To improve the feature representation of visual quality, Zhang \emph{et al.} \cite{IQA_SDS_Zhang20} developed a Sparse-Domain Similarity (SDS) index, where sparse representation was used to learn more effective quality features. Meanwhile, Divisive Normalization Transform (DNT) was used to remove statistical and perceptual redundancies. Due to the powerful capability of deep neural networks in feature representation and non-linear data fitting, Kim \emph{et al.} \cite{IQA_DeepQA} proposed a deep learning based full-reference IQA model without considering prior knowledge of HVS, but exploiting data statistics in databases. Fan \emph{et al.} \cite{IQA_MultiCNN_QA} utilized multi-expert CNNs to classify distortion types and predicted the quality in each distortion type. Deep IQA models usually outperform in quality prediction. However, they are highly data-dependent and their accuracy may degrade in cross-database validation. Moreover, large dataset with quality labels is required in training general and stable deep IQAs. These metrics are IQAs and temporal information is not considered.

\subsection{VQA Models}
In addition to the spatial artifacts, temporal artifacts introduced in video compression, such as flickering, jerkiness and floating, shall be considered in VQA. VQM and MOVIE are two classical VQAs. In VQM \cite{VQA_VQM}, seven key features, including spatial information loss, spread of chroma, spatial gain, temporal impairments, and local color impairments and edge shift, were calculated and fused with empirically weighted summation. In MOVIE \cite{VQA_MOVIE}, reference and distorted videos were decomposed with multiple Gabor filters and quality degradations were measured with the Gabor coefficient differences in spatial and temporal domains. Finally, the two degradations were multiplicatively fused. In \cite{VQA_HDR_VQM}, quality features of the sub-band distortion in HDR videos were extracted with log-Gabor filters. In \cite{VQA_PWMSE}, a perceptually weighted MSE metric was developed by considering low-pass filter based contrast sensitivity, visual attention and spatiotemporal randomness based masking effects. Li \emph{et al.} \cite{VQA_VMAF} proposed a practical perceptual VQA for video streaming, named Video Multi-method Assessment Fusion (VMAF). By learning a Support Vector Machine (SVM) model, it fused the scores from four existing metrics, including Anti-noise SNR (AN-SNR), Detail Loss Measure (DLM), Visual Information Fidelity (VIF), and Mean Co-Located Pixel Difference (MCPD). However, VMAF used the average frame difference as the only temporal feature, which highly related to video content rather than the temporal distortion. To handle this problem, Bampis \emph{et al.} \cite{VQA_ST_VMAF} proposed a Spatio-Temporal VMAF (ST-VMAF) to measure the temporal distortion by considering temporal masking. It modelled bandpass-filtered maps of frames differences and used entropy differences to predict video quality. In addition, an Ensemble VMAF (E-VMAF) scheme was proposed by aggregating the conventional VMAF \cite{VQA_VMAF} and the enhanced ST-VMAF. Kim \emph{et al.} \cite{VQA_DeepVQA} used a CNN to extract spatio-temporal visual features for each frame, which were temporally pooled with a Convolutional Neural Aggregation Network (CNAN). Xu \emph{et al.} \cite{VQA_C3DVQA} proposed a 3D-CNN based VQA model, called C3DVQA, which was composed of two-stream 2D convolutional layers to learn spatial features from the distorted and residual frames, 3D convolutional layers to learn spatiotemporal features and regression layers to predict the visual quality.

\subsection{Discussions on Quality Assessment}

Although the VQA is an open problem investigated for many years, there are still many challenging issues and unsolved problems. Firstly, many IQA/VQA models were developed based on some visual properties and characteristics in specific applications. Although SSIM and VMAF are often used in measuring quality of compressed video, there is still no perceptual visual quality model that is commonly recognized and widely used as the PSNR.

Secondly, visual quality databases shall be built based on subjective tests under a controllable environment, and it is very laborious to score large amount of distorted visual contents with dozens of human subjects. Thus, the available datasets are with small scale. As they cannot be fused to form a large one, it causes different properties of the datasets and inconsistences in evaluating the existing VQA models. A large dataset for compression distortions is highly demanded.


Thirdly, most of previous VQAs were motivated from modelling the perceptual factors, noted as perception based VQA schemes, such as SSIM and MOVIE, as shown in Fig. \ref{fig:PerVQA}. However, effective feature extraction and feature fusion are the keys to model accuracy. Recently, many learning based VQA schemes were developed as the learning schemes were applied to improve the features extraction and fusion models, as shown in Fig.\ref{fig:LearnVQA}. Since it is challenging to extract effective features for IQA, learning based representations, such as sparse representation \cite{IQA_SDS_Zhang20, VQA_SR3DVQA}, Singular Vector Decomposition (SVD)\cite{IQA_SVD_SPIC19} and tucker decomposition \cite{IQA_Tucker_TIP20}, were used to represent quality features more effectively. In addition, end-to-end deep IQAs/VQAs \cite{IQA_MultiCNN_QA,IQA_DeepQA, VQA_DeepVQA,VQA_C3DVQA} were capable of learning feature representation and fusion simultaneously. However, they require a large scale dataset with labeled subjective quality in training, which is laborious and expensive to collect. Generality and interpretability of the deep IQA/VQA models shall be improved.

Finally, perception $\mathbf{V}$ is actually affected by factors $F_{HVS}$, $F_D$, $F_E$, $F_R$ and $\mathbf{P}$ according to Eq. \ref{formula:VisualD}. Thus, the VQA model is a conditional function of $F_D$, $F_E$, $F_R$ and $\mathbf{P}$. 
It is important to improve the adaptability of models and transfer the models from one condition to another.

\section{Perceptual Video Coding Optimization}
\label{sec:2D-PVC}

The optimization objective of video coding is to maximize the visual quality of compressed videos $\mathbf{V}$ at a given target bit rate $R_T$, which can be formulated as
\begin{equation}
	\max(\mathbf{V}), s.t. R \leq R_T,
\label{formula:obj_fun}
\end{equation}
where $R$ and $R_T$ are coding bit rates. Suppose the original visual signal $\mathbf{I}$ has the highest visual quality. The coding optimization problem in Eq. \ref{formula:obj_fun} is equivalent to minimizing visual distortion at the given bit rate, which can be presented as
 \begin{equation}
	\min(\mathbf{D}_V), s.t. R \leq R_T,
\label{formula:obj_fun2}
\end{equation}
where $\mathbf{D}_V$ is the visual quality degradation defined in Eq. \ref{formula:VisualD}. However, in current video coding standards, the distortion $\mathbf{D}_V$ is mainly measured with MSE/PSNR, i.e., $\mathbf{D}_V\approx\mathbf{D}$, which cannot truly reflect the perceptual quality. 

\begin{figure}[!t]
	\centering
	\includegraphics[width=0.55\linewidth]{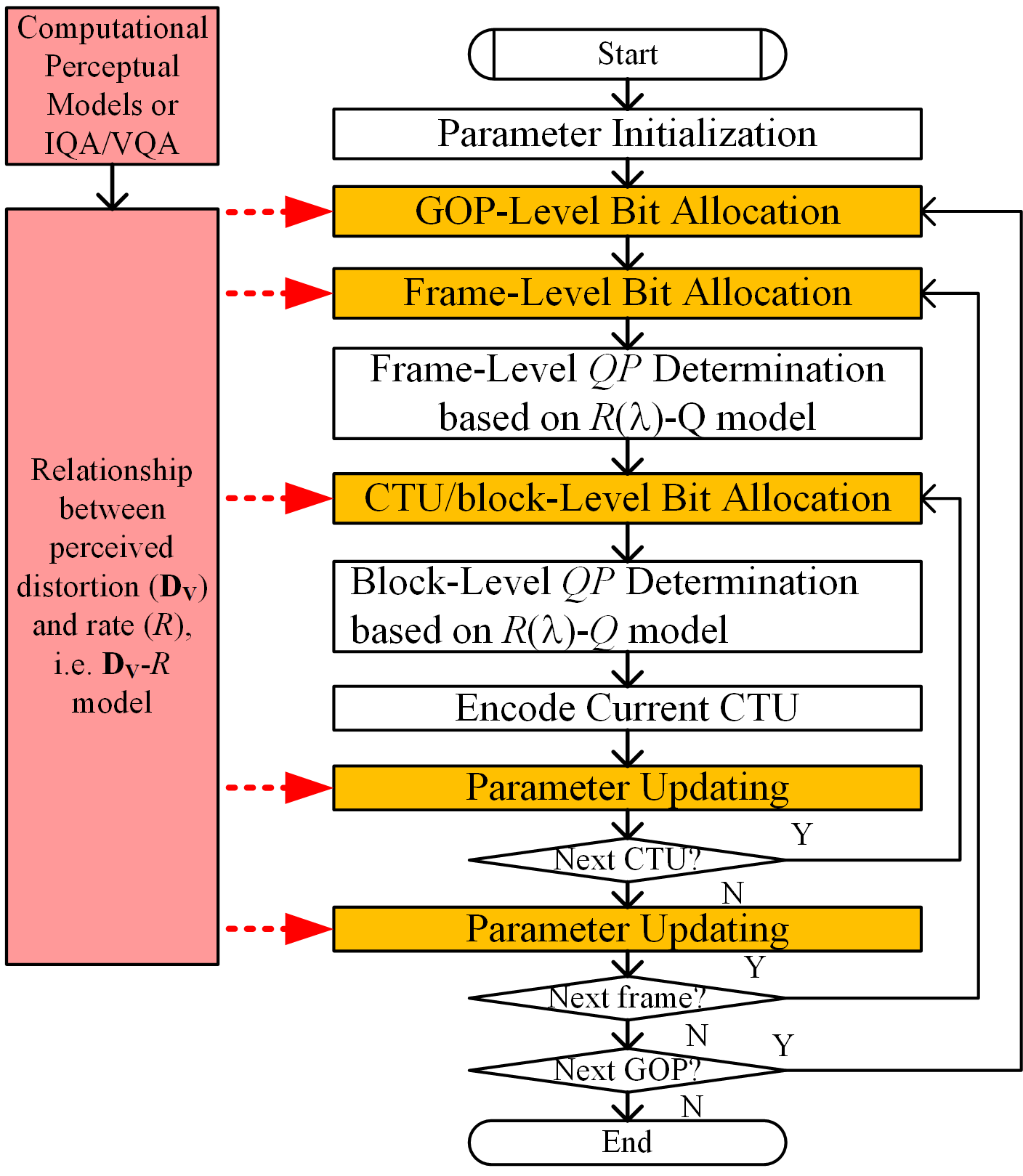}

	\caption{Flowchart of the perceptually optimized bit allocation and rate control.}
	\label{fig:PerceptualRC}
\end{figure}

\subsection{Perceptually Optimized Bit Allocation and Rate Control}
Video transmission is to transmit bitstream under the conditions of limited bandwidth and transmission delay to ensure the playback quality of video services, which has two key objectives. One is to accurately control the amount of encoded bit rate below the target bandwidth of networks to avoid buffering. The other is to maximize the visual quality of video services at the target bit rate, which requires a reasonable bit allocation. The optimization objective is formulated as
\begin{equation}
	\{QP^*_i\}=\mathop{\arg\min}\limits_{\{QP_i\}}\sum_{i=1}^{n}(\mathbf{\hat{D}_V}(QP_i)), s.t. \sum_{i=1}^{n}R_i \leq R_T,
\label{formula:obj_BA}
\end{equation}
where $QP_i$ and $R_i$ are Quantization Parameter (QP) and bit rate for unit $i$, $n$ is the number of coding units, $\mathbf{\hat{D}_V}(QP_i)$ is the visual quality degradation at $QP_i$. 

\begin{table*}[!t]
\tiny
	\renewcommand{\arraystretch}{1.1}
	\caption{Perceptually Optimized Bit Allocation and Rate Control Schemes.}
	\label{table_PBA}
	\centering
\begin{threeparttable}

	\begin{tabular}{|c|c|c|c|c|c|c|c|}
		\hline
		\multirow{2}{*}{Author/Year} &\multirow{2}{*} {Visual Features and Coding Optimizations}& \multirow{2}{*}{VQA, $Q$}  &\multicolumn{3}{c|}{BDBR($Q$)[\%]}&{Bit Error}&{$\Delta T$} \\
        \cline{4-6}
        & & &{AI}&{LD}&{RA}&[\%]&{[\%]} \\
		\hline
		{Gao'16\cite{RC_SSIM_Gao}}& {$D_{MSE}$ was replaced with $\frac{1}{SSIM}$ in game theory based bit allocation.} & {SSIM}	& {-2.74} & {/} & {/} & {0.1} & {1.0} \\
		\hline
	   \multirow{3}{*}{Zhou'19\cite{RC_SSIM_Zhou}} & {Image structural similarity from  mean, variance, and variance covariance.} &  {SSIM}	& {/}&{-14.0}&{/} & \multirow{3}{*}{0.08}&\multirow{3}{*}{$2.7$} \\
        \cline{3-6}
        &$D_{SSIM}$ was approximated as $\frac{D_{MSE}}{S^2}$ for CTU-level rate control. &{PSNR} &{/}&{-3.1}&{/}&& \\
        \cline{3-6}
        & &{MS-SSIM}&{/}&{-12.8}&{/}&& \\
	   \hline
		\multirow{2}{*}{Li'21 \cite{RC_SSIM_Li}}  & {Structural similarity $D_{SSIM}$ was approximated as $\alpha \Theta D_{MSE} + \beta$ for CTU} & {SSIM}&{-5.2}&{-12.3}&{-5.3}&\multirow{2}{*}{1.9}&\multirow{2}{*}{2.2} \\
       \cline{3-6}
        {}&{level bit allocation and RDO with joint $R$-$D_{SSIM}$-$\lambda_{SSIM}$ model}&{PSNR} &{2.7}&{2.3}&{-2.2}&& \\
       \hline
		\multirow{2}{*}{Xu'16 \cite{RC_FreeEnergy}} &{$\mathbf{\hat{D}_V}= K_p\times D_{MSE}$, where weight $K_p$ considered the strengths of } & \multirow{2}{*}{VQM}	& \multirow{2}{*}{/}&\multirow{2}{*}{-3.41}&\multirow{2}{*}{/}&	\multirow{2}{*}{/} &\multirow{2}{*}{/} \\
        &{motion ($k_{MS}$), structure ($k_{SS}$) and texture ($k_{TS}$) as $K_p=\frac{k_{MS}\times{k_{SS}}}{k_{TS}}$.}&{}&  & &  & & \\
	   \hline
		\multirow{2}{*}{Zeng'16 \cite{RC_Zeng}} &{$\mathbf{\hat{D}_V}=1+K_p \times D_{MSE}$, where weight $K_p$ to MSE was calculated}  &  \multirow{2}{*}{SSIM}	& \multirow{2}{*}{/}&  \multirow{2}{*}{-6.94}& \multirow{2}{*}{-5.71} &	\multirow{2}{*}{0.8}&\multirow{2}{*}{/} \\
        &{as spatial texture complexity multiplied by temporal motion activity.}& &  & & & &\\
        \hline
		\multirow{5}{*}{Liu'19 \cite{RC_PWMSE}}  & {}  & {PWMSE}	&{-6.27}&{/}&{/} &\multirow{5}{*}{0.1} &\multirow{5}{*}{/} \\
      \cline{3-6}
        &{PWMSE exploiting spatial and pattern masking effects, }&{PWMSE-V}&{-4.04}&{/}&{/}& & \\
       \cline{3-6}
        &{polynomial PWMSE-$Q$ and $R$-$Q$ models were used for bit }&{SSIM}&{-8.0}&{/}&{/}& & \\
        \cline{3-6}
         & {allocation and rate control. }&{VIF}&{-1.57}&{/}&{/}& {}&{} \\
        \cline{3-6}
         &  &{MS-SSIM}&{-7.85} &{/}&{/}& {}&{} \\
      \hline
		{Gao'22 \cite{RC_ConstVQ}} & {Consistent quality based rate control considering SSIM and MSE variations.}  &{SSIM}&{/} &{-8.12}&{/}&{0.5} &{/} \\
      \hline
		\multirow{2}{*}{Xiang'22\cite{RC_Xiang}}  &\multirow{2}{*} {$D_{MSE}$ was suppressed by a masking effect or JND threshold.} & {PSNR}	&{-3.6}&{/}&{/} &\multirow{2}{*}{0.1} &\multirow{2}{*}{5.05} \\
      \cline{3-6}
         &{}&{PSPNR}&{-6.4} &{/}&{/}& {}&{} \\
      \hline
		{Li'22 [54]}  & {ROI and JND for region-level Intra-period determination.}  & {PSNR}	&{/}&{-5.63}&{/} &{/} &{/} \\
      \hline
		\multirow{2}{*}{Zhou'20\cite{RC_Zhou_JND}}  &{Block level JND factor based on masking effect and brightness contrast.}  & {PSNR}	&{/}&{-3.30}&{/} &\multirow{2}{*}{0.08}&\multirow{2}{*}{15.10} \\
      \cline{3-6}
        &{was used as perceptual weight in RD model for rate control.}& {SSIM}&{/}&{-6.50}&{/}& & \\
        \hline
		{Lim'20\cite{RC_Lim}}  & {QP determination considering luminance adaptation effect based JND.}  & {MOS}	&{/}&{/}&{/} &{0.22} &{0.36} \\
       \hline
      {Yang'16\cite{RC_Yang}}  & {CTU-level bit allocation considering spatial GM and temporal GMSD.}  & {GMSD}	&{/}&{-7.43}&{-10.4} &{0.3} &{/} \\
      \hline
      \multirow{3}{*}{Wang'18\cite{RC_Mask_Wang}} &{Masking effects based rate control considering GMSD, } &  {PSNR}	& {/}&{-1.29}&{/} & \multirow{3}{*}{0.19}&\multirow{3}{*}{1.2} \\
        \cline{3-6}
        &{motion information and texture complexity.}&{SSIM} &{/}&{-3.81}&{/}&& \\
        \cline{3-6}
        & &{MOS}&{/}&{-4.67}&{/}&& \\
      \hline
		{Chao 16'\cite{RC_JND}}  & {CU level bit allocation based on local JND.}  & {DMOS}	&\multicolumn{3}{c|}{{-14.0}} &{/}&{/} \\
\hline
	\end{tabular}
   \end{threeparttable}
\end{table*}

\begin{figure}[!t]
	\centering
	\subfigure[]{
		\label{fig:SSIM-DMSE-I}
	\includegraphics[width=0.48\linewidth]{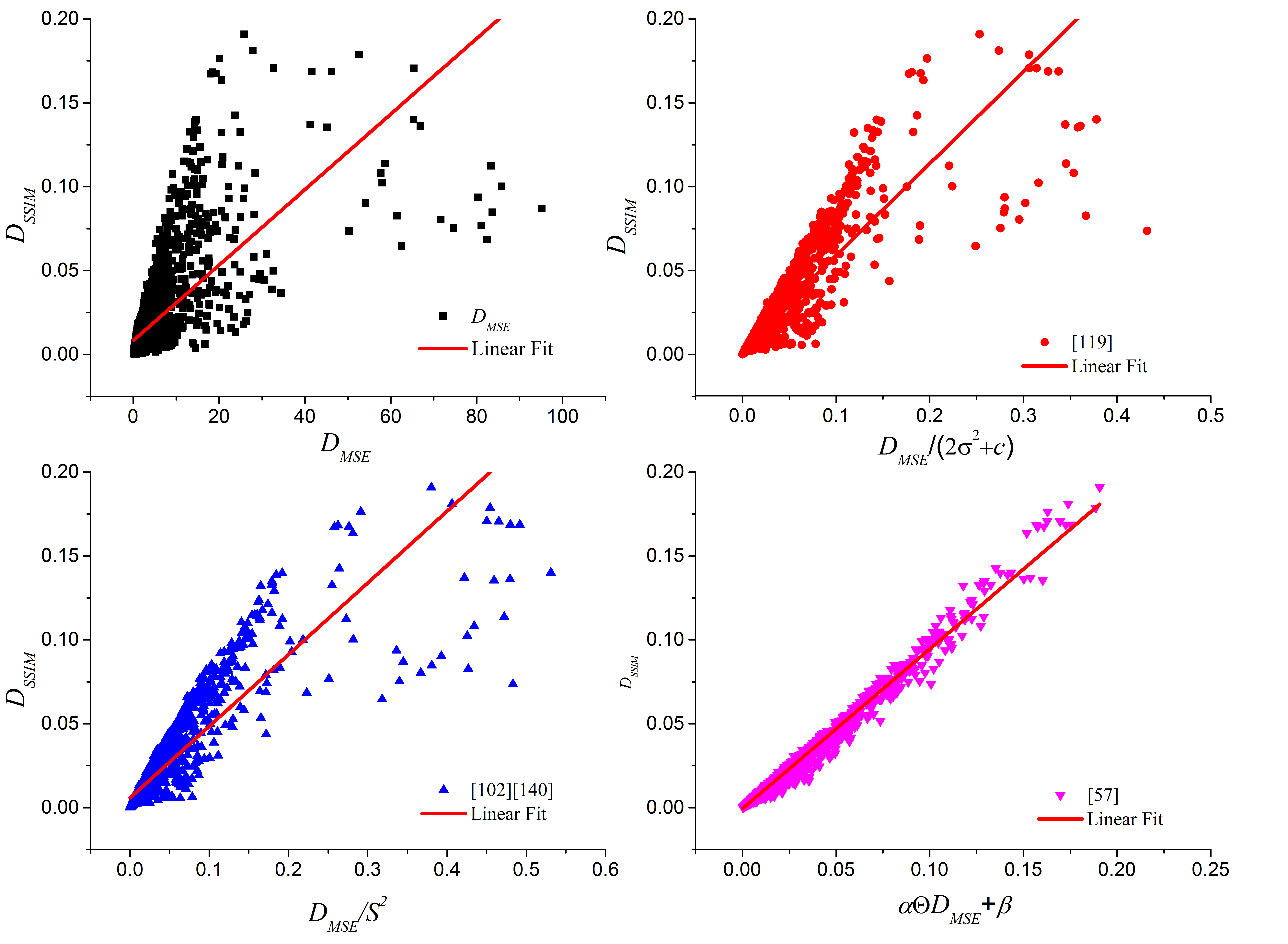}}
	\subfigure[]{
	   \label{fig:SSIM-DMSE-B}
	\includegraphics[width=0.48\linewidth]{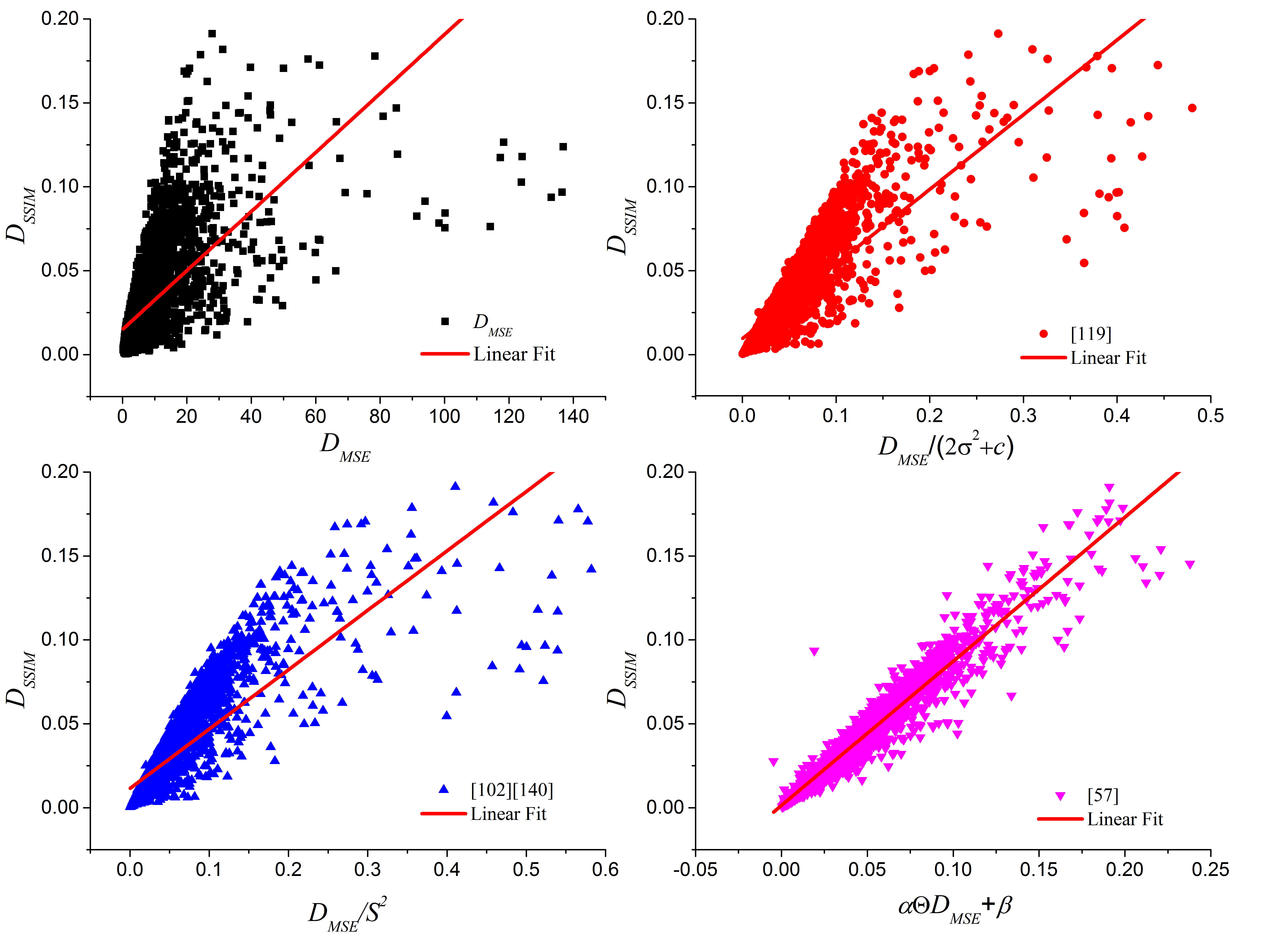}}

	\caption{Relationships between $D_{SSIM}$ and four $D_{MSE}$ based approximation models, i.e., $D_{MSE}$, \cite{RC_SSIM_SQWang,RC_SSIM_Zhou},\cite{RDO_Yeo}, and \cite{RC_SSIM_Li}, respectively.(a) Intra frames, $R^2$ of the four linearly fitted models are (0.482, 0.791, 0.787, 0.987). (b) Inter frames, $R^2$ are (0.458, 0.694, 0.686, 0.953).}
	\label{fig:SSIM-DMSE}
\end{figure}

Table \ref{table_PBA} illustrates representative works on perceptually optimized bit allocation and rate control algorithms, which includes key features, quality metric, Bj{\o}ntegaard Delta Bit Rate (BDBR) \cite{BDBRCalc}, rate accuracy and complexity overhead ($\Delta T$). These works can be divided into two types: quality metric-based and perceptual factor-based. In the quality metric-based category, Gao \emph{et al.} \cite{RC_SSIM_Gao} proposed a SSIM-based game theory approach for Coding Tree Unit (CTU)-level bit allocation in intra frames, where MSE was replaced with $\frac{1}{SSIM}$ in measuring the visual distortion $\mathbf{\hat{D}_V}$, i.e., $\mathbf{\hat{D}_V}=D_{SSIM}=\frac{1}{SSIM}$. Then, SSIM-$QP$ relationship was studied for bit allocation. However, coding modules in the original video encoder were implemented with the target of optimizing MSE/Mean Absolute Difference (MAD), it was complicated to implement all their optimization criteria to be SSIM. To handle this problem, Zhou \emph{et al.} \cite{RC_SSIM_Zhou} extended the divisive normalized SSIM-based RD model \cite{RC_SSIM_SQWang} from Discrete Cosine Transform (DCT)-domain, and the relationship between 1-SSIM and MSE was built as $D_{SSIM}\approx\frac{D_{MSE}}{S^2}$, where $S^2$ is a ratio of DCT energies between the current block and entire frame. Such that optimizing SSIM can be approximately achieved by using MSE as $\mathbf{\hat{D}_V}=D_{SSIM}\approx\frac{D_{MSE}}{S^2}$, which was then applied to CTU-level rate control and global $QP$ determination. Li \emph{et al.} \cite{RC_SSIM_Li} proposed a more accurate relationship between SSIM and MSE as $D_{SSIM}={\alpha}{\Theta}{D_{MSE}}+\beta$, where $\alpha$ and $\beta$ were linear model parameters updated every block, $\Theta$ related to the texture variance of image contents. Consequently, $D_{SSIM}-\lambda$ and $D_{SSIM}-QP$ models were derived for bit allocation.
Fig.\ref{fig:SSIM-DMSE} shows relationship analyses between $D_{SSIM}$ and four $D_{MSE}$ based approximation models, i.e., $D_{MSE}$,\cite{RC_SSIM_SQWang,RC_SSIM_Zhou},\cite{RDO_Yeo}, and \cite{RC_SSIM_Li}. The dot data was collected from encoding four sequences with HEVC at four $QP\in\{22,27,32,37\}$ and solid line was linear fit of the data. $R^2$ of the fitted $D_{SSIM}$ and $D_{MSE}$ are 0.482 and 0.458 for All Intra (AI) and Low Delay (LD) configurations, respectively, which are low. The models in \cite{RC_SSIM_SQWang,RC_SSIM_Zhou,RDO_Yeo} improved the $R^2$ to (0.791,0.787) and (0.694, 0.686), respectively. Li \emph{et al} \cite{RC_SSIM_Li} further improved the fitting accuracy ($R^2$) to 0.987 and 0.953, respectively, which was highly accurate. A more accurate relation between SSIM and MSE lead a better consistency between optimization objective and codec implementation.

In addition to SSIM, Xu \emph{et al.} \cite{RC_FreeEnergy} proposed a novel Free-energy Principle inspired Video Quality metric (FePVQ), in which strengths of motion, structure and texture were considered as a perceptual weight $K_p$ for local MSE, i.e., $\mathbf{\hat{D}_V}=K_p \times D_{MSE}$. Then, this FePVQ was used bit allocation and RDO to allocate more bits to quality sensitive regions. Similarly, perceptual weight to MSE was modelled with spatial texture complexity and temporal motion activity\cite{RC_Zeng}, which guided the frame and CTU-level bit allocations. Motivated by the PWMSE metric \cite{IQA_PWMSE}, Liu \emph{et al.} \cite{RC_PWMSE} applied PWMSE as the visual objective ($\mathbf{\hat{D}_V}=D_{PWMSE}$) and proposed a perceptual CTU level bit allocation algorithm for HEVC intra coding, thereby minimizing the perceptual distortion of each CTU under a given bit rate constraint. It achieved 6.27\% bit rate reduction and 0.38 dB Bj{\o}ntegaard Delta Perceptually Weighted Peak Signal to Noise Ratio (BD-PWPSNR) gain on average. Gao \emph{et al.} \cite{RC_ConstVQ} proposed a rate control scheme to maintain consistent image quality among adjacent frames, where PSNR variations among frames was used as an additional key objective in $\mathbf{\hat{D}_V}$. Xiang \emph{et al.}\cite{RC_Xiang} modeled the perceptual distortion $\mathbf{\hat{D}_V}$ as signal distortion $D_{MSE}$ subtracted by a masking effect or JND threshold $M$ , i.e., $\mathbf{\hat{D}_V}=D_{MSE}-M$, which was then applied to CTU-level $\lambda$-domain rate control. It achieved 6.4\% BDBR gain when the quality was measured with a JND based Peak Signal to Perceptual Noise Ratio (PSPNR). Meanwhile, the quality consistency measured by PSNR deviation among frames was also improved. In these schemes, more visually plausible quality metrics, such as SSIM, PWMSE, or PSNR variation, were directly used to replace the MSE in the bit allocation objective. There are two key problems to be solved. One is the accuracy of the used visual quality metric, i.e., $\mathbf{\hat{D}_V}$ is approximately measured with IQA, and the other is approximation accuracy between IQA and MSE for easy codec implementation. The two approximations lead to losses of accuracy and generality. Meanwhile, these quality metrics used in coding are IQAs. It will be more complicated to the VQAs considering temporal distortion.

In the other category, perceptual factors were exploited in optimizing video compression, which were regarded as indirect approaches. Li \emph{et al.}\cite{RC_Li} proposed a ROI based perceptual video coding scheme, where a region-level Intra-period was determined based on ROI and JND to reduce error propagation. Also, a ROI based bit allocation was used to adjust bit budgets for Intra blocks.
Since HVS is more sensitive to textural and motion regions, Yang \emph{et al.} \cite{RC_Yang} proposed to allocate more bits to perceptually sensitive regions with larger spatial Gradient Magnitude (GM) and temporal Gradient Magnitude Similarity Deviation (GMSD). In \cite{RC_Mask_Wang}, a masking effect based perceptual model was proposed by considering texture complexity and motion. This model was then used to allocate fewer bits to CTUs in the insensitive masking regions. Lim \emph{et al.}\cite{RC_Lim} proposed a perceptual rate control algorithm that assigned CTU-level bits by considering luminance adaptation effect based JND. The subjective MOS of decoded sequence was improved about 0.19. Based on the pixel-level JND of masking effect and brightness contrast, Zhou \emph{et al.}\cite{RC_Zhou_JND} derived a block-level JND factor, which was used as a perceptual weight in the RD model for frame and CTU-level rate control. Chao \emph{et al.} \cite{RC_JND} proposed a local JND model by considering contrast sensitivity, transparent masking, temporal masking, residual correlation adjustment and quantization distortion. Then, $QP$ was gradually increased to allocate fewer bits when the visual distortion was within the JND threshold. In this category, these approaches were not specifically designed for an IQA/VQA metric. Instead, perceptual factors, such as masking effect, visual sensitivity, JND or visual attention, were exploited for better generality since they were fundamental features in visual perception. However, the disadvantage is that only one or two key properties were exploited, which removed a partial of the visual redundancies. Moreover, inaccurate computational perceptual model and indirect objective may degrade the coding gain at a target IQA/VQA .

According to Table \ref{table_PBA}, average bit error of these perceptually optimized schemes was from 0.08\% to 1.9\%, which was highly accurate in rate control. In addition, their complexity overhead was from 0.36\% to 15.10\%, which was mainly from calculating the perceptual models. Figure \ref{fig:PerceptualRC} shows a general flowchart of perceptually optimized bit allocation and rate control, which includes visual perception models and enhanced bit allocation. Different from the pixel-by-pixel MSE, visual perception models, such as SSIM, ROI, JND and visual sensitivity, indicate the relative visual importance of Group-of-Picture (GOP), frame or region in videos. So, the relationship between perceptual distortion and rate ($\mathbf{D}_V$-$R$) shall be established, which is used to guide GOP/frame/CTU-level bit allocation algorithms by assigning more bits to perceptually sensitive regions and fewer bits to perceptually insensitive regions. Then, based on the assigned bits $R$, $QP$ is determined using $R$-$Q$ or $R$-$\lambda$-$Q$ models. Finally, parameters of these relationship models will be updated in encoding each unit. Consequently, the overall quality of compressed video becomes more visually plausible. The advantage of these works is that the number of bits can be more reasonably allocated with the guidance of visual perception. However, there are some aspects to be further improved: 1) HVS is a complicated non-linear system that has multiple visual properties with compound effects. Existing computational perceptual models usually modelled one or two key visual properties, which can hardly model the HVS accurately. Thus, developing more complete and accurate perceptual models will be helpful in exploiting visual redundancies further. 2) Although the perceptual factors and quality models optimize the bit allocation, some other coding modules, such block matching in predictive coding, mode decision, loop filtering and RDO still use MSE based distortion criteria, which does not conform with the perceptual bit allocation. 3) Different quality assessment metrics may not be consistent with each other. Also, they measure quality at image or video-level, which is different from coding unit processed at block-level. 4) It is difficult to determine the optimal bits and QPs for some perceptual models, such as learning based schemes that cannot be mathematically expressed and piece-wise JND function that is non-derivative. In these cases, convex is not guaranteed and Lagrange multiplier method is not applicable in solving their RD minimization problem. Further adaptations or approximations are required.

\subsection{Perceptual Rate Distortion Optimization (PRDO)}

Mainstream video coding is a hybrid framework, which includes predictive coding, transform coding, entropy coding and filtering/enhancement processing. Excluding the entropy coding, RDO \cite{OV_HEVC} is to select the optimal mode or parameter among a set of candidates by minimizing the Rate-Distortion (RD) cost consists of distortion $\mathbf{D}$ and coding bit rate $R$. The RDO has been used in many coding modules such as variable-size Coding Unit (CU)/Prediction Unit (PU) modes in the predictive coding, angular intra modes in intra prediction, variable Transform Unit (TU) sizes and transform kernels (e.g., Discrete Sine Transform (DST) and DCT kernels) in transform coding, and parameter determination in in-loop filter. Traditionally, the distortion $\mathbf{D}$ is measured by MSE or MAD. Since the ultimate objective of video coding is to minimize visual distortion $\mathbf{D_V}$ at a given bit rate according to Eq. \ref{formula:obj_fun2}, in PRDO, the optimal parameter $\alpha^*$ is selected by minimizing a perceptual RD cost $J(\mathbf{B},\mathbf{\hat{B}})$ as
\begin{equation}
 \begin{cases}
\alpha^* = \mathop {\min }\limits_{{{\alpha}} \in \mathbf{A}} J(\mathbf{B},\mathbf{\hat{B}}(\alpha))\\
J(\mathbf{B},\mathbf{\hat{B}}(\alpha))=\mathbf{\hat{D}_V}(\mathbf{B},\mathbf{\hat{B}}(\alpha))+\lambda R
\end{cases},
\label{formula:RPDO_fun}
\end{equation}
where $\mathbf{B}$ and $\mathbf{\hat{B}}$ are the reference and reconstructed blocks, $\alpha$ is a parameter from set $\mathbf{A}$; $\mathbf{\hat{D}_V}$ is the visual distortion between $\mathbf{B}$ and $\mathbf{\hat{B}}$, $R$ is the total encoding bits of block $\mathbf{B}$, which includes the bits of residual, mode index and other parameters, $\lambda$ is a Lagrange multiplier adjusting the relative importance between distortion and rate, which can be derived by $\frac{\partial\mathbf{\hat{D}_V}}{\partial R}$ with convex optimization.

\begin{table*}[!t]
\tiny
	\renewcommand{\arraystretch}{1.1}
	\caption{Representative PRDO Schemes.}
	\label{table_PRDO}
	\centering
\begin{threeparttable}

	\begin{tabular}{|c|c|c|c|c|c|c|c|}
		\hline
		\multirow{2}{*}{Author/Year} &\multirow{2}{*} {Visual Features}& \multirow{2}{*}{Codec}& \multirow{2}{*}{VQA}  &\multicolumn{3}{c|}{BDBR($Q$)[\%]}&{$\Delta T$} \\
        \cline{5-7}
        & & & &AI&LD&RA&[\%] \\
		\hline
		Wang'12 \cite{RDO_SSIM_Wang}   & {$D_{MSE}$ was replaced with 1-SSIM}& {H.264/JM15.1} & {SSIM}	& /&{-16.28}& {-7.74} &{6.18} \\
		\hline
	   \multirow{2}{*}{Yeo'13 \cite{RDO_Yeo}}  &\multirow{2}{*} {Approximated $D_{SSIM}$ with $\frac{D_{MSE}}{2\sigma^2+c}$}& \multirow{2}{*}{H.264/JM17.2} &  {SSIM }	&{-8.1}&{-9.8}&{-14.0} &\multirow{2}{*}{1.0} \\
        \cline{4-7}
        & & & PSNR &3.9&6.8&6.0& \\
	   \hline
		{Wang'13 \cite{RDO_Wang_TIP}}  & {Masking effect, SSIM-based divisive normalization}&{H.264/JM15.1} & {SSIM}	&/&{-15.8}&/ &{/} \\
	   \hline
		{Lee'18 \cite{RDO_Lee}}  & {$D_{SSIM}\approx\frac{D_{MSE}}{2\sigma^2+c}$, empirical Lagrange multiplier clip}&{HEVC/HM16.0} & {SSIM}	&/&{-8.1}&{-4.0} &{7.1} \\
  \hline
		\multirow{2}{*}{Wang'19 \cite{RDO_Wang_ICIP}} &\multirow{2}{*}{Weighted combination of $D_{MSE}$ and $D_{SSIM}$} &\multirow{2}{*}{HEVC/HM16.14} & {SSIM}	&/ &{-4.65}&/ &\multirow{2}{*}{/} \\
        \cline{4-7}
        &{}& & PSNR & /& 3.27 &/ & \\
	   \hline
		\multirow{2}{*}{Wu'20 \cite{RDO_PWMSE}} &{MSE was replaced with the PWMSE considering } &	\multirow{2}{*}{HEVC/HM16.0} & {PWMSE}	&/& {-14.92\tnote{*}}&{-27.12\tnote{*}} &\multirow{2}{*}{/} \\
        \cline{4-7}
        &{spatial and temporal masking effects}& & MOS &/ & \multicolumn{2}{c|}{-17.55} & \\
        \hline
		\multirow{4}{*}{Luo'21 \cite{RDO_Luo_VMAF}}  & {Built a linear relationship between VMAF difference} &\multirow{3}{*}{HEVC/HM16.20} & {VMAF}	& /&{-3.61}&{-2.67} &\multirow{4}{*}{/} \\
      \cline{4-7}
        &{ and MSE difference with pre-analysis, block-level   }& & PSNR  & /&4.72  & 3.60& \\
       \cline{4-7}
        &{Lagrange multiplier is adjusted based on the }& & SSIM  &/ &3.66  & 2.06& \\
        \cline{3-7}
         & {correlation}&{VVC/VTM10.0} & VMAF  &/ &-1.92  & -1.20& \\
        \hline
		Jung'15 \cite{RDO_Jung} & {FEJND from spatial and temporal masking effects} &{H.264/JM17.2} & {PSNR}	&/&/ &{-8.46}& {/} \\
      \hline
		{Bae'16 \cite{RDO_Bae} } & {DCT domain JND directed suppression for SSE } &{HEVC/HM11.0} & {MOS}	&/ &{-12.10}&{-9.90} &{/} \\
    \hline
		\multirow{2}{*}{Yang'17 \cite{RDO_Yang17}} & {GMR for spatial feature and GMSDR for} &\multirow{2}{*}{HEVC/HM10.0} & {SSIM}& /& -6.18 &-5.95 &\multirow{2}{*}{/} \\
        \cline{4-7}
        &{ temporal feature}& & GMSD  &/ &-19.80  & -16.24& \\
      \hline
		\multirow{2}{*}{Rouis'18 \cite{RDO_Rouis}} & {Temporal correction, spectral saliency } &\multirow{2}{*}{HEVC/HM16.12} & {PWMSE}& /& -6.14 &-4.41 &\multirow{2}{*}{3 to 5} \\
        \cline{4-7}
        &{and visual stationary tuned the Lagrange Multiplier }& & SSIM  &/ &-6.95  & -9.86& \\

   \hline
		{Cui'21 \cite{RDO_Cui_UHD}}  & {JND of UHD/HDR based on saliency, CSF, LM and GDE} &{HEVC/HM16.20} & {DMOS}	&/ &{-35.93}&{-24.93} &{19.26} \\
   \hline
		{Liu'18 \cite{RDO_Liu}}  & {binocular combination distortion from left and right view } &{MV-HEVC} & {Bino-PSNR}	& /&/&{-5.93} &{1.0} \\
 \hline
		\multirow{2}{*}{Zhang'16 \cite{RDO_3DSVQM}}  & \multirow{2}{*}{Flickering distortion in 3D synthesized video} &\multirow{2}{*}{3DVC/3D-HTM} & {SVQM}	&/ &/&{-15.27} &\multirow{2}{*}{/} \\
 \cline{4-7}
        & && SIAT-VQA  &/ & / & -14.58& \\
        \hline
	\end{tabular}
   \begin{tablenotes}
     \item[*] BDBR($Q$) subjects to quality degradation.
   \end{tablenotes}
   \end{threeparttable}
\end{table*}

A number of perceptual quality models \cite{RDO_Lee,RDO_SSIM_Wang,RDO_Yeo,RDO_Wang_TIP,RC_SSIM_Li,RDO_PWMSE,RDO_Wang_ICIP} and perceptual factors \cite{RDO_Jung,RDO_Bae,RDO_Rouis,RDO_Liu,RDO_3DSVQM} have been investigated to improve the PRDO in Eq.\ref{formula:RPDO_fun}. Table \ref{table_PRDO} shows their visual features and coding gains of the featured PRDO schemes. To improve the RDO with perceptual quality models, Wang \emph{et al.} \cite{RDO_SSIM_Wang} proposed SSIM-motivated RDO for block mode decision in video coding, where the MSE based distortion term in the cost function was replaced with 1-SSIM, i.e., $\mathbf{\hat{D}_V}=1-{SSIM}$. Then, frame-level Lagrange multiplier adaptation was developed based on a reduced-reference SSIM estimation since the whole distorted frame is not available during the coding process. Meanwhile, marcoblock-level Lagrange multiplier adaptation was developed by considering moving content and motion perception. Yeo \emph{et al.} \cite{RDO_Yeo} approximated SSIM between the reference and reconstructed blocks with their MSE $D_{MSE}$ divided by the variance of the current block ($\sigma^2$), i.e., $D_{SSIM}\approx\frac{D_{MSE}}{2\sigma^2+c}$, where $c$ is a constant. Then, MSE based RDO was modified to be SSIM based RDO by scaling the Lagrange multiplier with a weighting factor, which was calculated as local variance normalized by mean variance of the entire frame, i.e., $\frac{\sigma^2}{\bar{\sigma}^2}$). Meanwhile, QP adaption was used for each marcoblock. In \cite{RDO_Wang_TIP}, based on a DCT domain SSIM index, DCT coefficients were divisively normalized to a perceptually uniform space, which determined the relative perceptual importance of each marcoblock by exploiting the masking effect of HVS. Then, the distortion in RDO was measured with and MSE between the normalized DCT coefficients of the reference and distorted blocks. These works were proposed for H.264/AVC. Lee \emph{et al.} \cite{RDO_Lee} applied Yeo's $D_{SSIM}$-$D_{MSE}$ model \cite{RDO_Yeo} to HEVC and an empirical clip was designed to constrain Lagrange multiplier variations among frames. Similarly, approximated $D_{SSIM}$-$D_{MSE}$ models in \cite{RC_SSIM_Gao,RC_SSIM_Zhou,RC_SSIM_Li} can also be applicable to the PRDO. In \cite{RDO_Wang_ICIP}, input video was divided into two classes, i.e., simple/regular textural regions with high sensitivity and complex textural regions with low sensitivity, based on the free-energy principle. Then, a weighted combination of fidelity ($D_{SSE}$) and perceptual distortion ($D_{SSIM}$), i.e., $\mathbf{\hat{D}_V}=\omega D_{SSIM}+D_{MSE}$, was used as the final distortion metric in the PRDO, where a larger weight $\omega$ was given to the distortion in simple/regular texture regions. Similarly, in \cite{RDO_PWMSE}, PWMSE-V \cite{VQA_PWMSE} that considered spatial and temporal masking modulation was applied to the distortion term in RDO, and then Lagrange multiplier adaptation was derived accordingly. Luo \emph{et al.} \cite{RDO_Luo_VMAF} built a linear relationship between block-level VMAF degradation and MSE difference from multi-pass pre-coding, which derived a block-level Lagrange multiplier adaptation. However, when measured with VMAF, only 3.61\% and 2.67\% BDBR gains were achieved for LD and Random Access (RA) configurations, respectively. Meanwhile, BDBR lose if measured with PSNR and SSIM.

The second category is to improve the RDO by exploiting the perceptual factors, such as JND and masking effects.
Based on the spatiotemporal masking effects, Jung \emph{et al.} \cite{RDO_Jung} proposed a Free-Energy based JND (FEJND) model to identify the disorderliness of video content, which was applied to adjust the Lagrange multiplier and QP values in RDO process. Bae \emph{et al.} \cite{RDO_Bae} proposed a generalized JND (GJND) model in DCT domain which bridged the gap between 8$\times$8 DCT transform in JND calculation and variable-size transform in video coding. The GJND adapted transformed coefficients to different DCT kernel sizes. Then, GJND directed suppression was applied in calculating MSE based distortion at different $QP$s. It was reported that larger coding gains were mainly achieved from small $QP$ settings.
Yang \emph{et al.} \cite{RDO_Yang17} adopted Gradient Magnitude Ratio (GMR) for spatial feature and GMSD Ratio (GMSDR) for temporal feature, which were divisively combined as a perceptual weight for adjusting Lagrange multiplier in RDO. Rouis \emph{et al.} \cite{RDO_Rouis} adjusted CTU-level Lagrange multiplier in the RDO by jointly considering temporal correlation, visual stationary and spectral saliency. Similarly, Cui \emph{et al.} \cite{RDO_Cui_UHD} built a JND model for UHD and HDR video based on visual saliency, CSF, LM effect, and Gaussian Differential Entropy (GDE), which was incorporated into a weighting factor to scale the Lagrange multiplier in RDO.
Liu \emph{et al.} \cite{RDO_Liu} developed a binocular combination-oriented distortion model by binocularly weighting the distortions from left and right views based on stereoscopic visual perception of two eyes in HVS. Then, this binocular distortion was applied to optimize symmetrical and asymmetrical stereoscopic video coding, in which the Lagrange multiplier in encoding the right view was scaled. Zhang \emph{et al.} \cite{RDO_3DSVQM} proposed a full reference Synthesized Video Quality Metric (SVQM) to measure the perceptual quality of the synthesized video. Then, the synthesized video distortion and depth distortion were combined as a distortion term in 3D RDO for depth coding with the target of minimizing the perceptual distortion of synthesized view at the given bit rate.

	In addition to the RD performance, computational complexity is another important aspect of the PRDO. As shown in Table \ref{table_PRDO}, it was reported that the complexity overhead ($\Delta T$) of these PRDOs increases from 1.00\% to 19.26\% on average \cite{RDO_SSIM_Wang,RDO_Yeo,RDO_Lee,RDO_Cui_UHD,RDO_Rouis,RDO_Liu}. For most of these PRDO works, they built mapping relations between  $\mathbf{\hat{D}_V}$ and ${D_{MSE}}$ with perceptual weights. The computational complexity is mainly from computing the perceptual weights of $\mathbf{\hat{D}_V}$ or Lagrange multiplier adjustment for each block. Instead of being repeatedly calculated like ${D_{MSE}}$ in the coding loop, the perceptual weights were usually calculated only once or reused in CTU for the PRDO of each coding block, which reduced the complexity overhead.

The general workflow of the PRDO is adding perceptual quality metrics or perceptual factors, such as JND \cite{RDO_Jung,RDO_Bae}, masking effect \cite{RDO_Cui_UHD} and saliency \cite{RDO_Rouis}, to the MSE based distortion term of the RD cost function. Then, Lagrange multiplier is adjusted correspondingly based on the perceptual distortion analysis. For some methods, such as \cite{RDO_Rouis,RDO_Wang_ICIP}, QP is adjusted accordingly to allocate bits more reasonably. By using this PRDO criteria, smaller partitions, refined modes or motion parameters were allocated to perceptually significant regions, and coarse modes will be used in perceptually insensitive regions. However, there are two key problems to be addressed. Firstly, since the HVS is complex and not fully understood, the current available models only exploited partial properties of the HVS. It is challenging to build a perceptual model that can simulate HVS accurately. Besides,the available quality assessment models were usually proposed and validated in dataset with coarse-grain quality scores, e.g., five quality scales for all distortion levels. A fine-grain distortion measurement specified for compression distortion is required in coding optimization. Secondly, video encoder is a complex system consists of hundreds of coding algorithms, which have been tuning to their optimal based on MSE. When the distortion metric in objective function is changed, it requires unaffordable working load to modify coding modules one-by-one to their optimal. For example, MAD based criteria was used in motion estimation, which has scarcely been considered in the PRDO. In fact, one shortcut way is to establish a relationship between perceptual models and the conventional MSE/MAD, such as \cite{RDO_PWMSE,RDO_Liu,RDO_3DSVQM,RDO_Luo_VMAF}. Then, Lagrange multiplier is adjusted in practical implementation. However, perceptual models can hardly be presented by MSE/MAD and inaccurate approximation may degrade the compression efficiency.

\subsection{Perceptually Optimized Transform and Quantization}

\begin{figure}[!t]
	\centering
	\includegraphics[width=0.5\linewidth]{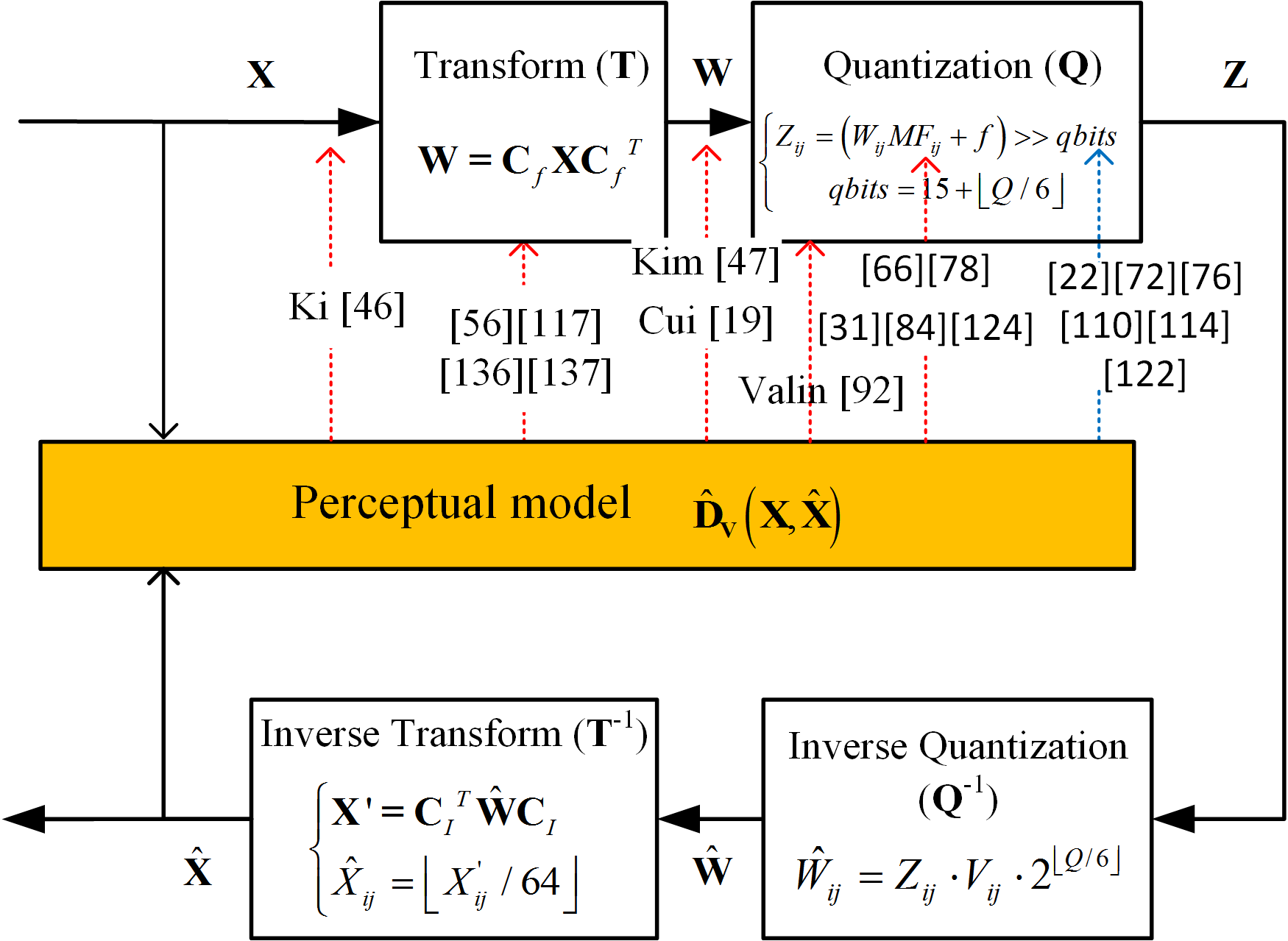}

	\caption{Framework of perceptually optimized transform and quantization.}
	\label{fig:Quantization}
\end{figure}

Let $\mathbf{X}$ be input signal, $\mathbf{A}$ and $\mathbf{A}_{inv}$ be kernels for forward and inverse transform. Forward transform and quantization, and their inverse operations can be presented as
\begin{equation}
 \begin{cases}
 \mathbf{Y}=\mathbf{A\times X}\\
 \mathbf{Z}=\mathbf{Y}./(\mathbf{Q}\times q_s) \\
 \mathbf{\hat{Y}}= \mathbf{Z}.\times\mathbf{Q}\times q_s \\
 \mathbf{\hat{X}}=\mathbf{A}_{inv}\times\mathbf{\hat{Y}}
\end{cases},
\label{formula:Transform}
\end{equation}
where $./$ and $.\times$ are element-wise division and multiplication, respectively. $\mathbf{\hat{X}}$ and $\mathbf{\hat{Y}}$ are reconstructed $\mathbf{X}$ and $\mathbf{Y}$, $\mathbf{Z}$ is a matrix of transformed coefficients input to entropy coding, $\mathbf{Q}$ is a quantization matrix, and $q_s$ is a scalar quantization step. The reconstructed coefficients after inverse quantization are no longer the same as the input ones $\mathbf{\hat{X}}\neq\mathbf{X}$, i.e., lossy coding. While using a larger QP, i.e., $q_s$, more coding bits can be saved, i.e., higher compression ratio, but larger quality degradation will be caused at the meantime, as shown in Fig. \ref{fig:Quantization}. Perceptually optimized transform and quantization are to minimize the perceptual loss at a given bit rate, which can be formulated as
\begin{equation}
\begin{cases}
	\{\mathbf{A^*,A^*}_{inv},\mathbf{Q^*}\}=\mathop{\arg\min}\limits_{\{\mathbf{A,A}_{inv},\mathbf{Q}\}}\sum_{i=1}^{n}(\mathbf{\hat{D}_V}(\mathbf{X}_i,\mathbf{\hat{X}}_i))\\
    \sum_{i=1}^{n}R(\mathbf{Z}_i)\leq R_T
\end{cases},
\label{formula:obj_Trans}
\end{equation}
where $\mathbf{X}_i$, $\mathbf{\hat{X}}_i$ and $\mathbf{Z}_i$ are $\mathbf{X}$, $\mathbf{\hat{X}}$ and $\mathbf{Z}$ of coding unit $i$, respectively; $\mathbf{\hat{D}_V}(.)$ is a visual quality measurement, $R(.)$ counts the coding bits. Framework of perceptually optimized transform and quantization, and their related works are shown in Fig.\ref{fig:Quantization}.

Transform coding is to transform the input residue from predictive coding to de-correlate or compact energy of coefficients. It transforms the input residue from spatial to frequency domain, which can be regarded as a dimension reduction. Usually, the input and the reconstructed signals from transform and inverse transform are lossless, such as wavelet transform, DCT, and KLT. DCT or its variants, such as Integer Cosine Transform (ICT), DCT type II, and DST, have been widely used in the video coding standards, such as H.264/AVC, HEVC and VVC. The ICT is the integer version of DCT, which is presented as $\mathbf{{Y}=\mathbf{W}\bigotimes\mathbf{E}}$. The ICT is composed of a orthogonal transform ($\mathbf{{W}}=\mathbf{{C_fXC^T_f}}$) and a scaling ($\bigotimes\mathbf{{E}}$), where $\mathbf{C_f}$ is a forward transform kernel, $\mathbf{E}$ is a scaling matrix, and $\bigotimes$ is an element-wise multiplication. The scaling operation ($\bigotimes\mathbf{{E}}$) is incorporated into quantization $\mathbf{Q}$, where multiplication is replaced with additions and shifts for low complexity. Based on the directional pattern of image blocks, Zhao \emph{et al.} \cite{TQ_EMT} proposed an Enhanced Multiple Transform (EMT) by selecting the optimal from a number of sinusoidal transform kernels. To further improve the energy compaction, a Non-Separable Secondary Transform (NSST) \cite{TQ_NSST} was proposed to transform the coefficients output from the primary EMT, which has been adopted in the latest VVC. Moreover, data-driven transform was further investigated to exploit data statistics. Saab transform \cite{TQ_Saab} learned multistage transform kernels by exploiting the directional properties of intra predicted residual data, while deep learning-based transform \cite{TQ_Yang2020} took advantage of the non-linear representation ability of CNN to improve the energy compaction. In summary, the current researches mainly focused on developing transform kernels or multistage transform to improve the capability of energy compaction or de-correlation. Since transform and inverse transform are lossless, perceptual distortion is scarcely concerned. In fact, since transform is followed by lossy quantization, lossy transform will be acceptable.

In terms of perceptual quantization, in JPEG and MPEG-1/2, non-uniform quantization was developed to give high frequency AC coefficients with larger quantization intervals and give DC and low frequency AC coefficients with smaller quantization intervals, because HVS has a low pass masking effect that is more sensitive to low frequency distortions than the high frequency ones. In H.264/AVC, HEVC and VVC, uniform quantization were adopted and meanwhile perceptual quantization matrix/table \cite{OV_VVC,OV_HEVC} was developed to give smaller quantization interval to sensitive coefficients. 
In \cite{TQ_QP_Papa}, Papadopoulos \emph{et al.} adjusted quantization step in assigning macroblock bits so as to reflect visual importance of each macroblock. Similarly, Zhang \emph{et al.} \cite{TQ_QP_Zhang} established a relationship between masking features in spatiotemporal domain and QPs, then selected a local QP according to the visual characteristics of video contents. In PROVISION project \cite{TQ_QP_Dias}, areas prone to contouring were identified, where smaller QPs were assigned to prevent contouring artifacts.

\begin{table*}[!t]
\tiny
	\renewcommand{\arraystretch}{1.1}
	\caption{Visual Perception Guided Transform and Quantization.}
	\label{table_PTQ}
	\centering
\begin{threeparttable}
	\begin{tabular}{|c|c|c|c|c|c|c|c|c|}
		\hline
	\multirow{2}{*}{Works} &\multirow{2}{*} {Visual Features}& \multirow{2}{*}{Codec}& \multirow{2}{*}{VQA,$Q$}  &\multicolumn{2}{c|}{BDBR($Q$)[\%]}&\multicolumn{2}{c|}{$\Delta T$[\%]} \\
      \cline{5-8}
       & & & &LD&RA&LD&RA \\
		\hline

        Papa.'17 \cite{TQ_QP_Papa}&{Textural masking and fovea to adjust $QP$} &HEVC/HM16.2& MOS &/&{-9.00} & /&/\\
		\hline
        Xiang'14 \cite{TQ_Xiang}& {Adding a $QP$ offset based on block-level JND}& {AVS baseline} & {MS-SSIM}	&/ &{-24.50\tnote{*}} &{/} &{/}\\
        \hline
       \multirow{2}{*}{Yan'20 \cite{TQ_Yan}} &{Adding a CTU level $QP$ offset based on the}&\multirow{2}{*}{AVS-P2/RD17.0}&\multirow{2}{*}{SSIM}&\multirow{2}{*}{-8.60}&\multirow{2}{*}{/}&\multirow{2}{*}{/}&\multirow{2}{*}{/}\\
        &  spatial and temporal masking effects& &  & &  & & \\
        \hline

        Kim'15  \cite{TQ_Kim}  & {Transform domain JND considering masking effects}& {HEVC/HM11.0} & {MS-SSIM}	 &{-16.10}& {-11.11} &{11.25}&{22.78} \\
        \hline
        Cui'19 \cite{TQ_Cui} & {Transform coefficients are suppressed with the JND}& {HEVC/HM 16.9} & {MOS}	& {-32.98}& {-28.61} &{12.94}&{22.45} \\
        \hline
		Ki'17  \cite{TQ_Ki}& {Learning-based JNQD models for preprocessing}& {HEVC/HM16.17} & {DMOS}	 &{-10.38\tnote{*}}& {-24.91\tnote{*}} &{34.65}&/ \\
		\hline
		Nami'22  \cite{TQ_Nami}& {CNN-based JND prediction and saliency for $QP$ determination}& {HEVC/HM16.2} & {DMOS}	 &{-25.44\tnote{*}}& {/} &{/}&/ \\
		\hline
         Zhang'17 \cite{TQ_Zhang}& {Exploit spatial contrast sensitivity in quantization matrix}& {H.264/Wyner-Ziv} & {MOS} &\multicolumn{2}{c|}{-11.96 / -9.92\tnote{\#}} &{/} &{/}\\
        \hline
        Luo'13 \cite{TQ_Luo} & {Adjust quantization matrices by reflecting JND values}& {H.264/JM14.2} & {MOS} &	/&{-28.32} &{/}&{/} \\
        \hline
        Prang.'16 \cite{TQ_Prangnell} &{Quantization matrix considering CSF of HD/UHD display}&{SHVC/SHM9.0}&{PSNR}&{-2.50}&{-3.00} & \multicolumn{2}{c|}{-0.75} \\
        \hline
        \multirow{2}{*}{Grois'20 \cite{TQ_Grois}} &{ Develop variable-size quantization matrices by  }&\multirow{2}{*}{X265}&{SSIMPlus}&{-11.30}&/&/&/\\
        \cline{4-8}
         & considering the CSF of 4K UHD HDR videos& &PSNR$_{YUV}$  & {-2.40}&/ &/ & /\\
        \hline
        Shang'19 \cite{TQ_Shang} &{Quantization matrices based on CSF of RGB videos}&{HEVC/HM16.0}&{PSNR$_{GBR}$}&{-20.51}&/ & /& \\
        \hline
        Valin'15 \cite{TQ_Valin}&{Luminance contrast masking used in vector quantization}&Daala Codec &SSIM&  \multicolumn{2}{c|}{-13.70}&/&{/}\\
        \hline

	\end{tabular}
   \begin{tablenotes}
     \item[*] BDBR($Q$) subjects to quality degradation. \item[\#] benchmark is a distributed coding scheme.
   \end{tablenotes}
   \end{threeparttable}
\end{table*}

Table \ref{table_PTQ} depicts representative schemes and performances of the perceptually optimized quantization. Since not every distortion can be perceived by HVS, visibility threshold between perceivable and non-perceivable distortion is regarded as JND, which could be exploited in quantization. Based on the JND suppression, $QP$ was enlarged in encoding the prediction residue while the distortion was within the JND range \cite{TQ_Xiang}. It improved compression ratio while maintaining the same visual quality. Meanwhile, spatial and temporal masking effects \cite{TQ_Xiang,TQ_Yan} were jointly considered in JND modelling and $QP$ adaptation. Kim \emph{et al.} \cite{TQ_Kim} proposed a transform domain JND model to improve the transform, quantization and RDO modules. Firstly, different JND scaling factors were developed to scale JND for variable-size transform and quantization. Then, transform coefficients were suppressed with scaled JND by subtracting the scaled JND value from the residual value. Finally, the distortion in RDO cost function for PU and TU mode decision was compensated by the JND-suppression, which was done by subtraction operation between transform coefficients and JND values. JND driven Rate-Distortion Optimized Quantization (RDOQ) accounting for the noticeable distortion was also investigated for mode decision in \cite{TQ_QP_Dias,TQ_Cui}. Furthermore, Cui \emph{et al.} \cite{TQ_Cui} exploited spatial JND characteristics of the HDR video considering contrast sensitivity, luminance adaptation and saliency, which then suppressed the transformed coefficients if the distortion was smaller than the JND. Ki \emph{et al.} \cite{TQ_Ki} proposed learning-based Just-Noticeable-Quantization-Distortion (JNQD) models, LR-JNQD and CNN-JNQD, for perceptual video coding, which were able to adjust JND levels according to quantization steps for preprocessing the input to video encoders. JNQD models caused about 34.65\% complexity overhead, which can be reduced with code optimization and GPU acceleration. Nami \emph{et al.}\cite{TQ_Nami} proposed a JND-based perceptual coding scheme, named BL-JUNIPER, where block-Level CNN based JND prediction and visual importance from visual attention models were used to adjust QP for each block. In these JND based perceptual coding schemes, more coding gains (up to 50\%) were achieved at high bit rate (low QP) since compression distortion is less perceivable when using small $QP$ and more visual redundancies can be exploited as compared with that of low bit rate. In these works, the $QP$ value was determined at block or frame level.

To exploit the visual redundancies inner a block, a number of perceptually optimized quantization matrices were proposed. Based on the spatial contrast sensitivity and transform domain correlation noise, Zhang \emph{et al.} \cite{TQ_Zhang} proposed a perceptually optimized Adaptive Quantization Matrix (AQM), which was learned online from key frames and applied to Wyner-Ziv frame coding. In \cite{TQ_Luo}, the element values of quantization matrices for 4 $\times$ 4 and 8 $\times$ 8 transforms were enlarged by considering the JND. Consequently,  transformed coefficients were suppressed by the JND values after quantization. Prangnell \emph{et al.} \cite{TQ_Prangnell} improved CSF model by considering the enlarged resolutions of HD and UHD displays and developed an AQM for SHVC, where weighting values for high frequency coefficients were reduced. Grois \emph{et al.} \cite{TQ_Grois} developed luma and chroma CSF tuned Frequency Weighting Matrices (FWMs) for 8$\times$8 TU by extending Barten's CSF model to HDR and UHD video. Then, 4$\times$4, 16$\times$16 and 32$\times$32 FWMs were derived with up and down-samplings from the 8$\times$8 FWMs. Shang \emph{et al.} \cite{TQ_Shang} proposed Weighting Quantization Matrices (WQMs) by considering the CSF of DCT subbands for RGB videos, where G channel was given higher priority and PSNRs of G, B, and R channels were combined with a ratio of 4:1:1. In addition to the quantization matrix for luma coding\cite{TQ_Zhang,TQ_Prangnell}, quantization matrix for chroma coding was investigated in \cite{TQ_Shang}. However, since VQA for chroma has been scarcely addressed, it is difficult to measure the coding gains achieved by the chroma quantization matrices. These schemes are perception based scalar quantization, where each transform coefficient is separately quantized to an integer index causing blocking and ringing artifacts. Perceptual vector quantization \cite{TQ_Valin} was developed by considering the CSF, where a visual codebook or perceptual vector was indexed in the quantization. However, it is challenging to build a widely applicable visual codebooks.

Transform is to exploit frequency or pattern similarity in video representation by mapping spatial coefficients into a more compact manner. Since the transform and the inverse transform are lossless, the perceptual redundancies are exploited when combining the transform with quantization. Most existing works are model based schemes, which optimize the QPs and quantization matrix based on visual models, such as the visual sensitivity and JND. Due to the strong non-linear representation and learning abilities of learning algorithms, it will be a promising direction to investigate learning based transform \cite{TQ_Saab,TQ_Yang2020} and quantization \cite{TQ_Ki} by considering perceptual loss. 

\subsection{Perception based Filtering and Visual Enhancement}
Due to the block based prediction and quantization, compression artifacts, including blurring, blocking and ringing, are introduced in reconstructed frames, which significantly reduce the perceptual quality. To solve this problem, filtering and visual enhancement are proposed, whose optimization objective can be formulated as
\begin{equation}
\begin{cases}
	\mathbf{H^*}=\mathop{\arg\min}\limits_{\{\mathbf{H}\}}\sum_{i=1}^{n}\mathbf{\hat{D}_V}(\hat{\mathbf{X}}_i,\mathbf{X}_i)\\
\mathbf{\hat{D}_V}(\hat{\mathbf{X}}_i,\mathbf{X}_i)=F_{HVS}(F_D(\mathbf{H}(\hat{\mathbf{X}}_i)))-F_{HVS}(F_D(\mathbf{X}_i))
\end{cases},
\label{formula:obj_Enhance}
\end{equation}
where $\mathbf{H}()$ is a visual enhancement operator, $\mathbf{X}_i$ and $\hat{\mathbf{X}}_i$ are the $i$-th reference and distorted visual signals, $n$ is the number of visual blocks. The related works can be divided into two categories, i.e., in-loop filtering and out-loop post/pre-processing.

The first category is block-level visual enhancement in the coding loop. Because of independent quantization of DCT coefficients in each block, block based DCT usually gives rise to visually annoying blocking, ringing and blurring compression artifacts, especially at low bit rates. Deblocking filter and Sample Adaptive Offset filter (SAO) were proposed to selectively smooth the discontinuities at the block boundaries, which have been adopted in HEVC. To further reduce the blocking artifacts, Adaptive Loop Filter (ALF) and cross-component ALF \cite{OV_VVC}, which were improved spatial domain Wiener filters, were adopted after the deblocking filter in VVC. These schemes considered the discontinuities at the block boundaries as artifacts and selectively smoothed. However, blur was introduced for textural boundaries.
To enhance the visual quality, Zhao \emph{et al.} \cite{EN_Zhao} proposed an image deblocking algorithm by using Structural Sparse Representation (SSR) prior and Quantization Constraint (QC) prior. Owning to the superior performance of deep learning in image enhancement, Zhang \emph{et al.} \cite{EN_Zhang_TIP} proposed a Residual Highway Convolutional Neural Network (RHCNN) consists of residual highway units and convolutional layers for in-loop filtering in HEVC, where a skip shortcut from the beginning to the end was introduced to reduce the complexity. Pan \emph{et al.} \cite{EN_Pan_TIP} proposed an Enhanced Deep CNNs (EDCNN) for in-loop filtering in HEVC, which included a weighted normalization replacing batch normalization, a feature fusion sub-network and a joint loss function combining MSE and MAD. It improved PSNR, PSNR smoothness and subjective visual quality.
The CNN based in-loop filtering further achieved about 7\% BDBR gain and improved the visual quality. However, the visual quality of these schemes was mainly measured with PSNR. It costs extremely high computational complexities at both encoder and decoder, which are about 247 \cite{EN_Pan_TIP} and 267 \cite{EN_Zhang_TIP} times respectively.

The second category is frame-level visual enhancement out of the coding loop. Jin \emph{et al.} \cite{EN_Jin_CSVT} proposed a dual stream Multi-Path Recursive Residual Network (MPRRN) to reduce the compression artifacts, where the MPRRN model was applied three times to enhance low frequency image content, high frequency residual map and aggregated images, respectively. Li \emph{et al.}\cite{EN_Li} proposed a single CNN model for handling a wide range of qualities, where quantization tables were used as a partial of input data in training network. In addition, the network had two parallel branches, where the restoration branch dealt with local ringing artifacts and the global branch dealt with global blocking and color shifting. Yang \emph{et al.} \cite{EN_Yang_CSVT} proposed Quality Enhancement CNN (QE-CNN) networks for coded I and B/P frames, respectively, which were extended from a four-layer Artifacts Reduction CNN (AR-CNN) \cite{EN_Dong_ARCNN} by considering frame types and QPs. An average of 8.31\% BDBR gain was reported. Meanwhile, time-constrained optimization was proposed to have a good trade-off between complexity and quality. However, these schemes still adopted L1 or L2 norm as the loss function, i.e., $\mathbf{\hat{D}_V}(\hat{\mathbf{X}}_i,\mathbf{X}_i)\approx\mathbf{D}(\hat{\mathbf{X}}_i,\mathbf{X}_i)=||\mathbf{H}(\hat{\mathbf{X}}_i)-\mathbf{X}_i||_k$, $k=\{1,2\}$, which aim to improve the PSNR between the reconstructed and the reference videos.

To tackle this problem, Guo \emph{et al.} \cite{EN_Guo} proposed a one-to-many neural network training scheme by jointly considering perceptual loss from high-level feature differences, natureness loss from a discriminative network and JPEG quantization loss. Jin \emph{et al.} \cite{EN_Jin_NC} proposed a post-processing for intra frame coding based on Deep Convolutional Generative Adversarial Network (DCGAN), where a progressive refine strategy was to refine residue prediction and a perceptual discriminative network was used to differentiate the difference between refined image and ground truth. In addition, feature map differences in the discriminator were used as perceptual loss. However, the perceptual factors in HVS were actually not well considered. These schemes were proposed for out-loop coding, which can also be used as post-processing to enhance the visual quality of reconstructed images. Guan \emph{et al.} \cite{EN_Guan} proposed a video quality enhancement method for compressed videos. This method included a high quality frame determination subnetwork and a multi-frame enhancement CNN, which referred multiple temporal neighboring frames with higher quality. Although video quality was improved, it was still measured with PSNR that cannot truly reflect human perception. Vidal \emph{et al.} \cite{EN_Vidal} proposed a perceptual filter (called BilAWA and TBil) based on bilateral and Adaptive Weighting Average (AWA) filters, where JND model was introduced to control filtering strength. Similarly, Bhat \emph{et al.} \cite{EN_Bhat} proposed a perceptual pre-processing scheme for luma and chroma components based on a multi-scale CSF \cite{CM_Barten-CSF} and masking effects. Chadha \emph{et al.} \cite{EN_Chadha} proposed a CNN based Deep Perceptual Preprocessing (DPP) scheme to enhance the visual quality of the frames input to the video encoder, where the perceptual loss included a no-reference IQA, a full-reference L1 and structural similarity. DPP plus encoder were able to further reduce bit rate up to 11\% for AVC, AV1 and VVC encoders. These filters were used as pre-processing prior to smooth the imperceptible visual details and thus reduced the bit rate.

The video quality enhancement in Eq. \ref{formula:obj_Enhance} is an ill-posed problem that requires prior information to reconstruct high quality images, such as spatial, temporal correlation and visual pattern. Deep learning based schemes are more effective solutions to learn the priors as compared with the handcrafted and empirical filters. In training a deep learning based visual quality enhancement model, the source video can be regarded as the ground truth and there will be sufficient labels for learning the priors. In addition, applying a visual quality metric or perceptual model that is more consistent with HVS to form a perceptual loss and learning deep model for temporal features become two important issues for future video quality enhancement. Using no-reference quality metric will be an interesting exploration. However, in deep learning based visual enhancement, the computational complexity increases significantly and may not be affordable as they were applied in the coding loop.

\subsection{Discussions}
There are two main ways to exploit the visual redundancies in video coding. One is based on the computational perceptual models, such as the CSF, JND, ROI and so on, which are regarded as common visual features to guide the perceptual coding, such as bit allocation and mode decision. Then, the quality of compressed videos are evaluated with the VQA metric. The other way is applying VQA model or its approximation to the distortion term of the RD cost function, which was used as the criteria in RDO while selecting the optimal coding modes or parameters. While applying the VQA metrics to optimize the coding process, there are several major difficulties must be aware. 1) The quality assessment models are generated from coarse-grain MOS which is 1-5 with five scales. It is not accurate to distinguish the quality difference when the differences of coding bit rate or matching modes are small, e.g., bit rate difference is less than 1\% \cite{IQA_Fine}. 2) The image and video are usually assessed at image or video level. However, when the objective metric is applied to the video coding, the basic unit is processed with block level \cite{CM_JND_TOMM}. So adaptation is required. 3) When the quality term is used in the RDO in video coding, convex may not be guaranteed for some perceptual or VQA models, such as non-monotonic learning based models or piecewise JND models. Thus, convex proof and further adaptation are required to achieve the optimality.

Generally, there are three challenging issues in perceptually optimized video coding. Firstly, since the HVS is complicated, it is challenging to build accurate perceptual models and a VQA that adapts to various video applications. Although many VQAs are superior in some aspects, it is challenging to improve the generality of the VQAs. Secondly, there are various coding modules in video coding, which are developed based on PSNR/MSE. Adaptations to the video coding modules are challenging. VQA approximations based on MSE may facilitate the adaptation, but reduce the accuracy. Thirdly, video coding requires a large amount of optimal parameter determination processes using RD cost comparison. The perceptual models and VQA models are much more complex than MSE/MAD, especially when using deep learning based VQAs. Computational complexity will be an important issue to be addressed in perceptual coding.

\section{Experimental Validations for Coding Standards and Tools}

\subsection{Performances for the Video Coding Standards}
Coding experiments were performed to validate the compression efficiency of the latest coding standards, including VVC \cite{OV_VVC}, HEVC \cite{OV_HEVC}, AVS3 \cite{OV_AVS}, AVS2, X265 and AV1, and their reference models were VTM-13.0, HM-16.20, HPM-12.0, RD17.0, X265 and AV1 v3.1.2, respectively. Encoders were configured with RA. Fourteen sequences from JVET were encoded, which were BasketballPass, BlowingBubbles, BQSqaure, RaceHorces, BasketballDrill, BQMall, PartyScene, RaceHorcesC, FourPeople, Johnny, KristenAndSara, BasketballDrive, BQTerrace and Cactus. The visual quality of the reconstructed videos was measured with PSNR, MS-SSIM, VQM \cite{VQA_VQM}, VMAF \cite{VQA_VMAF}, MOVIE \cite{VQA_MOVIE} and C3DVQA \cite{VQA_C3DVQA}.

\begin{figure*}[!t]
	\centering
	\subfigure[]{
		\label{fig:VC_PSNR}
	\includegraphics[width=0.3\linewidth]{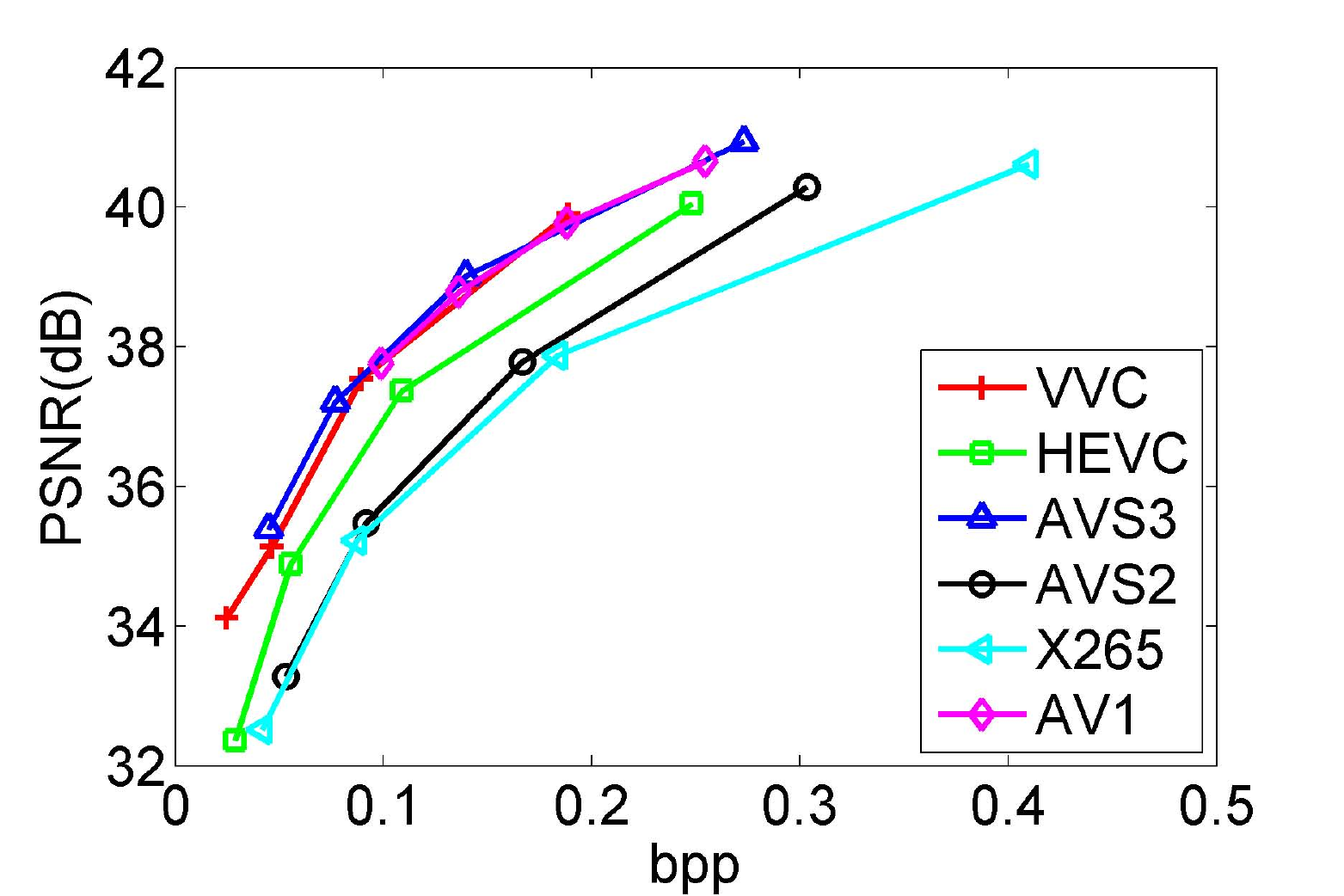}}
	\subfigure[]{
	   \label{fig:VC_SSIM}
	\includegraphics[width=0.3\linewidth]{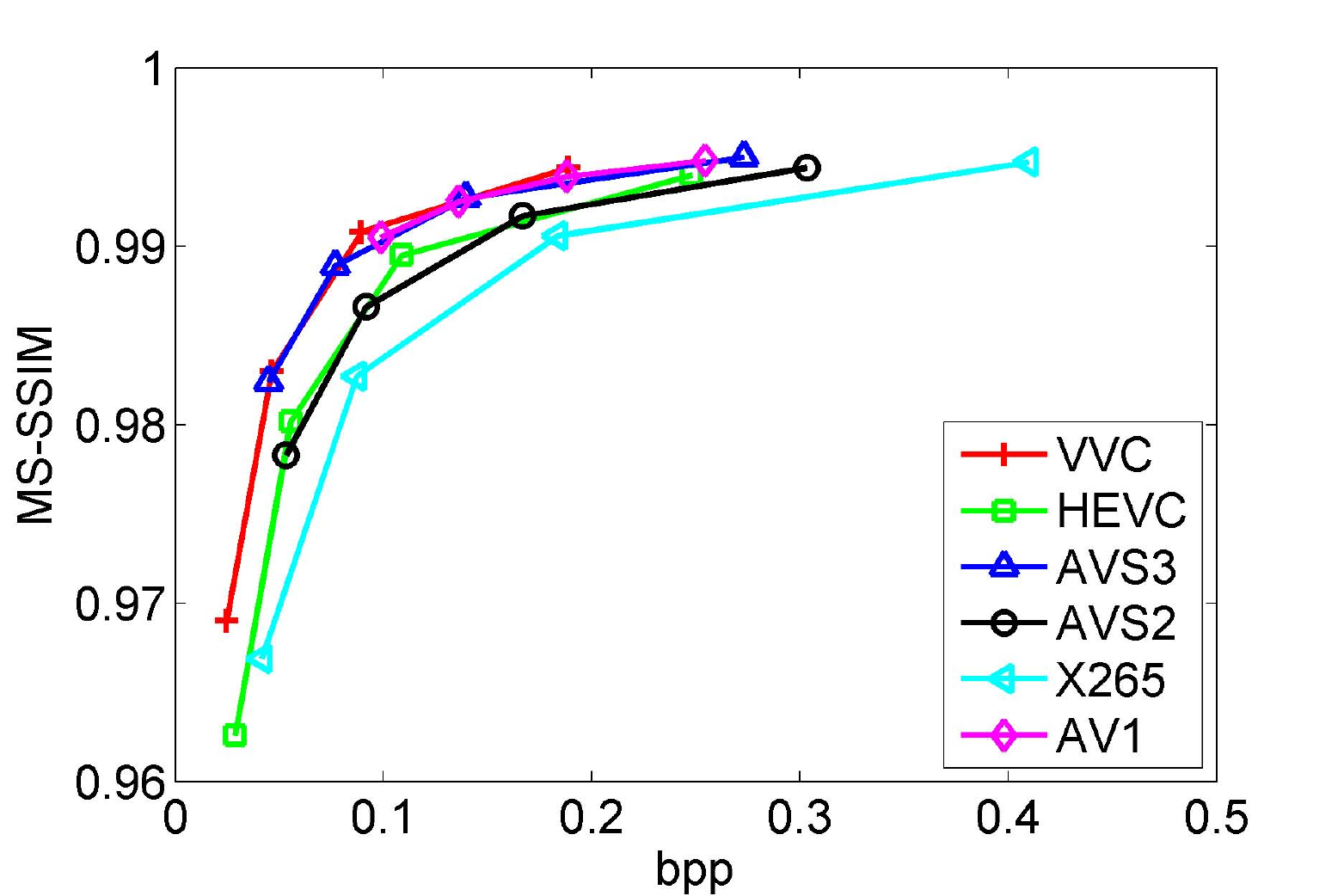}}
	\subfigure[]{
	   \label{fig:VC_SSIM}
	\includegraphics[width=0.3\linewidth]{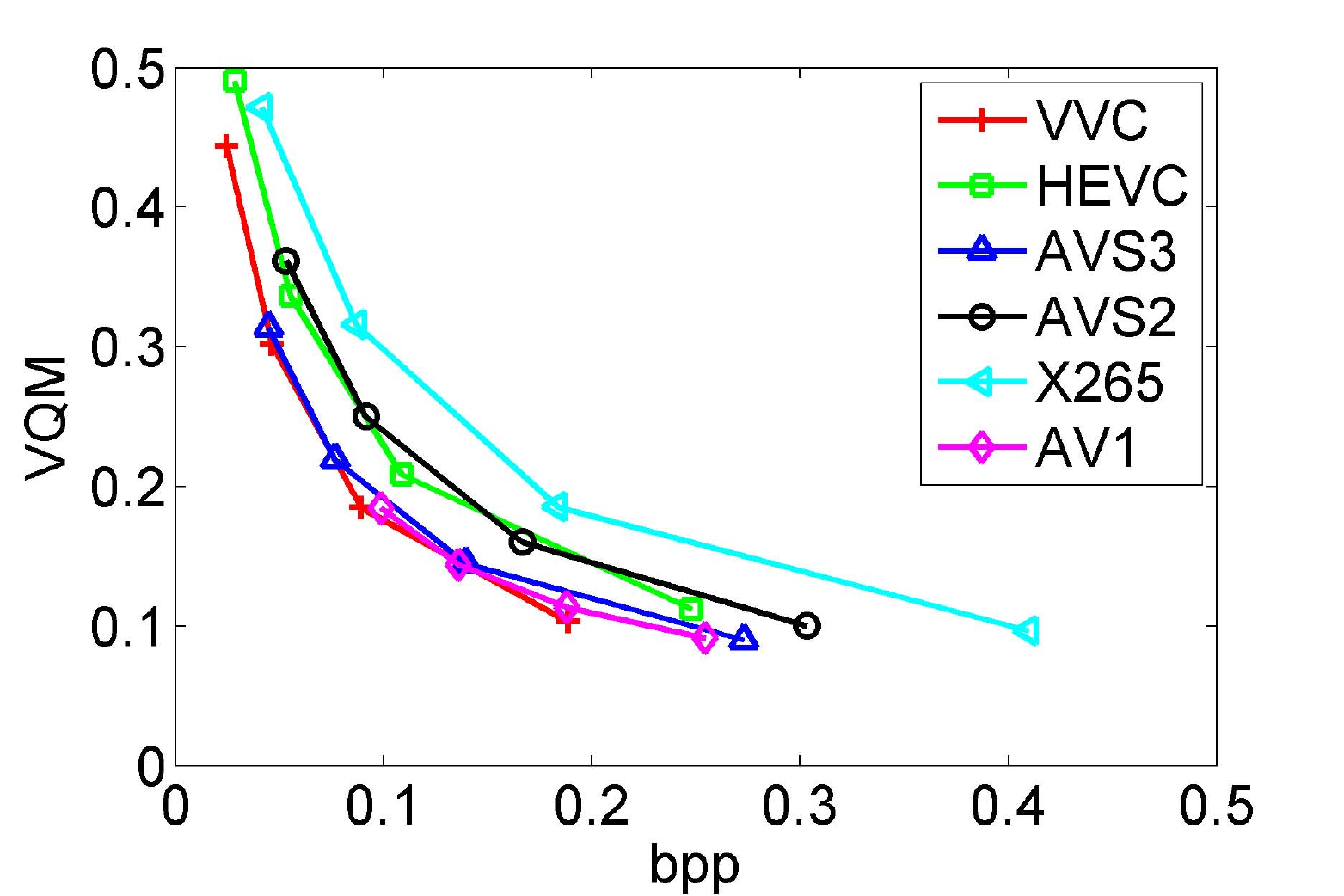}}
	\subfigure[]{
	   \label{fig:VC_SSIM}
	\includegraphics[width=0.3\linewidth]{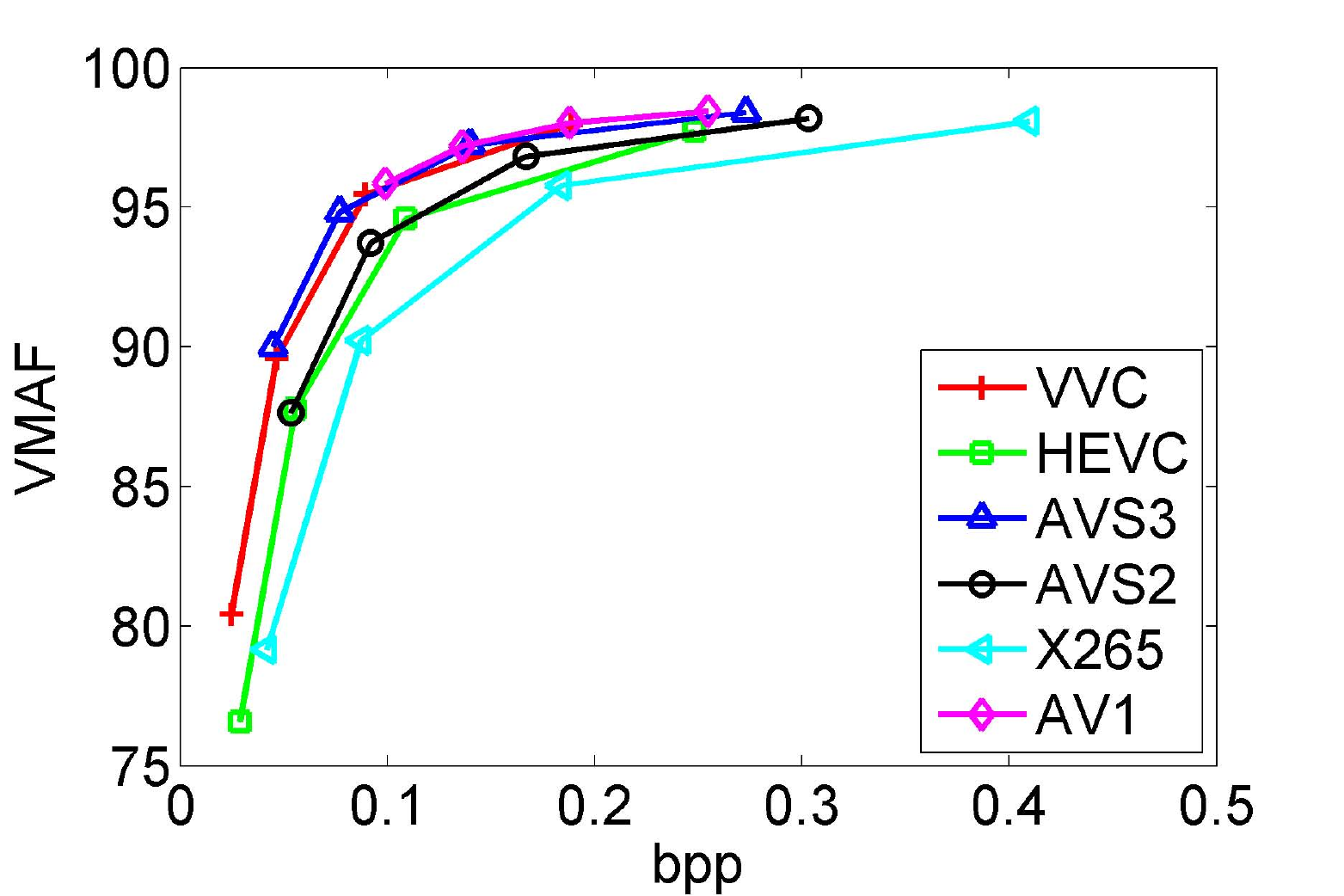}}
	\subfigure[]{
	   \label{fig:VC_SSIM}
	\includegraphics[width=0.3\linewidth]{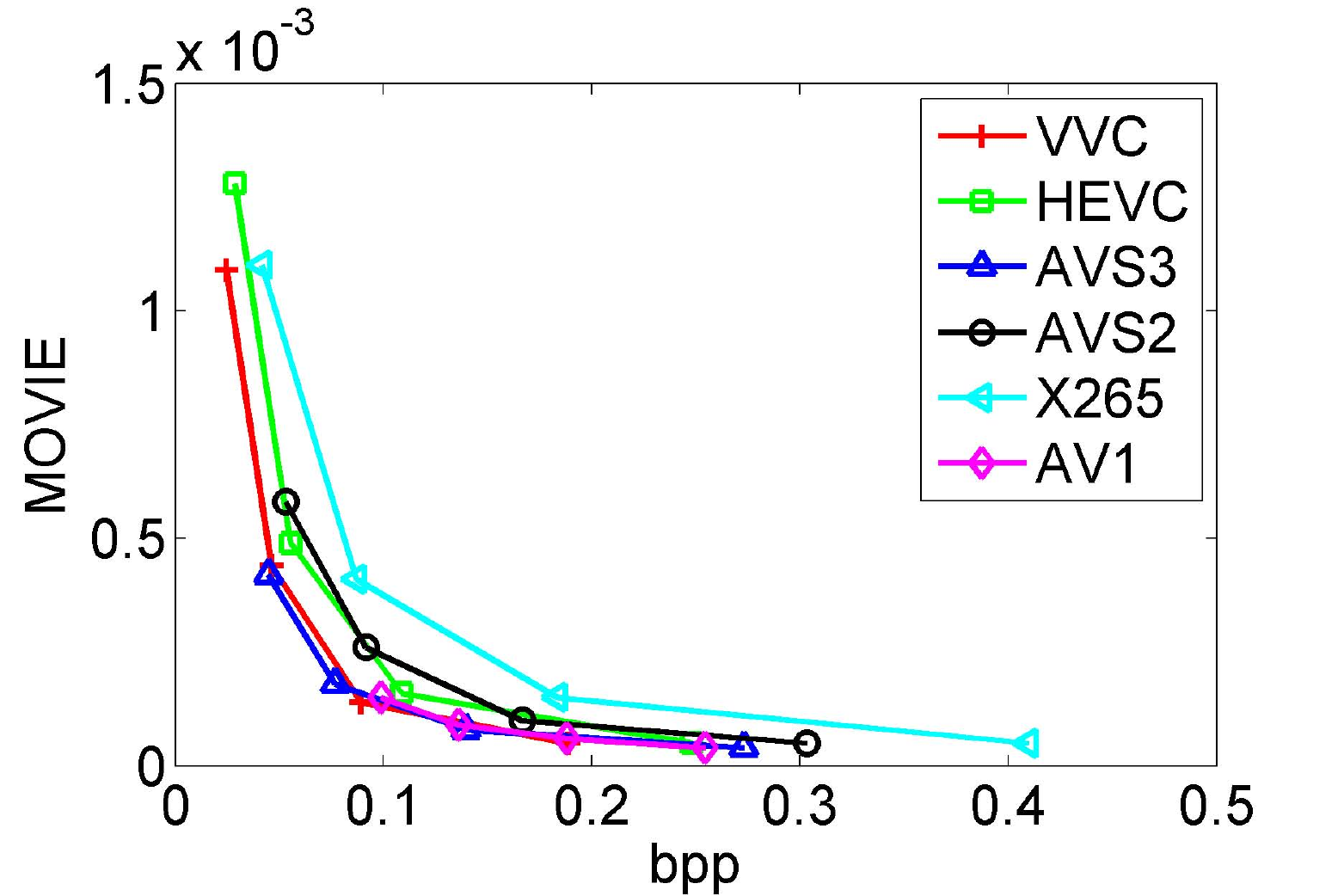}}
	\subfigure[]{
	   \label{fig:VC_SSIM}
	\includegraphics[width=0.3\linewidth]{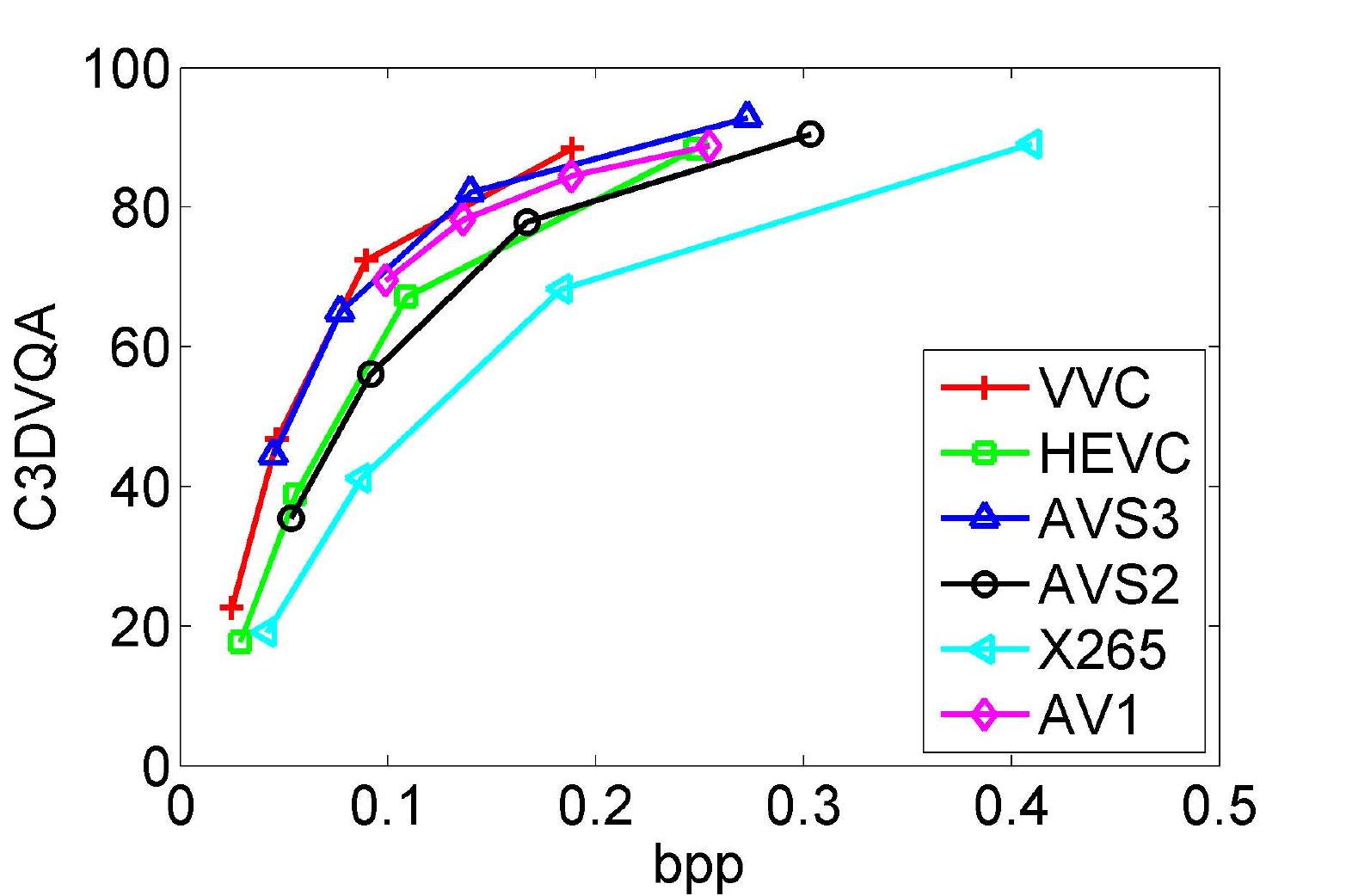}}
	\caption{Perceptual coding performance evaluations for the latest coding standards. (a) PSNR (b) MS-SSIM (c) VQM (d) VMAF (e) MOVIE (f) C3DVQA.}
	\label{fig:CodingStandards}
\end{figure*}

\begin{table*}[!t]
\footnotesize
	\renewcommand{\arraystretch}{1.1}
	\caption{Perceptual Coding Performance of Coding Tools in VVC \cite{OV_VVC,OV_VVC_tools}}
	\label{table_VVC}
	\centering
	\begin{tabular}{|c|c|c|c|c|c|c|c|}
		\hline
	\multirow{2}{*}{Abbr.}&\multirow{2}{*}{VVC Tool Descriptions} & \multicolumn{6}{c|} {BDBR gain with Different Quality Metrics[\%]} \\
      \cline{3-8}
       & &PSNR&MS-SSIM&VMAF&VQM &MOVIE&C3DVQA\\
		\hline
isp	&{Intra sub-partition mode}	&0.57	&0.60	&0.59&	0.64&0.69&1.47\\
\hline
mrl&	{Multiple reference lines}	&0.17	&0.27&	0.07&	0.50&-0.72&-0.42\\
\hline
mip	&{Matrix-based intra-picture prediction}&	0.33	&0.28&	0.95	&0.37&-0.23&0.94\\
\hline
lmchroma&	{Cross component linear model}	&0.50&	1.17	&0.79	&0.74&0.31&0.95\\
\hline
mts	&{Multiple transform selection}	&0.93	&0.81&	1.26	&1.12&0.50&2.17\\
\hline
sbt	&{Subblock transform mode}	&0.28	&0.19&	0.66&	0.14&0.18&0.11\\
\hline
lfnst&	{Low frequency non-separable transform}&	1.17&	1.56	&1.69&	1.60&0.74&1.59\\
\hline
mmvd&	{Merge with MVD}	&0.29	&0.23&	0.55	&0.38&-0.06&0.26\\
\hline
affine	&{Affine motion prediction}	&1.64&	1.56&	1.66&	1.12&0.32&1.54\\
\hline
sbtmvp&{	Subblock-based temporal MV prediction}	&0.35&	1.02&	0.55&	0.08&0.51&0.13\\
\hline
depquant&{	Dependent quantization}	&1.71	&1.91	&1.69	&2.17&1.44&1.69\\
\hline
alf+ccalf&{	ALF and Cross-component ALF}&	2.23&	-0.37&	3.67	&3.27&-1.17&0.93\\
\hline
bio	&{Bi-directional optical flow}&	0.77	&0.92	&1.42&	1.09&1.56&0.58\\
\hline
ciip	&{Combined intra/inter-picture prediction}&	0.21	&0.29&	0.72&	0.08&-0.11&-0.08\\
\hline
geo	&	{Geometric partitioning mode}&0.85&	0.95	&1.25	&0.75&1.28&0.66\\
\hline
dmvr	&{Decoder MV refinement}	&0.48	&0.66&	0.41	&0.35&0.09&0.28\\
\hline
smvd	&{Symmetric MVD	}&0.13&	0.12&	0.08	&-0.15&-0.39&0.13\\
\hline
jointcbcr&	{Joint coding of chroma residuals}&	0.88&	0.90&	0.95&	4.77&-0.04&1.37\\
\hline
prof&	{Prediction refinement with optical flow}&	0.23&	0.19	&0.28	&3.26&0.16&0.35\\
\hline
\multicolumn{2} {|c|} {\textbf{All new tools} }&	\textbf{19.54} &	\textbf{16.92}&	\textbf{23.18}	&\textbf{15.37}&\textbf{16.88}&\textbf{19.86}\\
        \hline

	\end{tabular}
\end{table*}

Fig.\ref{fig:CodingStandards} shows the average RD performances of the six tested codecs. We have the following three key observations: 1) AVS-3 and VVC are the state-of-the-art standards. They achieve comparable coding efficiency and are the top two in terms of the six metrics. 2) AV1 has comparable coding efficiency at high rate with the AVS-3 and VVC for all the metrics, and a little inferior to them if evaluated with the C3DVQA. 3) AVS-2 is inferior to the HEVC in this coding experiment, since the video resolution is small and test sequences are from JVET. X265 is a fast version of HEVC at the cost of a large efficiency loss. 4) Overall, the coding efficiency ranks of the tested standards are similar as measured with PSNR, MS-SSIM, VQM, VMAF, MOVIE and C3DVQA.

\subsection{Performances for Coding Tools in VVC and AVS-3}
Coding experiments were performed to validate the effectiveness of the coding tools in the latest VVC and AVS-3, where different coding tools were disabled one-by-one. The reference softwares of the VVC and AVS-3 are VTM-13.0 and HPM-12.0, which were configured with the default RA, respectively. The coding QP values were set as \{22, 27, 32, 37\} for VTM-13.0 and \{27, 32, 38, 45\} for HPM-12.0. According to the common test conditions, the sequences for VVC were BasketballPass, BlowingBubbles, BQSquare, RaceHorces, BasketballDrill, BQMall, PartyScene, RaceHorsesC, FourPeople, Johnny, KristenAndSara, BasketballDrive, BQTerrace, and Cactus. The test sequences for AVS3 were Crew, City, Vidyo1, Vidyo3, BasketballDirve, and Cactus. Besides the PSNR, another five image/video metrics were used to measure the visual quality of reconstructed videos, which included MS-SSIM, VMAF, VQM, MOVIE and C3DVQA. The compression efficiency was measured by BDBR \cite{BDBRCalc} under different quality metrics and reference softwares (VTM-13.0 and HPM-12.0) were anchors for comparison.

\begin{table*}[!t]
\footnotesize
	\renewcommand{\arraystretch}{1.1}
	\caption{Perceptual Coding Performance of Coding Tools in AVS3 \cite{OV_AVS}}
	\label{table_AVS3}
	\centering
	\begin{tabular}{|c|c|c|c|c|c|c|c|}
		\hline
	\multirow{2}{*}{Abbr.}&\multirow{2}{*}{AVS3 Tool Descriptions} & \multicolumn{6}{c|} {BDBR gain with Different Quality Metrics[\%]} \\
      \cline{3-8}
       & &PSNR&MS-SSIM&VMAF&VQM &MOVIE&C3DVQA\\
		\hline
ipf	&Intra prediction filter&	0.64	&0.88&	0.92	&1.01 &0.96&0.43\\
	\hline
iip	&Improved intra prediction	&0.08	&-0.01	&0.09&	0.34&0.03&-0.51\\
	\hline
sawp	&Spatial angle weighted prediction&	0.39&	0.42&	0.53	&0.40&0.60&-0.03\\
	\hline
ipc	&Inter prediction correction&	0.71&	0.72&	0.57&	1.12&1.16&1.01\\
	\hline
affine	&Affine motion prediction	&2.64	&2.64&	1.99&	2.31&0.96&2.16\\
	\hline
smvd	&Symmetric MV Difference	&0.03	&0.05&	0.12&	0.05&0.06&0.25\\
	\hline
dmvr&	Decoder side MV Refinement	&0.92	&1.26	&1.08&	1.29&1.04&0.69\\
	\hline
bio	&Bi-directional optical flow&	-0.19&	-0.10	&0.08&	0.01&-0.13&-0.26\\
	\hline
interpf	&Inter prediction filtering&	0.38	&0.29&	0.57	&0.04&0.71&0.22\\
	\hline
pmc&	Prediction from multiple cross-components&	0.01&	0.05	&0.04&	0.25&-0.15&-0.52\\
	\hline
esao&	Enhanced sample adaptive offset	&0.75&	0.32&	1.25&	0.17&0.20&0.15\\
	\hline
dbr	&Deblocking refinement&	0.06	&0.06&	0.17&	0.38&-0.26&-0.23\\
	\hline
alf	&Adaptive loop filter	&4.34	&0.99&	4.63&	0.97&3.98&5.12\\
	\hline
tscpm	&Two-step cross component prediction mode&	0.64	&0.88	&0.92&	1.01&-0.12&-0.46\\
	\hline
amvr	&Adaptive motion vector resolution&	0.83&	0.76	&0.84&	0.97&0.48&0.93\\
	\hline
emvr	&Extended AMVR&0.12&	0.11	&0.19	&-0.05&-0.16&-0.12\\
	\hline
umve	&Ultra MV expression	&1.00	&0.80&	1.13	&0.65&0.18&0.95\\
        \hline
\multicolumn{2} {|c|} {\textbf{All new tools}}&	\textbf{14.82}	&\textbf{11.25}&	\textbf{17.01}	&\textbf{6.21}&\textbf{11.71}&\textbf{14.81}\\
        \hline
	\end{tabular}
\end{table*}

\begin{figure*}[!t]
	\centering
	\subfigure[]{
		\label{fig:AVS3-CT}
	\includegraphics[width=0.47\linewidth]{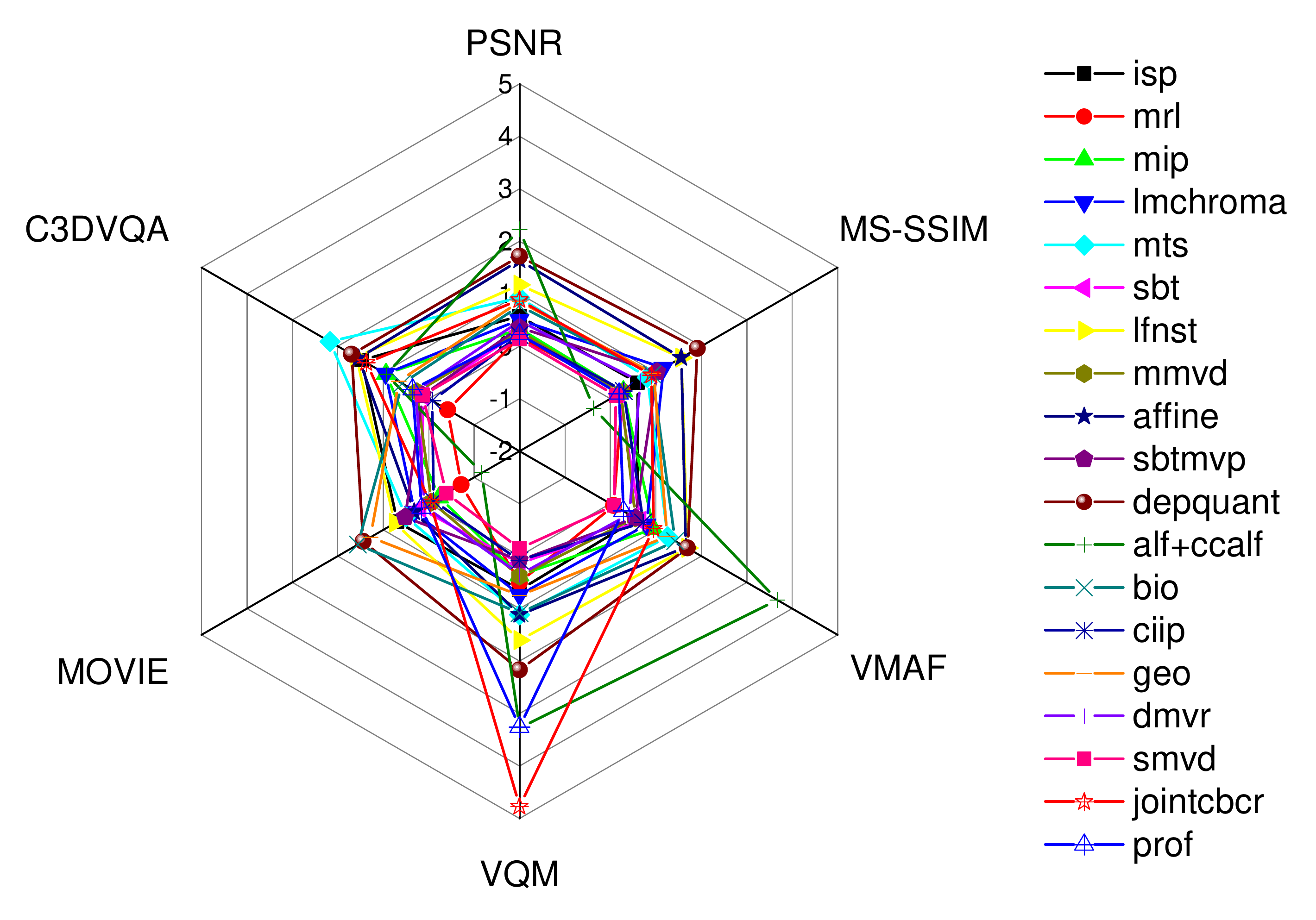}}
	\subfigure[]{
	   \label{fig:VVC-CT}
	\includegraphics[width=0.47\linewidth]{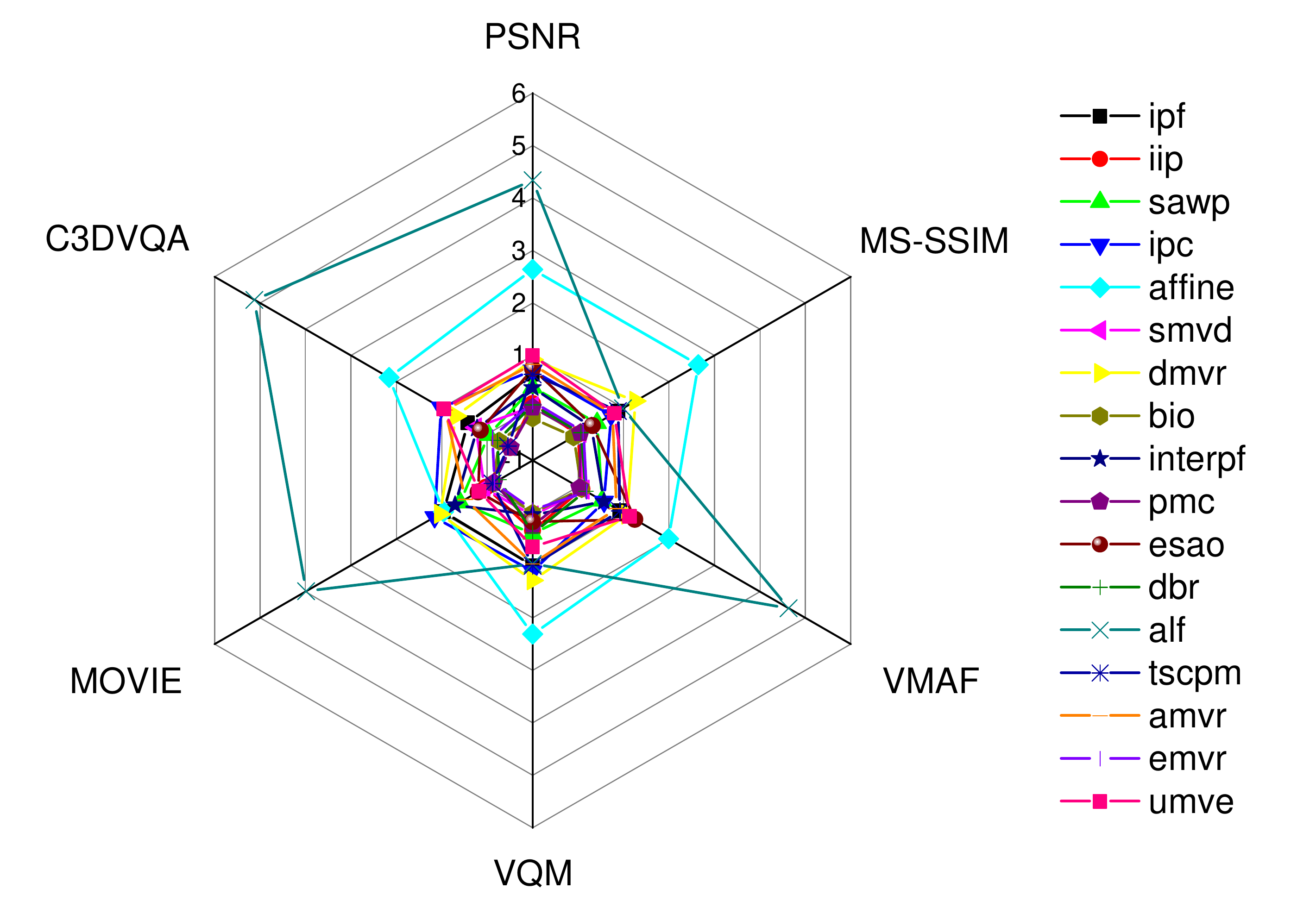}}
	\caption{Perceptual coding performance comparison among the coding tools in VVC and AVS-3. (a) VVC  \cite{OV_VVC} (b) AVS-3 \cite{OV_AVS}.}
	\label{fig:CodingTools}
\end{figure*}

Tables \ref{table_VVC} and \ref{table_AVS3} show the average perceptual coding gains of coding tools in VVC and AVS-3 for the test sequences, where the positive value indicates the BDBR gain. We can have the following three key observations: 1) the VVC coding tools are able to obtain BDBR gains from 0.13\% to 2.23\% and AVS-3 tools are able to achieve gains from -0.19\% to 4.34\%. The negative gain is achieved for the bio in AVS-3 mainly because the test sequences are a subset of the sequences in the Common Test Conditions (CTC), which means the bio is more content-aware. 2) alf+ccalf and alf achieve 2.23\% and 4.34\% BDBR gain, which are the best among the tested tools in the VVC and AVS-3. 3) In terms of different VQA metrics, BDBR gains vary significantly for some coding tools. For example, the alf+ccalf in VVC is able to achieve 2.23\%, 3.67\% and 3.27\% in terms of PSNR, VMAF and VQM. However, negative BDBR gain is achieved while reconstructed videos are measured with MS-SSIM. Similar situation can be found for AVS-3 tools, such as alf and esao. Fig.\ref{fig:CodingTools} shows radar chart of the BDBR gain of the coding tools in VVC and AVS-3. We can observe that mrl and alf+ccalf in VVC and alf and affine in AVS3 cover larger areas than others. Since the current coding tools are developed with the target of one specific metric, e.g., MSE/MAD, the results of the visual quality metrics may contradict with each other. Overall, these coding tools in VVC and AVS-3 are mainly developed for PSNR, there are still large room to remove the visual redundancies by considering the human perceptual properties.

\section{Conclusions and Future Directions}

Perceptually optimized video coding is to further improve coding efficiency by exploiting visual redundancies. In this paper, we do a systematic survey on the recent advances and challenges associated with the perceptually optimized video coding, including visual perception modelling, visual quality assessment and visual perception guided video coding optimization. In each part, problem formulation, workflows, recent advances, advantages and challenging issues are presented. As more psychological and physiological visual factors from HVS are revealed and understood, more accurate perceptual models are built. Meanwhile, visual signal representation, coding distortion and display shall be jointly considered to build an accurate perceptual model. Since the conventional PSNR cannot truly reflect the visual quality of HVS, developing VQA model to evaluate the visual quality of video becomes one of the most crucial parts. However, the perceptual factors of HVS are not fully revealed and the visual response correlates with multiple factors, including input $\mathbf{P}$, display $F_D$, representation $F_R$, and video processing $F_E$. Learning based VQA is a possible solution that transfers the VQA prediction to data fitting problem. However, to learn a stable VQA model, building a large scale video quality dataset is required, which is very laborious and time-consuming. Finally, due to different mechanisms of the video coding models, problems and solutions are identified while applying the perceptual models or VQA models to exploit the visual redundancies. Further investigations and adaptations are required to maximize their performance. In summary, perceptual video coding does have many advantages and potentials in improving the compression efficiency, which will be promising directions for visual signal communication.

Based on the review, there rise a number of promising research directions for future work:
\begin{itemize}
\item [1)] Deep Learning Optimized Perceptual Coding. Deep learning is a powerful tool in solving the prediction, classification, and regression problems in promoting video coding \cite{SV_DLVC} since it is able to discover the hidden statistics and knowledge in massive data. More investigations on deep learning based perceptual modelling, visual quality assessment, coding module optimizations and end-to-end video compression are highly desired. However, there are two critical issues to be solved. One is the sufficient data labelling for deep model training which will be critical in collecting sufficient number of visual labels, especially in visual modelling and visual quality assessment. The other is the high complexity of using the deep model, especially when it is applied to in-loop coding modules and temporal successive frames.

\item [2)] QoE Modelling. In recent works, the perceptual quality of video usually refers to the clarity. However, in addition to the clarity, some other factors, such as depth quality, visual comfort, interactivity, immersion, naturalness and low delay, are highly correlated the perceptual quality. Therefore, QoE related visual factors and their perceptual models shall be investigated, which will be helpful for perceptually optimized visual signal processing. On the other hand, videos are developing in the trends of representing more realistic visual contents, which increases the spatial, temporal resolutions, luma and chroma dynamics. QoE models for higher resolution, fidelity and bit depth shall be investigated. Learning based approaches could be an important direction to solve the feature representation and fusion problems in the QoE modelling. Moreover, QoE factors have mutual affects and may contradict with each other, how to apply the QoE model to the video coding and tackle multi-objective problems will be challenging.

\item [3)] Video Coding for Machine (VCM) \cite{VCM}. Conventional digital video was mainly developed for human vision. However, with the increasing demand of intelligent video applications, such as smart video survivance, visual search, autonomous vehicle and robotics, videos are used not only for human vision but also for machine vision and intelligent visual analysis. Therefore, visual features and quality of machine vision or related learning algorithm shall be exploited, where $F_D()$ and $F_{HVS}()$ relate to machine vision. Conventional video coding is based on the hybrid coding framework exploiting the statistical and visual redundancies. With the development of deep learning and different coding objectives in VCM, deep feature analysis, quality evaluation methods, coding framework, and feature compression algorithms for VCM are of high interests in academic and industrial communities.

\item [4)] Perceptual Coding for Hyper-realistic and High Dimensional Videos. Hyper-realistic videos extend the existing visual media from resolution, frame rate, color gammut and dynamic range and so on, which provide users with more realistic visual experience. The perceptual factors and response of HVS $F_{HVS}$ may vary with the visual representation $F_{R}$, display $F_{D}$, and processing distortion $F_{E}$. The current perceptual models, such as masking, JND, sensitivity and IQA models, and coding algorithms, including RDO, bit allocation, transform, and visual enhancements, deserve further adaptations for higher bit depth. High dimensional videos, such as mutliview video, 3D video, point cloud and light field, extend the video from 2D to 3D space by representing 3D scene with more viewpoints and depth information. There are more visual redundancies and inter-correlations to be explored. The visual properties of the high dimensional videos are more complicated and significantly different from those of the 2D video, as they are able to provide users with depth, view and angle interactions, immersion and so on. Further investigations on 3D visual perception, 3D VQA and 3D perceptual coding for high dimensional videos will be interesting.

\item [5)] Perceptually Optimized Deep Visual Model. The deep learning demonstrates powerful capabilities on visual feature representation and data fitting. However, it requires big visual data with labels and causes extremely high computational complexity as the feature dimension increases and the model goes deeper. With the development of visual psychology and neuroscience, the psycho-visual findings in HVS would be possible to improve the design of deep learning based visual models, including computational perceptual model, visual quality assessment, perceptually optimized deep architecture for visual representation, reconstruction and recognition.
\end{itemize}


\bibliographystyle{ACM-Reference-Format}
\bibliography{sample-base}


\begin{thebibliography}{141}


\ifx \showCODEN    \undefined \def \showCODEN     #1{\unskip}     \fi
\ifx \showDOI      \undefined \def \showDOI       #1{#1}\fi
\ifx \showISBNx    \undefined \def \showISBNx     #1{\unskip}     \fi
\ifx \showISBNxiii \undefined \def \showISBNxiii  #1{\unskip}     \fi
\ifx \showISSN     \undefined \def \showISSN      #1{\unskip}     \fi
\ifx \showLCCN     \undefined \def \showLCCN      #1{\unskip}     \fi
\ifx \shownote     \undefined \def \shownote      #1{#1}          \fi
\ifx \showarticletitle \undefined \def \showarticletitle #1{#1}   \fi
\ifx \showURL      \undefined \def \showURL       {\relax}        \fi
\providecommand\bibfield[2]{#2}
\providecommand\bibinfo[2]{#2}
\providecommand\natexlab[1]{#1}
\providecommand\showeprint[2][]{arXiv:#2}

\bibitem[Athar and Wang(2019)]%
        {SV_IQA}
\bibfield{author}{\bibinfo{person}{Shahrukh Athar} {and} \bibinfo{person}{Zhou
  Wang}.} \bibinfo{year}{2019}\natexlab{}.
\newblock \showarticletitle{A comprehensive performance evaluation of image
  quality assessment algorithms}.
\newblock \bibinfo{journal}{\emph{IEEE Access}}  \bibinfo{volume}{7}
  (\bibinfo{year}{2019}), \bibinfo{pages}{140030--140070}.
\newblock


\bibitem[Bae et~al\mbox{.}(2016)]%
        {RDO_Bae}
\bibfield{author}{\bibinfo{person}{Sung-Ho Bae}, \bibinfo{person}{Jaeil Kim},
  {and} \bibinfo{person}{Munchurl Kim}.} \bibinfo{year}{2016}\natexlab{}.
\newblock \showarticletitle{HEVC-based perceptually adaptive video coding using
  a DCT-based local distortion detection probability model}.
\newblock \bibinfo{journal}{\emph{IEEE Transactions on Image Processing}}
  \bibinfo{volume}{25}, \bibinfo{number}{7} (\bibinfo{year}{2016}),
  \bibinfo{pages}{3343--3357}.
\newblock


\bibitem[Bampis et~al\mbox{.}(2019)]%
        {VQA_ST_VMAF}
\bibfield{author}{\bibinfo{person}{Christos~G. Bampis}, \bibinfo{person}{Zhi
  Li}, {and} \bibinfo{person}{Alan~C. Bovik}.} \bibinfo{year}{2019}\natexlab{}.
\newblock \showarticletitle{Spatiotemporal feature integration and model fusion
  for full reference video quality assessment}.
\newblock \bibinfo{journal}{\emph{IEEE Transactions on Circuits and Systems for
  Video Technology}} \bibinfo{volume}{29}, \bibinfo{number}{8}
  (\bibinfo{year}{2019}), \bibinfo{pages}{2256--2270}.
\newblock


\bibitem[Barten(2003)]%
        {CM_Barten-CSF}
\bibfield{author}{\bibinfo{person}{Peter G.~J. Barten}.}
  \bibinfo{year}{2003}\natexlab{}.
\newblock \showarticletitle{{Formula for the contrast sensitivity of the human
  eye}}. In \bibinfo{booktitle}{\emph{Image Quality and System Performance}},
  Vol.~\bibinfo{volume}{5294}. \bibinfo{publisher}{SPIE}, \bibinfo{pages}{231
  -- 238}.
\newblock


\bibitem[Bertalmio(2020)]%
        {CM_S_CSF}
\bibfield{author}{\bibinfo{person}{Marcelo Bertalmio}.}
  \bibinfo{year}{2020}\natexlab{}.
\newblock \showarticletitle{Chapter 5 - Brightness perception and encoding
  curves}.
\newblock In \bibinfo{booktitle}{\emph{Vision Models for High Dynamic Range and
  Wide Colour Gamut Imaging}}, \bibfield{editor}{\bibinfo{person}{Marcelo
  Bertalmio}} (Ed.). \bibinfo{publisher}{Academic Press},
  \bibinfo{pages}{95--129}.
\newblock
\showISBNx{978-0-12-813894-6}


\bibitem[Bhat et~al\mbox{.}(2019)]%
        {EN_Bhat}
\bibfield{author}{\bibinfo{person}{Madhukar Bhat}, \bibinfo{person}{Jean-Marc
  Thiesse}, {and} \bibinfo{person}{Patrick~Le Callet}.}
  \bibinfo{year}{2019}\natexlab{}.
\newblock \showarticletitle{HVS based perceptual pre-processing for video
  coding}. In \bibinfo{booktitle}{\emph{2019 27th European Signal Processing
  Conference (EUSIPCO)}}. \bibinfo{pages}{1--5}.
\newblock


\bibitem[Bj{\o}ntegaard(2001)]%
        {BDBRCalc}
\bibfield{author}{\bibinfo{person}{Gisle Bj{\o}ntegaard}.}
  \bibinfo{year}{2001}\natexlab{}.
\newblock \showarticletitle{Calculation of average PSNR differences between
  RD-curves}. In \bibinfo{booktitle}{\emph{ITU-T Video Coding Experts Group,
  VCEG-M33}}.
\newblock


\bibitem[Bosse et~al\mbox{.}(2019)]%
        {CM_Bosse}
\bibfield{author}{\bibinfo{person}{Sebastian Bosse}, \bibinfo{person}{S{\"o}ren
  Becker}, \bibinfo{person}{Klaus-Robert M{\"u}ller}, \bibinfo{person}{Wojciech
  Samek}, {and} \bibinfo{person}{Thomas Wiegand}.}
  \bibinfo{year}{2019}\natexlab{}.
\newblock \showarticletitle{Estimation of distortion sensitivity for visual
  quality prediction using a convolutional neural network}.
\newblock \bibinfo{journal}{\emph{Digital Signal Processing}}
  \bibinfo{volume}{91} (\bibinfo{year}{2019}), \bibinfo{pages}{54--65}.
\newblock
\showISSN{1051-2004}


\bibitem[Bross et~al\mbox{.}(2021a)]%
        {OV_VVC_tools}
\bibfield{author}{\bibinfo{person}{Benjamin Bross}, \bibinfo{person}{Jianle
  Chen}, \bibinfo{person}{Jens-Rainer Ohm}, \bibinfo{person}{Gary~J. Sullivan},
  {and} \bibinfo{person}{Ye-Kui Wang}.} \bibinfo{year}{2021}\natexlab{a}.
\newblock \showarticletitle{Developments in international video coding
  standardization after AVC, with an overview of versatile video coding (VVC)}.
\newblock \bibinfo{journal}{\emph{Proc. IEEE}} \bibinfo{volume}{109},
  \bibinfo{number}{9} (\bibinfo{year}{2021}), \bibinfo{pages}{1463--1493}.
\newblock


\bibitem[Bross et~al\mbox{.}(2021b)]%
        {OV_VVC}
\bibfield{author}{\bibinfo{person}{Benjamin Bross}, \bibinfo{person}{Ye-Kui
  Wang}, \bibinfo{person}{Yan Ye}, \bibinfo{person}{Shan Liu},
  \bibinfo{person}{Jianle Chen}, \bibinfo{person}{Gary~J. Sullivan}, {and}
  \bibinfo{person}{Jens-Rainer Ohm}.} \bibinfo{year}{2021}\natexlab{b}.
\newblock \showarticletitle{Overview of the versatile video coding (VVC)
  standard and its applications}.
\newblock \bibinfo{journal}{\emph{IEEE Transactions on Circuits and Systems for
  Video Technology}} \bibinfo{volume}{31}, \bibinfo{number}{10}
  (\bibinfo{year}{2021}), \bibinfo{pages}{3736--3764}.
\newblock


\bibitem[BT.2100-2(2018)]%
        {BT2100}
\bibfield{author}{\bibinfo{person}{BT.2100-2}.}
  \bibinfo{year}{2018}\natexlab{}.
\newblock \showarticletitle{Image parameter values for high dynamic range
  television for use in production and international programme exchange}.
\newblock \bibinfo{journal}{\emph{ITU-R Recommendations}}
  (\bibinfo{year}{2018}).
\newblock


\bibitem[BT.500-14(2015)]%
        {BT500}
\bibfield{author}{\bibinfo{person}{BT.500-14}.}
  \bibinfo{year}{2015}\natexlab{}.
\newblock \showarticletitle{Methodologies for the subjective assessment of the
  quality of television images}.
\newblock \bibinfo{journal}{\emph{ITU-R Recommendations}}
  (\bibinfo{year}{2015}).
\newblock


\bibitem[BT.709-6(2015)]%
        {BT709}
\bibfield{author}{\bibinfo{person}{BT.709-6}.} \bibinfo{year}{2015}\natexlab{}.
\newblock \showarticletitle{Parameter values for the HDTV standards for
  production and international programme exchange}.
\newblock \bibinfo{journal}{\emph{ITU-R Recommendations}}
  (\bibinfo{year}{2015}).
\newblock


\bibitem[Chadha and Andreopoulos(2021)]%
        {EN_Chadha}
\bibfield{author}{\bibinfo{person}{Aaron Chadha} {and} \bibinfo{person}{Yiannis
  Andreopoulos}.} \bibinfo{year}{2021}\natexlab{}.
\newblock \showarticletitle{Deep perceptual preprocessing for video coding}. In
  \bibinfo{booktitle}{\emph{2021 IEEE/CVF Conference on Computer Vision and
  Pattern Recognition (CVPR)}}. \bibinfo{pages}{14847--14856}.
\newblock


\bibitem[Chao et~al\mbox{.}(2016)]%
        {RC_JND}
\bibfield{author}{\bibinfo{person}{Wen-Wei Chao}, \bibinfo{person}{Yen-Yu
  Chen}, {and} \bibinfo{person}{Shao-Yi Chien}.}
  \bibinfo{year}{2016}\natexlab{}.
\newblock \showarticletitle{Perceptual HEVC/H.265 system with local
  just-noticeable-difference model}. In \bibinfo{booktitle}{\emph{2016 IEEE
  International Symposium on Circuits and Systems (ISCAS)}}.
  \bibinfo{pages}{2679--2682}.
\newblock


\bibitem[Chen et~al\mbox{.}(2010)]%
        {SV_PVC2}
\bibfield{author}{\bibinfo{person}{Zhenzhong Chen}, \bibinfo{person}{Weisi
  Lin}, {and} \bibinfo{person}{King~Ngi Ngan}.}
  \bibinfo{year}{2010}\natexlab{}.
\newblock \showarticletitle{Perceptual video coding: challenges and
  approaches}. In \bibinfo{booktitle}{\emph{2010 IEEE International Conference
  on Multimedia and Expo}}. \bibinfo{pages}{784--789}.
\newblock


\bibitem[Chen and Wu(2020)]%
        {CM_AsyFoveatedJND}
\bibfield{author}{\bibinfo{person}{Zhenzhong Chen} {and} \bibinfo{person}{Wei
  Wu}.} \bibinfo{year}{2020}\natexlab{}.
\newblock \showarticletitle{Asymmetric foveated just-noticeable-difference
  model for images with visual field inhomogeneities}.
\newblock \bibinfo{journal}{\emph{IEEE Transactions on Circuits and Systems for
  Video Technology}} \bibinfo{volume}{30}, \bibinfo{number}{11}
  (\bibinfo{year}{2020}), \bibinfo{pages}{4064--4074}.
\newblock


\bibitem[Cong et~al\mbox{.}(2019)]%
        {SV_VA}
\bibfield{author}{\bibinfo{person}{Runmin Cong}, \bibinfo{person}{Jianjun Lei},
  \bibinfo{person}{Huazhu Fu}, \bibinfo{person}{Ming-Ming Cheng},
  \bibinfo{person}{Weisi Lin}, {and} \bibinfo{person}{Qingming Huang}.}
  \bibinfo{year}{2019}\natexlab{}.
\newblock \showarticletitle{Review of visual saliency detection with
  comprehensive information}.
\newblock \bibinfo{journal}{\emph{IEEE Transactions on Circuits and Systems for
  Video Technology}} \bibinfo{volume}{29}, \bibinfo{number}{10}
  (\bibinfo{year}{2019}), \bibinfo{pages}{2941--2959}.
\newblock


\bibitem[Cui et~al\mbox{.}(2019)]%
        {TQ_Cui}
\bibfield{author}{\bibinfo{person}{Xin Cui}, \bibinfo{person}{Zongju Peng},
  \bibinfo{person}{Gangyi Jiang}, \bibinfo{person}{Fen Chen}, {and}
  \bibinfo{person}{Mei Yu}.} \bibinfo{year}{2019}\natexlab{}.
\newblock \showarticletitle{Perceptual video coding scheme using just
  noticeable distortion model based on entropy filter}.
\newblock \bibinfo{journal}{\emph{Entropy}}  \bibinfo{volume}{21}
  (\bibinfo{date}{11} \bibinfo{year}{2019}), \bibinfo{pages}{1095}.
\newblock


\bibitem[Cui et~al\mbox{.}(2021)]%
        {RDO_Cui_UHD}
\bibfield{author}{\bibinfo{person}{Xin Cui}, \bibinfo{person}{Zongju Peng},
  \bibinfo{person}{Gangyi Jiang}, \bibinfo{person}{Fen Chen},
  \bibinfo{person}{Mei Yu}, {and} \bibinfo{person}{Dongrong Jiang}.}
  \bibinfo{year}{2021}\natexlab{}.
\newblock \showarticletitle{Perceptual coding scheme for ultra-high definition
  video based on perceptual noise channel model}.
\newblock \bibinfo{journal}{\emph{Digital Signal Processing}}
  \bibinfo{volume}{108} (\bibinfo{date}{01} \bibinfo{year}{2021}),
  \bibinfo{pages}{102903}.
\newblock


\bibitem[Dai et~al\mbox{.}(2019)]%
        {Imaging}
\bibfield{author}{\bibinfo{person}{Qionghai Dai}, \bibinfo{person}{Jiamin Wu},
  \bibinfo{person}{Jingtao Fan}, \bibinfo{person}{Feng Xu}, {and}
  \bibinfo{person}{Xun Cao}.} \bibinfo{year}{2019}\natexlab{}.
\newblock \showarticletitle{Recent advances in computational photography}.
\newblock \bibinfo{journal}{\emph{IEEE Journal of Selected Topics in Signal
  Processing}} \bibinfo{volume}{28}, \bibinfo{number}{1}
  (\bibinfo{year}{2019}), \bibinfo{pages}{1--5}.
\newblock


\bibitem[Dias et~al\mbox{.}(2015)]%
        {TQ_QP_Dias}
\bibfield{author}{\bibinfo{person}{Andre~Seixas Dias},
  \bibinfo{person}{Sebastian Schwarz}, \bibinfo{person}{Mischa Siekmann},
  \bibinfo{person}{Sebastian Bosse}, \bibinfo{person}{Heiko Schwarz},
  \bibinfo{person}{Detlev Marpe}, \bibinfo{person}{John Zubrzycki}, {and}
  \bibinfo{person}{Marta Mrak}.} \bibinfo{year}{2015}\natexlab{}.
\newblock \showarticletitle{Perceptually optimised video compression}. In
  \bibinfo{booktitle}{\emph{2015 IEEE International Conference on Multimedia
  Expo Workshops (ICMEW)}}. \bibinfo{pages}{1--4}.
\newblock


\bibitem[Dong et~al\mbox{.}(2015)]%
        {EN_Dong_ARCNN}
\bibfield{author}{\bibinfo{person}{Chao Dong}, \bibinfo{person}{Yubin Deng},
  \bibinfo{person}{Chen Change~Loy}, {and} \bibinfo{person}{Xiaoou Tang}.}
  \bibinfo{year}{2015}\natexlab{}.
\newblock \showarticletitle{Compression Artifacts Reduction by a Deep
  Convolutional Network}. In \bibinfo{booktitle}{\emph{Proceedings of the IEEE
  International Conference on Computer Vision (ICCV)}}.
  \bibinfo{pages}{576--584}.
\newblock


\bibitem[Fan et~al\mbox{.}(2018)]%
        {IQA_MultiCNN_QA}
\bibfield{author}{\bibinfo{person}{Chunling Fan}, \bibinfo{person}{Yun Zhang},
  \bibinfo{person}{Liangbing Feng}, {and} \bibinfo{person}{Qingshan Jiang}.}
  \bibinfo{year}{2018}\natexlab{}.
\newblock \showarticletitle{No reference image quality assessment based on
  multi-expert convolutional neural networks}.
\newblock \bibinfo{journal}{\emph{IEEE Access}}  \bibinfo{volume}{6}
  (\bibinfo{year}{2018}), \bibinfo{pages}{8934--8943}.
\newblock


\bibitem[Fan et~al\mbox{.}(2021)]%
        {CM_SUR_JND_Stereo}
\bibfield{author}{\bibinfo{person}{Chunling Fan}, \bibinfo{person}{Yun Zhang},
  \bibinfo{person}{Raouf Hamzaoui}, \bibinfo{person}{Djemel Ziou}, {and}
  \bibinfo{person}{Qingshan Jiang}.} \bibinfo{year}{2021}\natexlab{}.
\newblock \showarticletitle{Learning-based satisfied user ratio prediction for
  symmetrically and asymmetrically compressed stereoscopic images}.
\newblock \bibinfo{journal}{\emph{IEEE MultiMedia}} \bibinfo{volume}{28},
  \bibinfo{number}{3} (\bibinfo{year}{2021}), \bibinfo{pages}{8--20}.
\newblock


\bibitem[Fang et~al\mbox{.}(2017)]%
        {CM_Stereo_VA}
\bibfield{author}{\bibinfo{person}{Yuming Fang}, \bibinfo{person}{Chi Zhang},
  \bibinfo{person}{Jing Li}, \bibinfo{person}{Jianjun Lei},
  \bibinfo{person}{Matthieu Perreira Da~Silva}, {and} \bibinfo{person}{Patrick
  Le~Callet}.} \bibinfo{year}{2017}\natexlab{}.
\newblock \showarticletitle{Visual attention modeling for stereoscopic video: a
  benchmark and computational model}.
\newblock \bibinfo{journal}{\emph{IEEE Transactions on Image Processing}}
  \bibinfo{volume}{26}, \bibinfo{number}{10} (\bibinfo{year}{2017}),
  \bibinfo{pages}{4684--4696}.
\newblock


\bibitem[Francois et~al\mbox{.}(2020)]%
        {HDR_Francois_CSVT20}
\bibfield{author}{\bibinfo{person}{Edouard Francois},
  \bibinfo{person}{C.~Andrew Segall}, \bibinfo{person}{Alexis~M. Tourapis},
  \bibinfo{person}{P. Yin}, {and} \bibinfo{person}{D. Rusanovskyy}.}
  \bibinfo{year}{2020}\natexlab{}.
\newblock \showarticletitle{High dynamic range video coding technology in
  responses to the joint call for proposals on video compression with
  capability beyond HEVC}.
\newblock \bibinfo{journal}{\emph{IEEE Transactions on Circuits and Systems for
  Video Technology}} \bibinfo{volume}{30}, \bibinfo{number}{5}
  (\bibinfo{year}{2020}), \bibinfo{pages}{1253--1266}.
\newblock


\bibitem[Gao et~al\mbox{.}(2022)]%
        {RC_ConstVQ}
\bibfield{author}{\bibinfo{person}{Wei Gao}, \bibinfo{person}{Qiuping Jiang},
  \bibinfo{person}{Ronggang Wang}, \bibinfo{person}{Siwei Ma},
  \bibinfo{person}{Ge Li}, {and} \bibinfo{person}{Sam Kwong}.}
  \bibinfo{year}{2022}\natexlab{}.
\newblock \showarticletitle{Consistent quality oriented rate control in HEVC
  via balancing Intra and Inter frame coding}.
\newblock \bibinfo{journal}{\emph{IEEE Transactions on Industrial Informatics}}
  \bibinfo{volume}{18}, \bibinfo{number}{3} (\bibinfo{year}{2022}),
  \bibinfo{pages}{1594--1604}.
\newblock


\bibitem[Gao et~al\mbox{.}(2016)]%
        {RC_SSIM_Gao}
\bibfield{author}{\bibinfo{person}{Wei Gao}, \bibinfo{person}{Sam Kwong},
  \bibinfo{person}{Yu Zhou}, {and} \bibinfo{person}{Hui Yuan}.}
  \bibinfo{year}{2016}\natexlab{}.
\newblock \showarticletitle{SSIM-based game theory approach for rate-distortion
  optimized intra frame CTU-level bit allocation}.
\newblock \bibinfo{journal}{\emph{IEEE Transactions on Multimedia}}
  \bibinfo{volume}{18}, \bibinfo{number}{6} (\bibinfo{year}{2016}),
  \bibinfo{pages}{988--999}.
\newblock


\bibitem[Gao et~al\mbox{.}(2021)]%
        {VCM}
\bibfield{author}{\bibinfo{person}{Wen Gao}, \bibinfo{person}{Siwei Ma},
  \bibinfo{person}{Lingyu Duan}, \bibinfo{person}{Yonghong Tian},
  \bibinfo{person}{Peiyin Xing}, \bibinfo{person}{Yaowei Wang},
  \bibinfo{person}{Shanshe Wang}, \bibinfo{person}{Huizhu Jia}, {and}
  \bibinfo{person}{Tiejun Huang}.} \bibinfo{year}{2021}\natexlab{}.
\newblock \showarticletitle{Digital retina: A way to make the city brain more
  efficient by visual coding}.
\newblock \bibinfo{journal}{\emph{IEEE Transactions on Circuits and Systems for
  Video Technology}} \bibinfo{volume}{31}, \bibinfo{number}{11}
  (\bibinfo{year}{2021}), \bibinfo{pages}{4147--4161}.
\newblock


\bibitem[Grois and Giladi(2020)]%
        {TQ_Grois}
\bibfield{author}{\bibinfo{person}{Dan Grois} {and} \bibinfo{person}{Alex
  Giladi}.} \bibinfo{year}{2020}\natexlab{}.
\newblock \showarticletitle{{Perceptual quantization matrices for high dynamic
  range H.265/MPEG-HEVC video coding }}. In
  \bibinfo{booktitle}{\emph{Applications of Digital Image Processing XLII}},
  Vol.~\bibinfo{volume}{11137}. \bibinfo{publisher}{SPIE}, \bibinfo{pages}{164
  -- 177}.
\newblock


\bibitem[Guan et~al\mbox{.}(2021)]%
        {EN_Guan}
\bibfield{author}{\bibinfo{person}{Zhenyu Guan}, \bibinfo{person}{Qunliang
  Xing}, \bibinfo{person}{Mai Xu}, \bibinfo{person}{Ren Yang},
  \bibinfo{person}{Tie Liu}, {and} \bibinfo{person}{Zulin Wang}.}
  \bibinfo{year}{2021}\natexlab{}.
\newblock \showarticletitle{MFQE 2.0: A new approach for multi-frame quality
  enhancement on compressed video}.
\newblock \bibinfo{journal}{\emph{IEEE Transactions on Pattern Analysis and
  Machine Intelligence}} \bibinfo{volume}{43}, \bibinfo{number}{3}
  (\bibinfo{year}{2021}), \bibinfo{pages}{949--963}.
\newblock


\bibitem[Guo and Chao(2017)]%
        {EN_Guo}
\bibfield{author}{\bibinfo{person}{Jun Guo} {and} \bibinfo{person}{Hongyang
  Chao}.} \bibinfo{year}{2017}\natexlab{}.
\newblock \showarticletitle{One-to-many network for visually pleasing
  compression artifacts reduction}. In \bibinfo{booktitle}{\emph{2017 IEEE
  Conference on Computer Vision and Pattern Recognition (CVPR)}}.
  \bibinfo{pages}{4867--4876}.
\newblock


\bibitem[Gupta et~al\mbox{.}(2011)]%
        {IQA_PSNR-HVS}
\bibfield{author}{\bibinfo{person}{Prateek Gupta}, \bibinfo{person}{Priyanka
  Srivastava}, \bibinfo{person}{Satyam Bhardwaj}, {and}
  \bibinfo{person}{Vikrant Bhateja}.} \bibinfo{year}{2011}\natexlab{}.
\newblock \showarticletitle{A modified PSNR metric based on HVS for quality
  assessment of color images}. In \bibinfo{booktitle}{\emph{2011 International
  Conference on Communication and Industrial Application}}.
  \bibinfo{pages}{1--4}.
\newblock


\bibitem[Hosseini et~al\mbox{.}(2019)]%
        {CM_Deep_CSF}
\bibfield{author}{\bibinfo{person}{Mahdi~S. Hosseini}, \bibinfo{person}{Yueyang
  Zhang}, {and} \bibinfo{person}{Konstantinos~N. Plataniotis}.}
  \bibinfo{year}{2019}\natexlab{}.
\newblock \showarticletitle{Encoding visual sensitivity by maxpol convolution
  filters for image sharpness assessment}.
\newblock \bibinfo{journal}{\emph{IEEE Transactions on Image Processing}}
  \bibinfo{volume}{28}, \bibinfo{number}{9} (\bibinfo{year}{2019}),
  \bibinfo{pages}{4510--4525}.
\newblock


\bibitem[Hu et~al\mbox{.}(2015)]%
        {IQA_PWMSE}
\bibfield{author}{\bibinfo{person}{Sudeng Hu}, \bibinfo{person}{Lina Jin},
  \bibinfo{person}{Hanli Wang}, \bibinfo{person}{Yun Zhang},
  \bibinfo{person}{Sam Kwong}, {and} \bibinfo{person}{C.-C.~Jay Kuo}.}
  \bibinfo{year}{2015}\natexlab{}.
\newblock \showarticletitle{Compressed image quality metric based on
  perceptually weighted distortion}.
\newblock \bibinfo{journal}{\emph{IEEE Transactions on Image Processing}}
  \bibinfo{volume}{24}, \bibinfo{number}{12} (\bibinfo{year}{2015}),
  \bibinfo{pages}{5594--5608}.
\newblock


\bibitem[Hu et~al\mbox{.}(2017)]%
        {VQA_PWMSE}
\bibfield{author}{\bibinfo{person}{Sudeng Hu}, \bibinfo{person}{Lina Jin},
  \bibinfo{person}{Hanli Wang}, \bibinfo{person}{Yun Zhang},
  \bibinfo{person}{Sam Kwong}, {and} \bibinfo{person}{C.-C.~Jay Kuo}.}
  \bibinfo{year}{2017}\natexlab{}.
\newblock \showarticletitle{Objective video quality assessment based on
  perceptually weighted mean squared error}.
\newblock \bibinfo{journal}{\emph{IEEE Transactions on Circuits and Systems for
  Video Technology}} \bibinfo{volume}{27}, \bibinfo{number}{9}
  (\bibinfo{year}{2017}), \bibinfo{pages}{1844--1855}.
\newblock


\bibitem[Itti et~al\mbox{.}(1998)]%
        {CM_VA_Itti}
\bibfield{author}{\bibinfo{person}{L. Itti}, \bibinfo{person}{C. Koch}, {and}
  \bibinfo{person}{E. Niebur}.} \bibinfo{year}{1998}\natexlab{}.
\newblock \showarticletitle{A model of saliency-based visual attention for
  rapid scene analysis}.
\newblock \bibinfo{journal}{\emph{IEEE Transactions on Pattern Analysis and
  Machine Intelligence}} \bibinfo{volume}{20}, \bibinfo{number}{11}
  (\bibinfo{year}{1998}), \bibinfo{pages}{1254--1259}.
\newblock


\bibitem[Jaballah et~al\mbox{.}(2018)]%
        {MVC_Jaballah}
\bibfield{author}{\bibinfo{person}{Sami Jaballah},
  \bibinfo{person}{Mohamed-Chaker Larabi}, {and} \bibinfo{person}{Jamel~Belhadj
  Tahar}.} \bibinfo{year}{2018}\natexlab{}.
\newblock \showarticletitle{Asymmetric DCT-JND for luminance adaptation
  effects: an application to perceptual video coding in MV-HEVC}. In
  \bibinfo{booktitle}{\emph{2018 IEEE International Conference on Acoustics,
  Speech and Signal Processing (ICASSP)}}. \bibinfo{pages}{1797--1801}.
\newblock


\bibitem[Jiang et~al\mbox{.}(2022)]%
        {CM_JND_JiangTIP22}
\bibfield{author}{\bibinfo{person}{Qiuping Jiang}, \bibinfo{person}{Zhentao
  Liu}, \bibinfo{person}{Shiqi Wang}, \bibinfo{person}{Feng Shao}, {and}
  \bibinfo{person}{Weisi Lin}.} \bibinfo{year}{2022}\natexlab{}.
\newblock \showarticletitle{Toward Top-Down Just Noticeable Difference
  Estimation of Natural Images}.
\newblock \bibinfo{journal}{\emph{IEEE Transactions on Image Processing}}
  \bibinfo{volume}{31} (\bibinfo{year}{2022}), \bibinfo{pages}{3697--3712}.
\newblock


\bibitem[Jin et~al\mbox{.}(2016)]%
        {CM_ImageJND_Jin}
\bibfield{author}{\bibinfo{person}{Lina Jin}, \bibinfo{person}{Joe Yu-chieh
  Lin}, \bibinfo{person}{Sudeng Hu}, \bibinfo{person}{Haiqiang Wang},
  \bibinfo{person}{Ping Wang}, \bibinfo{person}{Ioannis Katsavounidis},
  \bibinfo{person}{Anne Aaron}, {and} \bibinfo{person}{C.-C.~Jay Kuo}.}
  \bibinfo{year}{2016}\natexlab{}.
\newblock \showarticletitle{Statistical study on perceived JPEG image quality
  via MCL-JCI dataset construction and analysis}.
\newblock \bibinfo{journal}{\emph{Electronic Imaging}}  \bibinfo{volume}{2016}
  (\bibinfo{date}{02} \bibinfo{year}{2016}), \bibinfo{pages}{1--9}.
\newblock


\bibitem[Jin et~al\mbox{.}(2020)]%
        {EN_Jin_NC}
\bibfield{author}{\bibinfo{person}{Zhipeng Jin}, \bibinfo{person}{Ping An},
  \bibinfo{person}{Chao Yang}, {and} \bibinfo{person}{Liquan Shen}.}
  \bibinfo{year}{2020}\natexlab{}.
\newblock \showarticletitle{Post-processing for intra coding through perceptual
  adversarial learning and progressive refinement}.
\newblock \bibinfo{journal}{\emph{Neurocomputing}}  \bibinfo{volume}{394}
  (\bibinfo{year}{2020}), \bibinfo{pages}{158--167}.
\newblock
\showISSN{0925-2312}


\bibitem[Jin et~al\mbox{.}(2021)]%
        {EN_Jin_CSVT}
\bibfield{author}{\bibinfo{person}{Zhi Jin}, \bibinfo{person}{Muhammad~Zafar
  Iqbal}, \bibinfo{person}{Wenbin Zou}, \bibinfo{person}{Xia Li}, {and}
  \bibinfo{person}{Eckehard Steinbach}.} \bibinfo{year}{2021}\natexlab{}.
\newblock \showarticletitle{Dual-stream multi-path recursive residual network
  for JPEG image compression artifacts reduction}.
\newblock \bibinfo{journal}{\emph{IEEE Transactions on Circuits and Systems for
  Video Technology}} \bibinfo{volume}{31}, \bibinfo{number}{2}
  (\bibinfo{year}{2021}), \bibinfo{pages}{467--479}.
\newblock


\bibitem[Jung and Chen(2015)]%
        {RDO_Jung}
\bibfield{author}{\bibinfo{person}{Cheolkon Jung} {and} \bibinfo{person}{Yao
  Chen}.} \bibinfo{year}{2015}\natexlab{}.
\newblock \showarticletitle{Perceptual rate distortion optimisation for video
  coding using free-energy principle}.
\newblock \bibinfo{journal}{\emph{Electronics Letters}} \bibinfo{volume}{51},
  \bibinfo{number}{21} (\bibinfo{date}{10} \bibinfo{year}{2015}),
  \bibinfo{pages}{1656--1658}.
\newblock


\bibitem[Kelly(1961)]%
        {CM_TCSF_Kelly}
\bibfield{author}{\bibinfo{person}{D.~H. Kelly}.}
  \bibinfo{year}{1961}\natexlab{}.
\newblock \showarticletitle{Visual responses to time-dependent stimuli.$\ast$
  I. amplitude sensitivity measurements}.
\newblock \bibinfo{journal}{\emph{Journal of the Optical Society of America}}
  \bibinfo{volume}{51}, \bibinfo{number}{4} (\bibinfo{date}{Apr}
  \bibinfo{year}{1961}), \bibinfo{pages}{422--429}.
\newblock


\bibitem[Ki et~al\mbox{.}(2018)]%
        {TQ_Ki}
\bibfield{author}{\bibinfo{person}{Sehwan Ki}, \bibinfo{person}{Sung-Ho Bae},
  \bibinfo{person}{Munchurl Kim}, {and} \bibinfo{person}{Hyunsuk Ko}.}
  \bibinfo{year}{2018}\natexlab{}.
\newblock \showarticletitle{Learning-based just-noticeable-quantization-
  distortion modeling for perceptual video coding}.
\newblock \bibinfo{journal}{\emph{IEEE Transactions on Image Processing}}
  \bibinfo{volume}{27}, \bibinfo{number}{7} (\bibinfo{year}{2018}),
  \bibinfo{pages}{3178--3193}.
\newblock


\bibitem[Kim et~al\mbox{.}(2015)]%
        {TQ_Kim}
\bibfield{author}{\bibinfo{person}{Jaeil Kim}, \bibinfo{person}{Sung-Ho Bae},
  {and} \bibinfo{person}{Munchurl Kim}.} \bibinfo{year}{2015}\natexlab{}.
\newblock \showarticletitle{An HEVC-compliant perceptual video coding scheme
  based on JND models for variable block-sized transform kernels}.
\newblock \bibinfo{journal}{\emph{IEEE Transactions on Circuits and Systems for
  Video Technology}} \bibinfo{volume}{25}, \bibinfo{number}{11}
  (\bibinfo{year}{2015}), \bibinfo{pages}{1786--1800}.
\newblock


\bibitem[Kim and Lee(2017)]%
        {IQA_DeepQA}
\bibfield{author}{\bibinfo{person}{Jongyoo Kim} {and} \bibinfo{person}{Sanghoon
  Lee}.} \bibinfo{year}{2017}\natexlab{}.
\newblock \showarticletitle{Deep learning of human visual sensitivity in image
  quality assessment framework}. In \bibinfo{booktitle}{\emph{2017 IEEE
  Conference on Computer Vision and Pattern Recognition (CVPR)}}.
  \bibinfo{pages}{1969--1977}.
\newblock


\bibitem[Kim et~al\mbox{.}(2020)]%
        {CM_CSF_HDR}
\bibfield{author}{\bibinfo{person}{Minjung Kim}, \bibinfo{person}{Maliha
  Ashraf}, \bibinfo{person}{Mar{\'i}a P{\'e}rez-Ortiz}, \bibinfo{person}{Jasna
  Martinovic}, \bibinfo{person}{Sophie Wuerger}, {and} \bibinfo{person}{Rafal
  Mantiuk}.} \bibinfo{year}{2020}\natexlab{}.
\newblock \showarticletitle{Contrast sensitivity functions for HDR displays}.
\newblock \bibinfo{journal}{\emph{London Imaging Meeting}}
  \bibinfo{volume}{2020} (\bibinfo{date}{09} \bibinfo{year}{2020}),
  \bibinfo{pages}{44--48}.
\newblock


\bibitem[Lambrecht and Kunt(1998)]%
        {CM_STCSF_Lambrecht}
\bibfield{author}{\bibinfo{person}{Christian J. Van Den~Branden Lambrecht}
  {and} \bibinfo{person}{Murat Kunt}.} \bibinfo{year}{1998}\natexlab{}.
\newblock \showarticletitle{Characterization of human visual sensitivity for
  video imaging applications}.
\newblock \bibinfo{journal}{\emph{Signal Processing}}  \bibinfo{volume}{67}
  (\bibinfo{year}{1998}), \bibinfo{pages}{255--269}.
\newblock


\bibitem[Lee and Choi(2018)]%
        {RDO_Lee}
\bibfield{author}{\bibinfo{person}{Bumshik Lee} {and}
  \bibinfo{person}{Jae~Young Choi}.} \bibinfo{year}{2018}\natexlab{}.
\newblock \showarticletitle{A rate perceptual-distortion optimized video coding
  HEVC}.
\newblock \bibinfo{journal}{\emph{IEICE Transactions on Information and
  Systems}} \bibinfo{volume}{101}, \bibinfo{number}{12} (\bibinfo{date}{Dec}
  \bibinfo{year}{2018}), \bibinfo{pages}{3158--3169}.
\newblock


\bibitem[Lee and Ebrahimi(2012)]%
        {SV_PVC}
\bibfield{author}{\bibinfo{person}{Jong-Seok Lee} {and}
  \bibinfo{person}{Touradj Ebrahimi}.} \bibinfo{year}{2012}\natexlab{}.
\newblock \showarticletitle{Perceptual video compression: a survey}.
\newblock \bibinfo{journal}{\emph{IEEE Journal of Selected Topics in Signal
  Processing}} \bibinfo{volume}{6}, \bibinfo{number}{6} (\bibinfo{year}{2012}),
  \bibinfo{pages}{684--697}.
\newblock


\bibitem[Lee et~al\mbox{.}(2021)]%
        {SV_DNN}
\bibfield{author}{\bibinfo{person}{Royson Lee}, \bibinfo{person}{Stylianos~I.
  Venieris}, {and} \bibinfo{person}{Nicholas~D. Lane}.}
  \bibinfo{year}{2021}\natexlab{}.
\newblock \showarticletitle{Deep neural network-based enhancement for image and
  video streaming systems: a survey and future directions}.
\newblock \bibinfo{journal}{\emph{ACM Computing Survery}} \bibinfo{volume}{54},
  \bibinfo{number}{8}, Article \bibinfo{articleno}{169} (\bibinfo{date}{oct}
  \bibinfo{year}{2021}), \bibinfo{numpages}{30}~pages.
\newblock


\bibitem[Li et~al\mbox{.}(2022a)]%
        {RC_Li}
\bibfield{author}{\bibinfo{person}{Hao Li}, \bibinfo{person}{Weimin Lei}, {and}
  \bibinfo{person}{Wei Zhang}.} \bibinfo{year}{2022}\natexlab{a}.
\newblock \showarticletitle{Perceptual video coding based on adaptive
  region-level intra-period}. In \bibinfo{booktitle}{\emph{7th International
  Conference on Computer and Communication Systems (ICCCS)}}.
  \bibinfo{pages}{387--392}.
\newblock


\bibitem[Li et~al\mbox{.}(2020)]%
        {EN_Li}
\bibfield{author}{\bibinfo{person}{Jianwei Li}, \bibinfo{person}{Yongtao Wang},
  \bibinfo{person}{Haihua Xie}, {and} \bibinfo{person}{Kai-Kuang Ma}.}
  \bibinfo{year}{2020}\natexlab{}.
\newblock \showarticletitle{Learning a single model with a wide range of
  quality factors for JPEG image artifacts removal}.
\newblock \bibinfo{journal}{\emph{IEEE Transactions on Image Processing}}
  \bibinfo{volume}{29} (\bibinfo{year}{2020}), \bibinfo{pages}{8842--8854}.
\newblock


\bibitem[Li et~al\mbox{.}(2022b)]%
        {TQ_Saab}
\bibfield{author}{\bibinfo{person}{Na Li}, \bibinfo{person}{Yun Zhang}, {and}
  \bibinfo{person}{C.-C.~Jay Kuo}.} \bibinfo{year}{2022}\natexlab{b}.
\newblock \showarticletitle{High efficiency intra video coding based on
  data-driven transform}.
\newblock \bibinfo{journal}{\emph{IEEE Transactions on Broadcasting}}
  \bibinfo{volume}{68}, \bibinfo{number}{2} (\bibinfo{year}{2022}),
  \bibinfo{pages}{383--396}.
\newblock


\bibitem[Li and Mou(2021)]%
        {RC_SSIM_Li}
\bibfield{author}{\bibinfo{person}{Yang Li} {and} \bibinfo{person}{Xuanqin
  Mou}.} \bibinfo{year}{2021}\natexlab{}.
\newblock \showarticletitle{Joint optimization for SSIM-based CTU-level bit
  allocation and rate distortion optimization}.
\newblock \bibinfo{journal}{\emph{IEEE Transactions on Broadcasting}}
  \bibinfo{volume}{67}, \bibinfo{number}{2} (\bibinfo{year}{2021}),
  \bibinfo{pages}{500--511}.
\newblock


\bibitem[Li et~al\mbox{.}(2016)]%
        {VQA_VMAF}
\bibfield{author}{\bibinfo{person}{Zhi Li}, \bibinfo{person}{Anne Aaron},
  \bibinfo{person}{Katsavounidis}, \bibinfo{person}{Ioannis~Moorthy A}, {and}
  \bibinfo{person}{Megha Manohara}.} \bibinfo{year}{2016}\natexlab{}.
\newblock \showarticletitle{Toward a practical perceptual video quality
  metric}. In \bibinfo{booktitle}{\emph{Netflix TechBlog}}.
\newblock


\bibitem[Lim and Sim(2020)]%
        {RC_Lim}
\bibfield{author}{\bibinfo{person}{Woong Lim} {and} \bibinfo{person}{Donggyu
  Sim}.} \bibinfo{year}{2020}\natexlab{}.
\newblock \showarticletitle{A perceptual rate control algorithm based on
  luminance adaptation for HEVC encoders}.
\newblock \bibinfo{journal}{\emph{Signal, Image and Video Processing}}
  \bibinfo{volume}{14} (\bibinfo{year}{2020}), \bibinfo{pages}{887--895}.
\newblock


\bibitem[Lin and Ghinea(2022)]%
        {SV_JND_Lin}
\bibfield{author}{\bibinfo{person}{Weisi Lin} {and} \bibinfo{person}{Gheorghita
  Ghinea}.} \bibinfo{year}{2022}\natexlab{}.
\newblock \showarticletitle{Progress and opportunities in modelling
  just-noticeable difference (JND) for multimedia}.
\newblock \bibinfo{journal}{\emph{IEEE Transactions on Multimedia}}
  \bibinfo{volume}{24} (\bibinfo{year}{2022}), \bibinfo{pages}{3706--3721}.
\newblock


\bibitem[Lin and {Jay Kuo}(2011)]%
        {SV_PVQA}
\bibfield{author}{\bibinfo{person}{Weisi Lin} {and} \bibinfo{person}{C.-C. {Jay
  Kuo}}.} \bibinfo{year}{2011}\natexlab{}.
\newblock \showarticletitle{Perceptual visual quality metrics: A survey}.
\newblock \bibinfo{journal}{\emph{Journal of Visual Communication and Image
  Representation}} \bibinfo{volume}{22}, \bibinfo{number}{4}
  (\bibinfo{year}{2011}), \bibinfo{pages}{297--312}.
\newblock
\showISSN{1047-3203}


\bibitem[Liu et~al\mbox{.}(2020a)]%
        {SV_DLVC}
\bibfield{author}{\bibinfo{person}{Dong Liu}, \bibinfo{person}{Yue Li},
  \bibinfo{person}{Jianping Lin}, \bibinfo{person}{Houqiang Li}, {and}
  \bibinfo{person}{Feng Wu}.} \bibinfo{year}{2020}\natexlab{a}.
\newblock \showarticletitle{Deep learning-based video coding: a review and a
  case study}.
\newblock \bibinfo{journal}{\emph{ACM Computing Survery}} \bibinfo{volume}{53},
  \bibinfo{number}{1}, Article \bibinfo{articleno}{11} (\bibinfo{date}{Feb}
  \bibinfo{year}{2020}), \bibinfo{numpages}{35}~pages.
\newblock
\showISSN{0360-0300}


\bibitem[Liu et~al\mbox{.}(2020b)]%
        {CM_PWJND}
\bibfield{author}{\bibinfo{person}{Huanhua Liu}, \bibinfo{person}{Yun Zhang},
  \bibinfo{person}{Huan Zhang}, \bibinfo{person}{Chunling Fan},
  \bibinfo{person}{Sam Kwong}, \bibinfo{person}{C.-C.~Jay Kuo}, {and}
  \bibinfo{person}{Xiaoping Fan}.} \bibinfo{year}{2020}\natexlab{b}.
\newblock \showarticletitle{Deep learning-based picture-wise just noticeable
  distortion prediction model for image compression}.
\newblock \bibinfo{journal}{\emph{IEEE Transactions on Image Processing}}
  \bibinfo{volume}{29} (\bibinfo{year}{2020}), \bibinfo{pages}{641--656}.
\newblock


\bibitem[Liu et~al\mbox{.}(2019)]%
        {RC_PWMSE}
\bibfield{author}{\bibinfo{person}{Xiaoyan Liu}, \bibinfo{person}{Yun Zhang},
  \bibinfo{person}{Linwei Zhu}, {and} \bibinfo{person}{Huanhua Liu}.}
  \bibinfo{year}{2019}\natexlab{}.
\newblock \showarticletitle{Perception-based CTU level bit allocation for Intra
  high efficiency video coding}.
\newblock \bibinfo{journal}{\emph{IEEE Access}}  \bibinfo{volume}{7}
  (\bibinfo{year}{2019}), \bibinfo{pages}{154959--154970}.
\newblock


\bibitem[Liu et~al\mbox{.}(2018)]%
        {RDO_Liu}
\bibfield{author}{\bibinfo{person}{Yanwei Liu}, \bibinfo{person}{Jinxia Liu},
  \bibinfo{person}{Antonios Argyriou}, {and} \bibinfo{person}{Song Ci}.}
  \bibinfo{year}{2018}\natexlab{}.
\newblock \showarticletitle{Binocular-combination-oriented perceptual
  rate-distortion optimization for stereoscopic video coding}.
\newblock \bibinfo{journal}{\emph{IEEE Transactions on Circuits and Systems for
  Video Technology}} \bibinfo{volume}{28}, \bibinfo{number}{8}
  (\bibinfo{year}{2018}), \bibinfo{pages}{1949--1959}.
\newblock


\bibitem[Luo et~al\mbox{.}(2013)]%
        {TQ_Luo}
\bibfield{author}{\bibinfo{person}{Zhengyi Luo}, \bibinfo{person}{Li Song},
  \bibinfo{person}{Shibao Zheng}, {and} \bibinfo{person}{Nam Ling}.}
  \bibinfo{year}{2013}\natexlab{}.
\newblock \showarticletitle{H.264/Advanced video control perceptual
  optimization coding based on JND-directed coefficient suppression}.
\newblock \bibinfo{journal}{\emph{IEEE Transactions on Circuits and Systems for
  Video Technology}} \bibinfo{volume}{23}, \bibinfo{number}{6}
  (\bibinfo{year}{2013}), \bibinfo{pages}{935--948}.
\newblock


\bibitem[Luo et~al\mbox{.}(2021)]%
        {RDO_Luo_VMAF}
\bibfield{author}{\bibinfo{person}{Zhengyi Luo}, \bibinfo{person}{Chen Zhu},
  \bibinfo{person}{Yan Huang}, \bibinfo{person}{Rong Xie}, \bibinfo{person}{Li
  Song}, {and} \bibinfo{person}{C.-C.~Jay Kuo}.}
  \bibinfo{year}{2021}\natexlab{}.
\newblock \showarticletitle{VMAF oriented perceptual coding based on piecewise
  metric coupling}.
\newblock \bibinfo{journal}{\emph{IEEE Transactions on Image Processing}}
  \bibinfo{volume}{30} (\bibinfo{year}{2021}), \bibinfo{pages}{5109--5121}.
\newblock


\bibitem[Mansouri and Mahmoudi-Aznaveh(2019)]%
        {IQA_SVD_SPIC19}
\bibfield{author}{\bibinfo{person}{Azadeh Mansouri} {and}
  \bibinfo{person}{Ahmad Mahmoudi-Aznaveh}.} \bibinfo{year}{2019}\natexlab{}.
\newblock \showarticletitle{SSVD: Structural SVD-based image quality
  assessment}.
\newblock \bibinfo{journal}{\emph{Signal Processing: Image Communication}}
  \bibinfo{volume}{74} (\bibinfo{year}{2019}), \bibinfo{pages}{54--63}.
\newblock
\showISSN{0923-5965}


\bibitem[Mantiuk et~al\mbox{.}(2011)]%
        {IQA_HDR_VDP2}
\bibfield{author}{\bibinfo{person}{Rafa\l{} Mantiuk},
  \bibinfo{person}{Kil~Joong Kim}, \bibinfo{person}{Allan~G. Rempel}, {and}
  \bibinfo{person}{Wolfgang Heidrich}.} \bibinfo{year}{2011}\natexlab{}.
\newblock \showarticletitle{HDR-VDP-2: A calibrated visual metric for
  visibility and quality predictions in all luminance conditions}.
\newblock \bibinfo{journal}{\emph{ACM Transactions on Graphics}}
  \bibinfo{volume}{30}, \bibinfo{number}{4}, Article \bibinfo{articleno}{40}
  (\bibinfo{year}{2011}), \bibinfo{numpages}{14}~pages.
\newblock


\bibitem[Min et~al\mbox{.}(2021)]%
        {SV_SCIQA}
\bibfield{author}{\bibinfo{person}{Xiongkuo Min}, \bibinfo{person}{Ke Gu},
  \bibinfo{person}{Guangtao Zhai}, \bibinfo{person}{Xiaokang Yang},
  \bibinfo{person}{Wenjun Zhang}, \bibinfo{person}{Patrick Le~Callet}, {and}
  \bibinfo{person}{Chang~Wen Chen}.} \bibinfo{year}{2021}\natexlab{}.
\newblock \showarticletitle{Screen content quality assessment: overview,
  benchmark, and beyond}.
\newblock \bibinfo{journal}{\emph{ACM Computing Survery}} \bibinfo{volume}{54},
  \bibinfo{number}{9}, Article \bibinfo{articleno}{187} (\bibinfo{date}{oct}
  \bibinfo{year}{2021}), \bibinfo{numpages}{36}~pages.
\newblock
\showISSN{0360-0300}


\bibitem[Mullen(1985)]%
        {CM_Color-CSF}
\bibfield{author}{\bibinfo{person}{Kathy Mullen}.}
  \bibinfo{year}{1985}\natexlab{}.
\newblock \showarticletitle{The contrast sensitivity of human color vision to
  red-green and blue-yellow chromatic gratings}.
\newblock \bibinfo{journal}{\emph{The Journal of Physiology}}
  \bibinfo{volume}{359} (\bibinfo{date}{03} \bibinfo{year}{1985}),
  \bibinfo{pages}{381--400}.
\newblock


\bibitem[Nami et~al\mbox{.}(2022)]%
        {TQ_Nami}
\bibfield{author}{\bibinfo{person}{Sanaz Nami}, \bibinfo{person}{Farhad
  Pakdaman}, \bibinfo{person}{Mahmoud~Reza Hashemi}, {and}
  \bibinfo{person}{Shervin Shirmohammadi}.} \bibinfo{year}{2022}\natexlab{}.
\newblock \showarticletitle{BL-JUNIPER: A CNN-Assisted Framework for Perceptual
  Video Coding Leveraging Block-Level JND}.
\newblock \bibinfo{journal}{\emph{IEEE Transactions on Multimedia}}
  (\bibinfo{year}{2022}), \bibinfo{pages}{1--16}.
\newblock


\bibitem[Narwaria et~al\mbox{.}(2015)]%
        {VQA_HDR_VQM}
\bibfield{author}{\bibinfo{person}{Manish Narwaria}, \bibinfo{person}{Matthieu
  Perreira Da~Silva}, {and} \bibinfo{person}{Patrick Le~Callet}.}
  \bibinfo{year}{2015}\natexlab{}.
\newblock \showarticletitle{HDR-VQM: An objective quality measure for high
  dynamic range video}.
\newblock \bibinfo{journal}{\emph{Signal Processing: Image Communication}}
  \bibinfo{volume}{35} (\bibinfo{date}{05} \bibinfo{year}{2015}).
\newblock


\bibitem[P.910(2022)]%
        {P910}
\bibfield{author}{\bibinfo{person}{P.910}.} \bibinfo{year}{2022}\natexlab{}.
\newblock \showarticletitle{Subjective video quality assessment methods for
  multimedia applications}.
\newblock \bibinfo{journal}{\emph{ITU-T Recommendations}}
  (\bibinfo{year}{2022}).
\newblock


\bibitem[Pan et~al\mbox{.}(2020)]%
        {EN_Pan_TIP}
\bibfield{author}{\bibinfo{person}{Zhaoqing Pan}, \bibinfo{person}{Xiaokai Yi},
  \bibinfo{person}{Yun Zhang}, \bibinfo{person}{Byeungwoo Jeon}, {and}
  \bibinfo{person}{Sam Kwong}.} \bibinfo{year}{2020}\natexlab{}.
\newblock \showarticletitle{Efficient in-loop filtering based on enhanced deep
  convolutional neural networks for HEVC}.
\newblock \bibinfo{journal}{\emph{IEEE Transactions on Image Processing}}
  \bibinfo{volume}{29} (\bibinfo{year}{2020}), \bibinfo{pages}{5352--5366}.
\newblock


\bibitem[Papadopoulos et~al\mbox{.}(2017)]%
        {TQ_QP_Papa}
\bibfield{author}{\bibinfo{person}{M.~A. Papadopoulos}, \bibinfo{person}{Y.
  Rai}, \bibinfo{person}{A.~V. Katsenou}, \bibinfo{person}{D. Agrafiotis},
  \bibinfo{person}{P. Le~Callet}, {and} \bibinfo{person}{D.~R. Bull}.}
  \bibinfo{year}{2017}\natexlab{}.
\newblock \showarticletitle{Video quality enhancement via QP adaptation based
  on perceptual coding maps}. In \bibinfo{booktitle}{\emph{2017 IEEE
  International Conference on Image Processing (ICIP)}}.
  \bibinfo{pages}{2741--2745}.
\newblock


\bibitem[Pinson and Wolf(2004)]%
        {VQA_VQM}
\bibfield{author}{\bibinfo{person}{M.H. Pinson} {and} \bibinfo{person}{S.
  Wolf}.} \bibinfo{year}{2004}\natexlab{}.
\newblock \showarticletitle{A new standardized method for objectively measuring
  video quality}.
\newblock \bibinfo{journal}{\emph{IEEE Transactions on Broadcasting}}
  \bibinfo{volume}{50}, \bibinfo{number}{3} (\bibinfo{year}{2004}),
  \bibinfo{pages}{312--322}.
\newblock


\bibitem[Prangnell and Sanchez(2016)]%
        {TQ_Prangnell}
\bibfield{author}{\bibinfo{person}{Lee Prangnell} {and} \bibinfo{person}{Victor
  Sanchez}.} \bibinfo{year}{2016}\natexlab{}.
\newblock \showarticletitle{Adaptive quantization matrices for HD and UHD
  resolutions in scalable HEVC}. In \bibinfo{booktitle}{\emph{2016 Data
  Compression Conference (DCC)}}. \bibinfo{pages}{626--626}.
\newblock


\bibitem[Robson(1966)]%
        {CM_ST-CSF}
\bibfield{author}{\bibinfo{person}{J.~G. Robson}.}
  \bibinfo{year}{1966}\natexlab{}.
\newblock \showarticletitle{Spatial and temporal contrast-sensitivity functions
  of the visual system}.
\newblock \bibinfo{journal}{\emph{Journal of the Optical Society of America}}
  \bibinfo{volume}{56}, \bibinfo{number}{8} (\bibinfo{date}{Aug}
  \bibinfo{year}{1966}), \bibinfo{pages}{1141--1142}.
\newblock


\bibitem[Rouis et~al\mbox{.}(2018)]%
        {RDO_Rouis}
\bibfield{author}{\bibinfo{person}{Kais Rouis}, \bibinfo{person}{Mohamed-Chaker
  Larabi}, {and} \bibinfo{person}{Jamel Belhadj~Tahar}.}
  \bibinfo{year}{2018}\natexlab{}.
\newblock \showarticletitle{Perceptually adaptive lagrangian multiplier for
  HEVC guided rate-distortion optimization}.
\newblock \bibinfo{journal}{\emph{IEEE Access}}  \bibinfo{volume}{6}
  (\bibinfo{year}{2018}), \bibinfo{pages}{33589--33603}.
\newblock


\bibitem[Rovamo et~al\mbox{.}(1978)]%
        {CM_CSF_Rovamo}
\bibfield{author}{\bibinfo{person}{Jyrki Rovamo}, \bibinfo{person}{Veijo
  Virsu}, {and} \bibinfo{person}{Risto N{\"a}s{\"a}nen}.}
  \bibinfo{year}{1978}\natexlab{}.
\newblock \showarticletitle{Cortical magnification factor predicts the photopic
  contrast sensitivity of peripheral vision}.
\newblock \bibinfo{journal}{\emph{Natrue}}  \bibinfo{volume}{271}
  (\bibinfo{year}{1978}), \bibinfo{pages}{54--56}.
\newblock


\bibitem[Seshadrinathan and Bovik(2010)]%
        {VQA_MOVIE}
\bibfield{author}{\bibinfo{person}{Kalpana Seshadrinathan} {and}
  \bibinfo{person}{Alan~Conrad Bovik}.} \bibinfo{year}{2010}\natexlab{}.
\newblock \showarticletitle{Motion tuned spatio-temporal quality assessment of
  natural videos}.
\newblock \bibinfo{journal}{\emph{IEEE Transactions on Image Processing}}
  \bibinfo{volume}{19}, \bibinfo{number}{2} (\bibinfo{year}{2010}),
  \bibinfo{pages}{335--350}.
\newblock


\bibitem[Shang et~al\mbox{.}(2019a)]%
        {IQA_CSPSNR}
\bibfield{author}{\bibinfo{person}{Xiwu Shang}, \bibinfo{person}{Jie Liang},
  \bibinfo{person}{Guozhong Wang}, \bibinfo{person}{Haiwu Zhao},
  \bibinfo{person}{Chengjia Wu}, {and} \bibinfo{person}{Chang Lin}.}
  \bibinfo{year}{2019}\natexlab{a}.
\newblock \showarticletitle{Color-sensitivity-based combined PSNR for objective
  video quality assessment}.
\newblock \bibinfo{journal}{\emph{IEEE Transactions on Circuits and Systems for
  Video Technology}} \bibinfo{volume}{29}, \bibinfo{number}{5}
  (\bibinfo{year}{2019}), \bibinfo{pages}{1239--1250}.
\newblock


\bibitem[Shang et~al\mbox{.}(2019b)]%
        {TQ_Shang}
\bibfield{author}{\bibinfo{person}{Xiwu Shang}, \bibinfo{person}{Guozhong
  Wang}, \bibinfo{person}{Xiaoli Zhao}, \bibinfo{person}{Yifan Zuo},
  \bibinfo{person}{Jie Liang}, {and} \bibinfo{person}{Ivan~V. Bajic}.}
  \bibinfo{year}{2019}\natexlab{b}.
\newblock \showarticletitle{Weighting quantization matrices for
  HEVC/H.265-coded RGB videos}.
\newblock \bibinfo{journal}{\emph{IEEE Access}}  \bibinfo{volume}{7}
  (\bibinfo{year}{2019}), \bibinfo{pages}{36019--36032}.
\newblock


\bibitem[Shen et~al\mbox{.}(2021)]%
        {CM_PatchJND}
\bibfield{author}{\bibinfo{person}{Xuelin Shen}, \bibinfo{person}{Zhangkai Ni},
  \bibinfo{person}{Wenhan Yang}, \bibinfo{person}{Xinfeng Zhang},
  \bibinfo{person}{Shiqi Wang}, {and} \bibinfo{person}{Sam Kwong}.}
  \bibinfo{year}{2021}\natexlab{}.
\newblock \showarticletitle{Just noticeable distortion profile inference: a
  patch-level structural visibility learning approach}.
\newblock \bibinfo{journal}{\emph{IEEE Transactions on Image Processing}}
  \bibinfo{volume}{30} (\bibinfo{year}{2021}), \bibinfo{pages}{26--38}.
\newblock


\bibitem[Stockman and Sharpe(1998)]%
        {CM_Stochman}
\bibfield{author}{\bibinfo{person}{Andrew Stockman} {and}
  \bibinfo{person}{Lindsay~T. Sharpe}.} \bibinfo{year}{1998}\natexlab{}.
\newblock \showarticletitle{Human cone spectral sensitivities: a progress
  report}.
\newblock \bibinfo{journal}{\emph{Vision Research}} \bibinfo{volume}{38},
  \bibinfo{number}{21} (\bibinfo{year}{1998}), \bibinfo{pages}{3193--3206}.
\newblock
\showISSN{0042-6989}


\bibitem[Stuart(1998)]%
        {CM_PeriAcuity}
\bibfield{author}{\bibinfo{person}{Anstis Stuart}.}
  \bibinfo{year}{1998}\natexlab{}.
\newblock \showarticletitle{Picturing peripheral acuity}.
\newblock \bibinfo{journal}{\emph{Perception}}  \bibinfo{volume}{27}
  (\bibinfo{year}{1998}), \bibinfo{pages}{817--825}.
\newblock


\bibitem[Sullivan and Minoo(2012)]%
        {IQA_WTPSNR}
\bibfield{author}{\bibinfo{person}{Gary Sullivan} {and}
  \bibinfo{person}{Koohyar Minoo}.} \bibinfo{year}{2012}\natexlab{}.
\newblock \showarticletitle{JCT-VC AHG report: objective quality metric and
  alternative methods for measuring coding efficiency (AHG12)}. In
  \bibinfo{booktitle}{\emph{doc. JCT-VC-H0012, ITU-T/ISO/IEC JCT-VC}}.
\newblock


\bibitem[Sullivan et~al\mbox{.}(2012)]%
        {OV_HEVC}
\bibfield{author}{\bibinfo{person}{Gary~J. Sullivan},
  \bibinfo{person}{Jens-Rainer Ohm}, \bibinfo{person}{Woo-Jin Han}, {and}
  \bibinfo{person}{Thomas Wiegand}.} \bibinfo{year}{2012}\natexlab{}.
\newblock \showarticletitle{Overview of the high efficiency video coding (HEVC)
  standard}.
\newblock \bibinfo{journal}{\emph{IEEE Transactions on Circuits and Systems for
  Video Technology}} \bibinfo{volume}{22}, \bibinfo{number}{12}
  (\bibinfo{year}{2012}), \bibinfo{pages}{1649--1668}.
\newblock


\bibitem[Tech et~al\mbox{.}(2016)]%
        {OV_3DVC}
\bibfield{author}{\bibinfo{person}{Gerhard Tech}, \bibinfo{person}{Ying Chen},
  \bibinfo{person}{Karsten M{\"u}ller}, \bibinfo{person}{Jens-Rainer Ohm},
  \bibinfo{person}{Anthony Vetro}, {and} \bibinfo{person}{Ye-Kui Wang}.}
  \bibinfo{year}{2016}\natexlab{}.
\newblock \showarticletitle{Overview of the multiview and 3D extensions of high
  efficiency video coding}.
\newblock \bibinfo{journal}{\emph{IEEE Transactions on Circuits and Systems for
  Video Technology}} \bibinfo{volume}{26}, \bibinfo{number}{1}
  (\bibinfo{year}{2016}), \bibinfo{pages}{35--49}.
\newblock


\bibitem[Tian et~al\mbox{.}(2021)]%
        {CM_JND_TOMM}
\bibfield{author}{\bibinfo{person}{Tao Tian}, \bibinfo{person}{Hanli Wang},
  \bibinfo{person}{Sam Kwong}, {and} \bibinfo{person}{C.-C.~Jay Kuo}.}
  \bibinfo{year}{2021}\natexlab{}.
\newblock \showarticletitle{Perceptual image compression with block-Level just
  noticeable difference prediction}.
\newblock \bibinfo{journal}{\emph{ACM Transactions on Multimedia Computing,
  Communications, and Applications}} \bibinfo{volume}{16}, \bibinfo{number}{4},
  Article \bibinfo{articleno}{126} (\bibinfo{year}{2021}),
  \bibinfo{numpages}{15}~pages.
\newblock


\bibitem[Tolhurst(1975)]%
        {CM_ST-Nature}
\bibfield{author}{\bibinfo{person}{J.~Anthony Tolhurst, David Jand~Movshon}.}
  \bibinfo{year}{1975}\natexlab{}.
\newblock \showarticletitle{Spatial and temporal contrast sensitivity of
  striate cortical neurones}.
\newblock \bibinfo{journal}{\emph{Nature}}  \bibinfo{volume}{257}
  (\bibinfo{date}{11} \bibinfo{year}{1975}), \bibinfo{pages}{674--5}.
\newblock


\bibitem[Valin and Terriberry(2015)]%
        {TQ_Valin}
\bibfield{author}{\bibinfo{person}{Jean-Marc Valin} {and}
  \bibinfo{person}{Timothy~B. Terriberry}.} \bibinfo{year}{2015}\natexlab{}.
\newblock \showarticletitle{{Perceptual vector quantization for video coding}}.
  In \bibinfo{booktitle}{\emph{Visual Information Processing and Communication
  VI}}, Vol.~\bibinfo{volume}{9410}. \bibinfo{publisher}{SPIE},
  \bibinfo{pages}{65 -- 75}.
\newblock


\bibitem[van Nes and Bouman(1967)]%
        {CM_ContrastSen}
\bibfield{author}{\bibinfo{person}{Floris~L. van Nes} {and}
  \bibinfo{person}{Maarten~A. Bouman}.} \bibinfo{year}{1967}\natexlab{}.
\newblock \showarticletitle{Spatial modulation transfer in the human eye}.
\newblock \bibinfo{journal}{\emph{Journal of the Optical Society of America}}
  \bibinfo{volume}{57} (\bibinfo{year}{1967}), \bibinfo{pages}{401--406}.
\newblock


\bibitem[Vidal et~al\mbox{.}(2017)]%
        {EN_Vidal}
\bibfield{author}{\bibinfo{person}{Elo{\"i}se Vidal}, \bibinfo{person}{Nicolas
  Sturmel}, \bibinfo{person}{Christine Guillemot}, \bibinfo{person}{Patrick
  Corlay}, {and} \bibinfo{person}{Francois-Xavier Coudoux}.}
  \bibinfo{year}{2017}\natexlab{}.
\newblock \showarticletitle{New adaptive filters as perceptual preprocessing
  for rate-quality performance optimization of video coding}.
\newblock \bibinfo{journal}{\emph{Signal Processing: Image Communication}}
  \bibinfo{volume}{52} (\bibinfo{year}{2017}), \bibinfo{pages}{124--137}.
\newblock


\bibitem[Wang et~al\mbox{.}(2017a)]%
        {CM_VideoJND_Wang}
\bibfield{author}{\bibinfo{person}{Haiqiang Wang}, \bibinfo{person}{Ioannis
  Katsavounidis}, \bibinfo{person}{Jiantong Zhou}, \bibinfo{person}{Jeonghoon
  Park}, \bibinfo{person}{Shawmin Lei}, \bibinfo{person}{Xin Zhou},
  \bibinfo{person}{Man-On Pun}, \bibinfo{person}{Xin Jin},
  \bibinfo{person}{Ronggang Wang}, \bibinfo{person}{Xu Wang},
  \bibinfo{person}{Yun Zhang}, \bibinfo{person}{Jiwu Huang},
  \bibinfo{person}{Sam Kwong}, {and} \bibinfo{person}{C.-C.~Jay Kuo}.}
  \bibinfo{year}{2017}\natexlab{a}.
\newblock \showarticletitle{VideoSet: a large-scale compressed video quality
  dataset based on JND measurement}.
\newblock \bibinfo{journal}{\emph{Journal of Visual Communication and Image
  Representation}}  \bibinfo{volume}{46} (\bibinfo{date}{01}
  \bibinfo{year}{2017}).
\newblock


\bibitem[Wang et~al\mbox{.}(2018)]%
        {RC_Mask_Wang}
\bibfield{author}{\bibinfo{person}{Hao Wang}, \bibinfo{person}{Li Song},
  \bibinfo{person}{Rong Xie}, \bibinfo{person}{Zhengyi Luo}, {and}
  \bibinfo{person}{Xiangwen Wang}.} \bibinfo{year}{2018}\natexlab{}.
\newblock \showarticletitle{Masking effects based rate control scheme for high
  efficiency video coding}. In \bibinfo{booktitle}{\emph{2018 IEEE
  International Symposium on Circuits and Systems (ISCAS)}}.
  \bibinfo{pages}{1--5}.
\newblock


\bibitem[Wang et~al\mbox{.}(2021)]%
        {CM_JND_WangTIP21}
\bibfield{author}{\bibinfo{person}{Hongkui Wang}, \bibinfo{person}{Li Yu},
  \bibinfo{person}{Junhui Liang}, \bibinfo{person}{Haibing Yin},
  \bibinfo{person}{Tiansong Li}, {and} \bibinfo{person}{Shengwei Wang}.}
  \bibinfo{year}{2021}\natexlab{}.
\newblock \showarticletitle{Hierarchical Predictive Coding-Based JND Estimation
  for Image Compression}.
\newblock \bibinfo{journal}{\emph{IEEE Transactions on Image Processing}}
  \bibinfo{volume}{30} (\bibinfo{year}{2021}), \bibinfo{pages}{487--500}.
\newblock


\bibitem[Wang et~al\mbox{.}(2019)]%
        {RDO_Wang_ICIP}
\bibfield{author}{\bibinfo{person}{Qun Wang}, \bibinfo{person}{Hui Yuan},
  \bibinfo{person}{Junyan Huo}, {and} \bibinfo{person}{Peng Li}.}
  \bibinfo{year}{2019}\natexlab{}.
\newblock \showarticletitle{A fidelity-assured rate distortion optimization
  method for perceptual-based video coding}. In \bibinfo{booktitle}{\emph{2019
  IEEE International Conference on Image Processing (ICIP)}}.
  \bibinfo{pages}{4135--4139}.
\newblock


\bibitem[Wang et~al\mbox{.}(2012)]%
        {RDO_SSIM_Wang}
\bibfield{author}{\bibinfo{person}{Shiqi Wang}, \bibinfo{person}{Abdul Rehman},
  \bibinfo{person}{Zhou Wang}, \bibinfo{person}{Siwei Ma}, {and}
  \bibinfo{person}{Wen Gao}.} \bibinfo{year}{2012}\natexlab{}.
\newblock \showarticletitle{SSIM-motivated rate-distortion optimization for
  video coding}.
\newblock \bibinfo{journal}{\emph{IEEE Transactions on Circuits and Systems for
  Video Technology}} \bibinfo{volume}{22}, \bibinfo{number}{4}
  (\bibinfo{year}{2012}), \bibinfo{pages}{516--529}.
\newblock


\bibitem[Wang et~al\mbox{.}(2013)]%
        {RDO_Wang_TIP}
\bibfield{author}{\bibinfo{person}{Shiqi Wang}, \bibinfo{person}{Abdul Rehman},
  \bibinfo{person}{Zhou Wang}, \bibinfo{person}{Siwei Ma}, {and}
  \bibinfo{person}{Wen Gao}.} \bibinfo{year}{2013}\natexlab{}.
\newblock \showarticletitle{Perceptual video coding based on SSIM-inspired
  divisive normalization}.
\newblock \bibinfo{journal}{\emph{IEEE Transactions on Image Processing}}
  \bibinfo{volume}{22}, \bibinfo{number}{4} (\bibinfo{year}{2013}),
  \bibinfo{pages}{1418--1429}.
\newblock


\bibitem[Wang et~al\mbox{.}(2017b)]%
        {RC_SSIM_SQWang}
\bibfield{author}{\bibinfo{person}{Shiqi Wang}, \bibinfo{person}{Abdul Rehman},
  \bibinfo{person}{Kai Zeng}, \bibinfo{person}{Jiheng Wang}, {and}
  \bibinfo{person}{Zhou Wang}.} \bibinfo{year}{2017}\natexlab{b}.
\newblock \showarticletitle{SSIM-motivated two-pass VBR coding for HEVC}.
\newblock \bibinfo{journal}{\emph{IEEE Transactions on Circuits and Systems for
  Video Technology}} \bibinfo{volume}{27}, \bibinfo{number}{10}
  (\bibinfo{year}{2017}), \bibinfo{pages}{2189--2203}.
\newblock


\bibitem[Wang et~al\mbox{.}(2004)]%
        {IQA_SSIM}
\bibfield{author}{\bibinfo{person}{Zhou Wang}, \bibinfo{person}{A.C. Bovik},
  \bibinfo{person}{H.R. Sheikh}, {and} \bibinfo{person}{E.P. Simoncelli}.}
  \bibinfo{year}{2004}\natexlab{}.
\newblock \showarticletitle{Image quality assessment: from error visibility to
  structural similarity}.
\newblock \bibinfo{journal}{\emph{IEEE Transactions on Image Processing}}
  \bibinfo{volume}{13}, \bibinfo{number}{4} (\bibinfo{year}{2004}),
  \bibinfo{pages}{600--612}.
\newblock


\bibitem[Wang et~al\mbox{.}(2003)]%
        {IQA_MSSSIM}
\bibfield{author}{\bibinfo{person}{Zhou Wang}, \bibinfo{person}{E.P.
  Simoncelli}, {and} \bibinfo{person}{A.C. Bovik}.}
  \bibinfo{year}{2003}\natexlab{}.
\newblock \showarticletitle{Multiscale structural similarity for image quality
  assessment}. In \bibinfo{booktitle}{\emph{The Thrity-Seventh Asilomar
  Conference on Signals, Systems Computers, 2003}}, Vol.~\bibinfo{volume}{2}.
  \bibinfo{pages}{1398--1402 Vol.2}.
\newblock


\bibitem[Wien et~al\mbox{.}(2019)]%
        {OV_IVC}
\bibfield{author}{\bibinfo{person}{Mathias Wien}, \bibinfo{person}{Jill~M.
  Boyce}, \bibinfo{person}{Thomas Stockhammer}, {and}
  \bibinfo{person}{Wen-Hsiao Peng}.} \bibinfo{year}{2019}\natexlab{}.
\newblock \showarticletitle{Standardization status of immersive video coding}.
\newblock \bibinfo{journal}{\emph{IEEE Journal on Emerging and Selected Topics
  in Circuits and Systems}} \bibinfo{volume}{9}, \bibinfo{number}{1}
  (\bibinfo{year}{2019}), \bibinfo{pages}{5--17}.
\newblock


\bibitem[Woojae et~al\mbox{.}(2018)]%
        {VQA_DeepVQA}
\bibfield{author}{\bibinfo{person}{Kim Woojae}, \bibinfo{person}{Jongyoo Kim},
  \bibinfo{person}{Sewoong Ahn}, \bibinfo{person}{Jinwoo Kim}, {and}
  \bibinfo{person}{Sanghoon Lee}.} \bibinfo{year}{2018}\natexlab{}.
\newblock \showarticletitle{Deep video quality assessor: From spatio-temporal
  visual sensitivity to a convolutional neural aggregation network}. In
  \bibinfo{booktitle}{\emph{15th European Conference on Computer Vision}}.
  \bibinfo{pages}{224--241}.
\newblock


\bibitem[Wu et~al\mbox{.}(2017)]%
        {CM_JND_Pattern}
\bibfield{author}{\bibinfo{person}{Jinjian Wu}, \bibinfo{person}{Leida Li},
  \bibinfo{person}{Weisheng Dong}, \bibinfo{person}{Guangming Shi},
  \bibinfo{person}{Weisi Lin}, {and} \bibinfo{person}{C.-C.~Jay Kuo}.}
  \bibinfo{year}{2017}\natexlab{}.
\newblock \showarticletitle{Enhanced just noticeable difference model for
  images with pattern complexity}.
\newblock \bibinfo{journal}{\emph{IEEE Transactions on Image Processing}}
  \bibinfo{volume}{26}, \bibinfo{number}{6} (\bibinfo{year}{2017}),
  \bibinfo{pages}{2682--2693}.
\newblock


\bibitem[Wu et~al\mbox{.}(2019)]%
        {SV_JND}
\bibfield{author}{\bibinfo{person}{Jinjian Wu}, \bibinfo{person}{Guangming
  Shi}, {and} \bibinfo{person}{Weisi Lin}.} \bibinfo{year}{2019}\natexlab{}.
\newblock \showarticletitle{Survey of visual just noticeable difference
  estimation}.
\newblock \bibinfo{journal}{\emph{Frontiers of Computer Science}}
  \bibinfo{volume}{13}, \bibinfo{number}{1} (\bibinfo{year}{2019}),
  \bibinfo{pages}{4--15}.
\newblock


\bibitem[Wu et~al\mbox{.}(2020)]%
        {RDO_PWMSE}
\bibfield{author}{\bibinfo{person}{Xiuzhe Wu}, \bibinfo{person}{Hanli Wang},
  \bibinfo{person}{Sudeng Hu}, \bibinfo{person}{Sam Kwong}, {and}
  \bibinfo{person}{C.-C.~Jay Kuo}.} \bibinfo{year}{2020}\natexlab{}.
\newblock \showarticletitle{Perceptually weighted mean squared error based
  rate-distortion optimization for HEVC}.
\newblock \bibinfo{journal}{\emph{IEEE Transactions on Broadcasting}}
  \bibinfo{volume}{66}, \bibinfo{number}{4} (\bibinfo{year}{2020}),
  \bibinfo{pages}{824--834}.
\newblock


\bibitem[Xiang et~al\mbox{.}(2014)]%
        {TQ_Xiang}
\bibfield{author}{\bibinfo{person}{Guoqing Xiang}, \bibinfo{person}{Xiaodong
  Xie}, \bibinfo{person}{Huizhu Jia}, \bibinfo{person}{Xiaofeng Huang},
  \bibinfo{person}{Janny Liu}, \bibinfo{person}{Wei Kaijin},
  \bibinfo{person}{Yuanchao Bai}, \bibinfo{person}{Pei Liao}, {and}
  \bibinfo{person}{Wen Gao}.} \bibinfo{year}{2014}\natexlab{}.
\newblock \showarticletitle{An adaptive perceptual quantization algorithm based
  on block-level JND for video coding}. In \bibinfo{booktitle}{\emph{Pacific
  Rim Conference on Multimedia (PCM): Advanceds in Multimedia Information
  Processing}}. \bibinfo{pages}{54--63}.
\newblock
\showISBNx{978-3-319-13167-2}


\bibitem[Xiang et~al\mbox{.}(2022)]%
        {RC_Xiang}
\bibfield{author}{\bibinfo{person}{Guoqing Xiang}, \bibinfo{person}{Xinfeng
  Zhang}, \bibinfo{person}{Xiaofeng Huang}, \bibinfo{person}{Fan Yang},
  \bibinfo{person}{Chuang Zhu}, \bibinfo{person}{Huizhu Jia}, {and}
  \bibinfo{person}{Xiaodong Xie}.} \bibinfo{year}{2022}\natexlab{}.
\newblock \showarticletitle{Perceptual quality consistency oriented CTU level
  rate control for HEVC Intra coding}.
\newblock \bibinfo{journal}{\emph{IEEE Transactions on Broadcasting}}
  \bibinfo{volume}{68}, \bibinfo{number}{1} (\bibinfo{year}{2022}),
  \bibinfo{pages}{69--82}.
\newblock


\bibitem[Xu et~al\mbox{.}(2016)]%
        {RC_FreeEnergy}
\bibfield{author}{\bibinfo{person}{Long Xu}, \bibinfo{person}{Weisi Lin},
  \bibinfo{person}{Lin Ma}, \bibinfo{person}{Yongbing Zhang},
  \bibinfo{person}{Yuming Fang}, \bibinfo{person}{King~Ngi Ngan},
  \bibinfo{person}{Songnan Li}, {and} \bibinfo{person}{Yihua Yan}.}
  \bibinfo{year}{2016}\natexlab{}.
\newblock \showarticletitle{Free-energy principle inspired video quality metric
  and its use in video coding}.
\newblock \bibinfo{journal}{\emph{IEEE Transactions on Multimedia}}
  \bibinfo{volume}{18}, \bibinfo{number}{4} (\bibinfo{year}{2016}),
  \bibinfo{pages}{590 -- 602}.
\newblock


\bibitem[Xu et~al\mbox{.}(2020)]%
        {VQA_C3DVQA}
\bibfield{author}{\bibinfo{person}{Munan Xu}, \bibinfo{person}{Junming Chen},
  \bibinfo{person}{Haiqiang Wang}, \bibinfo{person}{Shan Liu},
  \bibinfo{person}{Ge Li}, {and} \bibinfo{person}{Zhiqiang Bai}.}
  \bibinfo{year}{2020}\natexlab{}.
\newblock \showarticletitle{C3DVQA: Full-reference video quality assessment
  with 3D convolutional neural network}. In \bibinfo{booktitle}{\emph{IEEE
  International Conference on Acoustics, Speech and Signal Processing
  (ICASSP)}}. \bibinfo{pages}{4447--4451}.
\newblock


\bibitem[Yan et~al\mbox{.}(2020)]%
        {TQ_Yan}
\bibfield{author}{\bibinfo{person}{Yunyao Yan}, \bibinfo{person}{Guoqing
  Xiang}, \bibinfo{person}{Yuan Li}, \bibinfo{person}{Xiaodong Xie},
  \bibinfo{person}{Wei Yan}, {and} \bibinfo{person}{Yungang Bao}.}
  \bibinfo{year}{2020}\natexlab{}.
\newblock \showarticletitle{Spatiotemporal perception aware quantization
  algorithm for video coding}. In \bibinfo{booktitle}{\emph{IEEE International
  Conference on Multimedia and Expo (ICME)}}. \bibinfo{pages}{1--6}.
\newblock


\bibitem[Yang et~al\mbox{.}(2017)]%
        {RDO_Yang17}
\bibfield{author}{\bibinfo{person}{Aisheng Yang}, \bibinfo{person}{Huanqiang
  Zeng}, \bibinfo{person}{Jing Chen}, \bibinfo{person}{Jianqing Zhu}, {and}
  \bibinfo{person}{Cai Canhui}.} \bibinfo{year}{2017}\natexlab{}.
\newblock \showarticletitle{Perceptual feature guided rate distortion
  optimization for high efficiency video coding}.
\newblock \bibinfo{journal}{\emph{Multidimensional Systems and Signal
  Processing}} \bibinfo{volume}{28}, \bibinfo{number}{4}
  (\bibinfo{year}{2017}), \bibinfo{pages}{1249--1266}.
\newblock


\bibitem[Yang et~al\mbox{.}(2016)]%
        {RC_Yang}
\bibfield{author}{\bibinfo{person}{Aisheng Yang}, \bibinfo{person}{Huanqiang
  Zeng}, \bibinfo{person}{Lin Ma}, \bibinfo{person}{Jing Chen},
  \bibinfo{person}{Canhui Cai}, {and} \bibinfo{person}{Kai-Kuang Ma}.}
  \bibinfo{year}{2016}\natexlab{}.
\newblock \showarticletitle{A perceptual-based rate control for HEVC}. In
  \bibinfo{booktitle}{\emph{2016 Sixth International Conference on Image
  Processing Theory, Tools and Applications (IPTA)}}. \bibinfo{pages}{1--5}.
\newblock


\bibitem[Yang et~al\mbox{.}(2020)]%
        {TQ_Yang2020}
\bibfield{author}{\bibinfo{person}{Kun Yang}, \bibinfo{person}{Dong Liu}, {and}
  \bibinfo{person}{Feng Wu}.} \bibinfo{year}{2020}\natexlab{}.
\newblock \showarticletitle{Deep learning-based nonlinear transform for HEVC
  intra coding}. In \bibinfo{booktitle}{\emph{2020 IEEE International
  Conference on Visual Communications and Image Processing (VCIP)}}.
  \bibinfo{pages}{387--390}.
\newblock


\bibitem[Yang et~al\mbox{.}(2019)]%
        {EN_Yang_CSVT}
\bibfield{author}{\bibinfo{person}{Ren Yang}, \bibinfo{person}{Mai Xu},
  \bibinfo{person}{Tie Liu}, \bibinfo{person}{Zulin Wang}, {and}
  \bibinfo{person}{Zhenyu Guan}.} \bibinfo{year}{2019}\natexlab{}.
\newblock \showarticletitle{Enhancing quality for HEVC compressed videos}.
\newblock \bibinfo{journal}{\emph{IEEE Transactions on Circuits and Systems for
  Video Technology}} \bibinfo{volume}{29}, \bibinfo{number}{7}
  (\bibinfo{year}{2019}), \bibinfo{pages}{2039--2054}.
\newblock


\bibitem[Yeo et~al\mbox{.}(2013)]%
        {RDO_Yeo}
\bibfield{author}{\bibinfo{person}{Chuohao Yeo}, \bibinfo{person}{Hui~Li Tan},
  {and} \bibinfo{person}{Yih~Han Tan}.} \bibinfo{year}{2013}\natexlab{}.
\newblock \showarticletitle{On rate distortion optimization using SSIM}.
\newblock \bibinfo{journal}{\emph{IEEE Transactions on Circuits and Systems for
  Video Technology}} \bibinfo{volume}{23}, \bibinfo{number}{7}
  (\bibinfo{year}{2013}), \bibinfo{pages}{1170--1181}.
\newblock


\bibitem[Yuan et~al\mbox{.}(2019)]%
        {SV_PQA_VC}
\bibfield{author}{\bibinfo{person}{Di Yuan}, \bibinfo{person}{Tiesong Zhao},
  \bibinfo{person}{Yiwen Xu}, \bibinfo{person}{Hong Xue}, {and}
  \bibinfo{person}{Liqun Lin}.} \bibinfo{year}{2019}\natexlab{}.
\newblock \showarticletitle{Visual JND: a perceptual measurement in video
  coding}.
\newblock \bibinfo{journal}{\emph{IEEE Access}}  \bibinfo{volume}{7}
  (\bibinfo{year}{2019}), \bibinfo{pages}{29014--29022}.
\newblock


\bibitem[Zeng et~al\mbox{.}(2016)]%
        {RC_Zeng}
\bibfield{author}{\bibinfo{person}{Huanqiang Zeng}, \bibinfo{person}{Aisheng
  Yang}, \bibinfo{person}{King~Ngi Ngan}, {and} \bibinfo{person}{Wang
  Miaohui}.} \bibinfo{year}{2016}\natexlab{}.
\newblock \showarticletitle{Perceptual sensitivity-based rate control method
  for high efficiency video coding}.
\newblock \bibinfo{journal}{\emph{Multimedia Tools and Applications}}
  \bibinfo{volume}{75}, \bibinfo{number}{17} (\bibinfo{year}{2016}),
  \bibinfo{pages}{10383 -- 10396}.
\newblock


\bibitem[Zhang and Bull(2016)]%
        {TQ_QP_Zhang}
\bibfield{author}{\bibinfo{person}{Fan Zhang} {and} \bibinfo{person}{David~R.
  Bull}.} \bibinfo{year}{2016}\natexlab{}.
\newblock \showarticletitle{HEVC enhancement using content-based local QP
  selection}. In \bibinfo{booktitle}{\emph{2016 IEEE International Conference
  on Image Processing (ICIP)}}. \bibinfo{pages}{4215--4219}.
\newblock


\bibitem[Zhang et~al\mbox{.}(2019a)]%
        {OV_AVS}
\bibfield{author}{\bibinfo{person}{Jiaqi Zhang}, \bibinfo{person}{Chuanmin
  Jia}, \bibinfo{person}{Meng Lei}, \bibinfo{person}{Shanshe Wang},
  \bibinfo{person}{Siwei Ma}, {and} \bibinfo{person}{Wen Gao}.}
  \bibinfo{year}{2019}\natexlab{a}.
\newblock \showarticletitle{Recent development of AVS video coding standard:
  AVS3}. In \bibinfo{booktitle}{\emph{2019 Picture Coding Symposium (PCS)}}.
  \bibinfo{pages}{1--5}.
\newblock


\bibitem[Zhang et~al\mbox{.}(2017)]%
        {TQ_Zhang}
\bibfield{author}{\bibinfo{person}{Lei Zhang}, \bibinfo{person}{Qiang Peng},
  {and} \bibinfo{person}{Xiao Wu}.} \bibinfo{year}{2017}\natexlab{}.
\newblock \showarticletitle{Perception-based adaptive quantization for
  transform-domain Wyner-Ziv video coding}.
\newblock \bibinfo{journal}{\emph{Multimedia Tools and Applications}}
  \bibinfo{volume}{76} (\bibinfo{date}{08} \bibinfo{year}{2017}),
  \bibinfo{pages}{16699--16725}.
\newblock


\bibitem[Zhang et~al\mbox{.}(2014)]%
        {IQA_VSI}
\bibfield{author}{\bibinfo{person}{Lin Zhang}, \bibinfo{person}{Ying Shen},
  {and} \bibinfo{person}{Hongyu Li}.} \bibinfo{year}{2014}\natexlab{}.
\newblock \showarticletitle{VSI: A visual saliency-induced index for perceptual
  image quality assessment}.
\newblock \bibinfo{journal}{\emph{IEEE Transactions on Image Processing}}
  \bibinfo{volume}{23}, \bibinfo{number}{10} (\bibinfo{year}{2014}),
  \bibinfo{pages}{4270--4281}.
\newblock


\bibitem[Zhang et~al\mbox{.}(2019c)]%
        {CM_DL_StereoVA}
\bibfield{author}{\bibinfo{person}{Qiudan Zhang}, \bibinfo{person}{Xu Wang},
  \bibinfo{person}{Shiqi Wang}, \bibinfo{person}{Shikai Li},
  \bibinfo{person}{Sam Kwong}, {and} \bibinfo{person}{Jianmin Jiang}.}
  \bibinfo{year}{2019}\natexlab{c}.
\newblock \showarticletitle{Learning to explore intrinsic saliency for
  stereoscopic video}. In \bibinfo{booktitle}{\emph{2019 IEEE/CVF Conference on
  Computer Vision and Pattern Recognition (CVPR)}}.
  \bibinfo{pages}{9741--9750}.
\newblock


\bibitem[Zhang et~al\mbox{.}(2019b)]%
        {IQA_Fine}
\bibfield{author}{\bibinfo{person}{Xinfeng Zhang}, \bibinfo{person}{Weisi Lin},
  \bibinfo{person}{Shiqi Wang}, \bibinfo{person}{Jiaying Liu},
  \bibinfo{person}{Siwei Ma}, {and} \bibinfo{person}{Wen Gao}.}
  \bibinfo{year}{2019}\natexlab{b}.
\newblock \showarticletitle{Fine-grained quality assessment for compressed
  images}.
\newblock \bibinfo{journal}{\emph{IEEE Transactions on Image Processing}}
  \bibinfo{volume}{28}, \bibinfo{number}{3} (\bibinfo{year}{2019}),
  \bibinfo{pages}{1163--1175}.
\newblock


\bibitem[Zhang et~al\mbox{.}(2021b)]%
        {IQA_SDS_Zhang20}
\bibfield{author}{\bibinfo{person}{Xiang Zhang}, \bibinfo{person}{Siwei Ma},
  \bibinfo{person}{Shiqi Wang}, \bibinfo{person}{Jian Zhang},
  \bibinfo{person}{Huifang Sun}, {and} \bibinfo{person}{Wen Gao}.}
  \bibinfo{year}{2021}\natexlab{b}.
\newblock \showarticletitle{Divisively normalized sparse coding: toward
  perceptual visual signal representation}.
\newblock \bibinfo{journal}{\emph{IEEE Transactions on Cybernetics}}
  \bibinfo{volume}{51}, \bibinfo{number}{8} (\bibinfo{year}{2021}),
  \bibinfo{pages}{4237--4250}.
\newblock


\bibitem[Zhang et~al\mbox{.}(2020b)]%
        {CM_SUR_SVR}
\bibfield{author}{\bibinfo{person}{Xinfeng Zhang}, \bibinfo{person}{Chao Yang},
  \bibinfo{person}{Haiqiang Wang}, \bibinfo{person}{Wei Xu}, {and}
  \bibinfo{person}{C.-C.~Jay Kuo}.} \bibinfo{year}{2020}\natexlab{b}.
\newblock \showarticletitle{Satisfied-user-ratio modeling for compressed
  video}.
\newblock \bibinfo{journal}{\emph{IEEE Transactions on Image Processing}}
  \bibinfo{volume}{29} (\bibinfo{year}{2020}), \bibinfo{pages}{3777--3789}.
\newblock


\bibitem[Zhang et~al\mbox{.}(2020a)]%
        {SV_MLVC}
\bibfield{author}{\bibinfo{person}{Yun Zhang}, \bibinfo{person}{Sam Kwong},
  {and} \bibinfo{person}{Shiqi Wang}.} \bibinfo{year}{2020}\natexlab{a}.
\newblock \showarticletitle{Machine learning based video coding optimizations:
  a survey}.
\newblock \bibinfo{journal}{\emph{Information Sciences}}  \bibinfo{volume}{506}
  (\bibinfo{year}{2020}), \bibinfo{pages}{395--423}.
\newblock
\showISSN{0020-0255}


\bibitem[Zhang et~al\mbox{.}(2021a)]%
        {CM_DL_JND_Video}
\bibfield{author}{\bibinfo{person}{Yun Zhang}, \bibinfo{person}{Huanhua Liu},
  \bibinfo{person}{You Yang}, \bibinfo{person}{Xiaoping Fan},
  \bibinfo{person}{Sam Kwong}, {and} \bibinfo{person}{C.~C.~Jay Kuo}.}
  \bibinfo{year}{2021}\natexlab{a}.
\newblock \showarticletitle{Deep learning based just noticeable difference and
  perceptual quality prediction models for compressed video}.
\newblock \bibinfo{journal}{\emph{IEEE Transactions on Circuits and Systems for
  Video Technology}} (\bibinfo{year}{2021}).
\newblock


\bibitem[Zhang et~al\mbox{.}(2018)]%
        {EN_Zhang_TIP}
\bibfield{author}{\bibinfo{person}{Yongbing Zhang}, \bibinfo{person}{Tao Shen},
  \bibinfo{person}{Xiangyang Ji}, \bibinfo{person}{Yun Zhang},
  \bibinfo{person}{Ruiqin Xiong}, {and} \bibinfo{person}{Qionghai Dai}.}
  \bibinfo{year}{2018}\natexlab{}.
\newblock \showarticletitle{Residual highway convolutional neural networks for
  in-loop filtering in HEVC}.
\newblock \bibinfo{journal}{\emph{IEEE Transactions on Image Processing}}
  \bibinfo{volume}{27}, \bibinfo{number}{8} (\bibinfo{year}{2018}),
  \bibinfo{pages}{3827--3841}.
\newblock


\bibitem[Zhang et~al\mbox{.}(2016)]%
        {RDO_3DSVQM}
\bibfield{author}{\bibinfo{person}{Yun Zhang}, \bibinfo{person}{Xiaoxiang
  Yang}, \bibinfo{person}{Xiangkai Liu}, \bibinfo{person}{Yongbing Zhang},
  \bibinfo{person}{Gangyi Jiang}, {and} \bibinfo{person}{Sam Kwong}.}
  \bibinfo{year}{2016}\natexlab{}.
\newblock \showarticletitle{High-efficiency 3D depth coding based on perceptual
  quality of synthesized video}.
\newblock \bibinfo{journal}{\emph{IEEE Transactions on Image Processing}}
  \bibinfo{volume}{25}, \bibinfo{number}{12} (\bibinfo{year}{2016}),
  \bibinfo{pages}{5877--5891}.
\newblock


\bibitem[Zhang et~al\mbox{.}(2020c)]%
        {VQA_SR3DVQA}
\bibfield{author}{\bibinfo{person}{Yun Zhang}, \bibinfo{person}{Huan Zhang},
  \bibinfo{person}{Mei Yu}, \bibinfo{person}{Sam Kwong}, {and}
  \bibinfo{person}{Yo-Sung Ho}.} \bibinfo{year}{2020}\natexlab{c}.
\newblock \showarticletitle{Sparse representation-based video quality
  assessment for synthesized 3D videos}.
\newblock \bibinfo{journal}{\emph{IEEE Transactions on Image Processing}}
  \bibinfo{volume}{29} (\bibinfo{year}{2020}), \bibinfo{pages}{509--524}.
\newblock


\bibitem[Zhao et~al\mbox{.}(2017)]%
        {EN_Zhao}
\bibfield{author}{\bibinfo{person}{Chen Zhao}, \bibinfo{person}{Jian Zhang},
  \bibinfo{person}{Siwei Ma}, \bibinfo{person}{Xiaopeng Fan},
  \bibinfo{person}{Yongbing Zhang}, {and} \bibinfo{person}{Wen Gao}.}
  \bibinfo{year}{2017}\natexlab{}.
\newblock \showarticletitle{Reducing image compression artifacts by structural
  sparse representation and quantization constraint prior}.
\newblock \bibinfo{journal}{\emph{IEEE Transactions on Circuits and Systems for
  Video Technology}} \bibinfo{volume}{27}, \bibinfo{number}{10}
  (\bibinfo{year}{2017}), \bibinfo{pages}{2057--2071}.
\newblock


\bibitem[Zhao et~al\mbox{.}(2018)]%
        {TQ_NSST}
\bibfield{author}{\bibinfo{person}{Xin Zhao}, \bibinfo{person}{Jianle Chen},
  \bibinfo{person}{Marta Karczewicz}, \bibinfo{person}{Amir Said}, {and}
  \bibinfo{person}{Vadim Seregin}.} \bibinfo{year}{2018}\natexlab{}.
\newblock \showarticletitle{Joint Separable and Non-Separable Transforms for
  Next-Generation Video Coding}.
\newblock \bibinfo{journal}{\emph{IEEE Transactions on Image Processing}}
  \bibinfo{volume}{27}, \bibinfo{number}{5} (\bibinfo{year}{2018}),
  \bibinfo{pages}{2514--2525}.
\newblock


\bibitem[Zhao et~al\mbox{.}(2016)]%
        {TQ_EMT}
\bibfield{author}{\bibinfo{person}{Xin Zhao}, \bibinfo{person}{Jianle Chen},
  \bibinfo{person}{Marta Karczewicz}, \bibinfo{person}{Li Zhang},
  \bibinfo{person}{Xiang Li}, {and} \bibinfo{person}{Wei-Jung Chien}.}
  \bibinfo{year}{2016}\natexlab{}.
\newblock \showarticletitle{Enhanced Multiple Transform for Video Coding}. In
  \bibinfo{booktitle}{\emph{2016 Data Compression Conference (DCC)}}.
  \bibinfo{pages}{73--82}.
\newblock


\bibitem[Zhao et~al\mbox{.}(2011)]%
        {CM_BJND_Zhao}
\bibfield{author}{\bibinfo{person}{Yin Zhao}, \bibinfo{person}{Zhenzhong Chen},
  \bibinfo{person}{Ce Zhu}, \bibinfo{person}{Yap-Peng Tan}, {and}
  \bibinfo{person}{Lu Yu}.} \bibinfo{year}{2011}\natexlab{}.
\newblock \showarticletitle{Binocular just-noticeable-difference model for
  stereoscopic images}.
\newblock \bibinfo{journal}{\emph{IEEE Signal Processing Letters}}
  \bibinfo{volume}{18}, \bibinfo{number}{1} (\bibinfo{year}{2011}),
  \bibinfo{pages}{19--22}.
\newblock


\bibitem[Zhou et~al\mbox{.}(2020b)]%
        {RC_Zhou_JND}
\bibfield{author}{\bibinfo{person}{Mingliang Zhou}, \bibinfo{person}{Xuekai
  Wei}, \bibinfo{person}{Sam Kwong}, \bibinfo{person}{Weijia Jia}, {and}
  \bibinfo{person}{Bin Fang}.} \bibinfo{year}{2020}\natexlab{b}.
\newblock \showarticletitle{Just noticeable distortion-based perceptual rate
  control in HEVC}.
\newblock \bibinfo{journal}{\emph{IEEE Transactions on Image Processing}}
  \bibinfo{volume}{29} (\bibinfo{year}{2020}), \bibinfo{pages}{7603--7614}.
\newblock


\bibitem[Zhou et~al\mbox{.}(2019)]%
        {RC_SSIM_Zhou}
\bibfield{author}{\bibinfo{person}{Mingliang Zhou}, \bibinfo{person}{Xuekai
  Wei}, \bibinfo{person}{Shiqi Wang}, \bibinfo{person}{Sam Kwong},
  \bibinfo{person}{Chi-Keung Fong}, \bibinfo{person}{Peter H.~W. Wong},
  \bibinfo{person}{Wilson Y.~F. Yuen}, {and} \bibinfo{person}{Wei Gao}.}
  \bibinfo{year}{2019}\natexlab{}.
\newblock \showarticletitle{SSIM-based global optimization for CTU-level rate
  control in HEVC}.
\newblock \bibinfo{journal}{\emph{IEEE Transactions on Multimedia}}
  \bibinfo{volume}{21}, \bibinfo{number}{8} (\bibinfo{year}{2019}),
  \bibinfo{pages}{1921--1933}.
\newblock


\bibitem[Zhou et~al\mbox{.}(2020a)]%
        {IQA_Tucker_TIP20}
\bibfield{author}{\bibinfo{person}{Wei Zhou}, \bibinfo{person}{Likun Shi},
  \bibinfo{person}{Zhibo Chen}, {and} \bibinfo{person}{Jinglin Zhang}.}
  \bibinfo{year}{2020}\natexlab{a}.
\newblock \showarticletitle{Tensor Oriented No-Reference Light Field Image
  Quality Assessment}.
\newblock \bibinfo{journal}{\emph{IEEE Transactions on Image Processing}}
  \bibinfo{volume}{29} (\bibinfo{year}{2020}), \bibinfo{pages}{4070--4084}.
\newblock


\end{thebibliography}


\end{document}